\documentclass[apjl]{emulateapj}
\usepackage{apjfonts}

\begin{document}
\title{Exploring Halo Substructure with Giant Stars 
       X. Extended Dark Matter or Tidal Disruption?: 
       The Case for the Leo~I Dwarf Spheroidal Galaxy}
\author{Sangmo Tony Sohn\altaffilmark{1,2,3,4},
Steven R. Majewski\altaffilmark{1}, 
Ricardo R. Mu\~{n}oz\altaffilmark{1}, 
William E. Kunkel\altaffilmark{5},\\ 
Kathryn V. Johnston\altaffilmark{6},
James C. Ostheimer\altaffilmark{1,3,7},  
Puragra Guhathakurta\altaffilmark{8}, 
Richard J. Patterson\altaffilmark{1,3},\\ 
Michael H. Siegel\altaffilmark{1,3,9},
and Michael C. Cooper\altaffilmark{10}}
\altaffiltext{1}{Department of Astronomy, University of Virginia, P.O. Box 3818, Charlottesville, VA 22903
(tonysohn@srl.caltech.edu; srm4n, rrm8f, jco9w, rjp0i, mhs4p@virginia.edu)}
\altaffiltext{2}{Korea Astronomy and Space Science Institute, 61-1, Hwaam-dong, Yuseong-gu, Daejeon 305-348, Korea (tonysohn@srl.caltech.edu)}
\altaffiltext{3}{Visiting Astronomer, Kitt Peak National Observatory, National Optical Astronomy Observatories (NOAO). NOAO is operated by the Association of Universities for Research in Astronomy, Inc., under cooperative agreement with the National Science Foundation.}
\altaffiltext{4}{Current address: California Institute of Technology, MC 405-47, 1200 East California Boulevard, Pasadena, CA 91125 (tonysohn@srl.caltech.edu)}
\altaffiltext{5}{Las Campanas Observatory, Casilla 601, La Serena, Chile (kunkel@lcoeps1.lco.cl)}
\altaffiltext{6}{Van Vleck Observatory, Wesleyan University, Middletown, CT 06459;
current address: Department of Astronomy, Columbia University, New York, New York 10027 
(kvj@astro.columbia.edu)}
\altaffiltext{7}{Current address: 1810 Kalorama Road NW, A3, Washington, DC 20009
(jostheim@alumni.virginia.edu)}
\altaffiltext{8}{UCO/Lick Observatory, University of California at Santa Cruz, Santa Cruz, CA 95064
(raja@ucolick.org)}
\altaffiltext{9}{Current address: Department of Astronomy, University of Texas, Austin, TX 78712
(siegel@astro.as.utexas.edu)}
\altaffiltext{10}{Department of Astronomy, University of California, Berkeley, CA 94720-3411
(cooper@astron.berkeley.edu)}

\begin{abstract}

We present a wide-field (4.5 deg$^2$) photometric and spectroscopic 
survey of the Leo~I dwarf spheroidal (dSph) galaxy to explore its 
extended morphology and dynamics.  As in previous papers in this 
series, we take advantage of photometry in the $M$, $T_{2}$, and 
$DDO51$ filter system to select Leo~I red giant branch star 
candidates, and, so far, this selection technique has proven 100\% 
reliable in selecting actual Leo~I members among more than 100 
$M < 21.5$ Leo I giant candidates having previous or new Keck DEIMOS
spectroscopy to a radius $>1.3$ times the limiting radius of the 
fitted, central King profile.  The two-dimensional distribution of all 
similarly-selected Leo~I giant candidates is well fitted by a central 
single-component King profile of limiting radius 13.3 arcmin, but 
many giant stars are found outside this newly derived King limiting 
radius.  The density profile thus shows a break at a major axis 
radial distance of $\sim 10$ arcmin produced by an excess of stars at 
and beyond the King limiting radius (spectroscopically confirmed to be 
made of true Leo~ I members), and primarily along the major axis of 
the main body of the rather elongated satellite.  This spatial 
configuration, a rather flat velocity dispersion profile and an 
asymmetric radial velocity (RV) distribution among the Leo~I members 
at large radii together support a picture where Leo~I has been 
tidally disrupted on at least one, but at most two, perigalactic 
passages of a massive Local Group member.
We demonstrate this hypothesis using mass-follows-light, $N$-body 
simulations of satellites in a Milky Way-like potential that reproduce 
the observed structural and dynamical properties of Leo~I remarkably well.
These models include $\sim3\times10^{7}$ solar mass, tidally disrupting 
dSph systems on bound orbits with rather high eccentricity (0.93--0.96) 
and small perigalactica (10-15 kpc).  The simulations allow the first 
observationally constrained orbit for Leo~I {\it without the 
measurement of its proper motion} and show that the observed RV 
distribution is more consistent with a two Milky Way orbit history 
for the satellite while ruling out a Leo~I orbit that includes a 
previous association with M31 within the last 10 Gyr.
Given the overall success of tidally disrupting mass-follows-light 
satellite models to account for the observed properties of Leo~I, we 
conclude that there is no need to invoke an extended dark matter 
halo around the satellite (e.g., as one explanation of the velocity 
dispersion and radial profiles at large radii), and that an overall 
modest $M/L$ for the satellite is consistent with the available data.
That a satellite on such a large (apogalacticon of $\sim450$ kpc),
long period ($P \sim 6$ Gyr) orbit as Leo~I can
experience tidal disruption suggests that similarly structured 
satellites with even smaller (eccentric) orbits will
experience even greater tidally-induced mass loss rates.

\end{abstract}


\section{Introduction}

\subsection{Motivations for a New Study of the Leo~I System}

Modern theories of the evolution of structure in the Universe that 
include cold dark matter (CDM) grow galaxies and clusters of galaxies 
and their dark halos through the accumulation of smaller subunits 
\citep[e.g.,][]{White78,Davis85,Navarro96,Navarro97,Moore99}.  
But the notion that the Milky Way (MW) halo was built up by 
protracted infall of ``protogalactic fragments'' after an initial 
central collapse had already been established by \citet{Searle78} 
based purely on stellar population arguments.  In CDM scenarios 
dwarf satellite galaxies represent the visible parts of (predominantly 
dark) subhalos; however, the number of dwarf galaxies discovered so 
far is several orders of magnitude less than the predicted number of 
subhalos made by CDM simulations \citep{Kauffman93,Klypin99,Moore99}.
This may be an indication that the majority of (especially smaller) 
subhalos have either not formed stars \citep{Bullock01} or have been 
destroyed \citep{Hayashi03}, and that the visible satellites of today
represent the high mass end of the mass spectrum 
of DM subhalos \citep{Stoehr02,Hayashi03} or those that 
were able to accrete substantial amounts of gas before reionization
\citep*{Bullock2000, Benson2002, Taylor2004}.  

On the other hand, dwarf galaxies exhibit some properties that may 
be inconsistent with the expected properties of CDM subhalos.  
For example, the flat central density profiles of dwarf spheroidal
(dSph) galaxies are at odds with the cuspy interiors predicted by CDM 
\citep*[e.g.,][]{Navarro96, Navarro97, Moore1998}, although this 
might be reconciled by appealing to triaxial halos \citep{Lokas2002, 
Navarro2004}.  Alternatively, warm dark matter (WDM) allows smaller 
central phase-space densities, and studies of dSph systems have 
provided important constraints on the properties of WDM species 
\citep{Lin1983, Gerhard1992, Goerdt2006, Strigari2006}.
In addition to the problem of the central concentrations of satellites, 
their apparent alignments around parent halos --- as has been observed in 
the Milky Way and argued to be related to ``dynamical families'' of 
satellites   \citep*{Kunkel1979, Lynden-Bell1982,
Majewski1994, FusiPecci1995, Lynden-Bell1995, Palma02} ---
have also been used to question the viability of dwarf galaxies as 
putative subhalos \citep{Kroupa05, Kang2005}.  On the other hand, 
such alignments have also been reported in high resolution N-body, 
hierarchical structure simulations, where subhalos accrete along 
filaments \citep*{Libeskind2005, Zentner2005, Wang2005, Libeskind2007}.

Despite uncertainties over the precise connection of observed 
satellite galaxies to the prevailing theoretical CDM models, it has 
become observationally clear that satellite galaxies \citep[e.g.,][]
{Ibata95,Newberg02,Majewski03b,Yanny03,Crane03,Rocha-Pinto03} 
and even star clusters \citep{Odenkirchen01,Odenkirchen03,Rockosi02,
Lee04,Grillmair2006a,Belokurov2006c,Grillmair2006c} could be significant 
contributors to the {\it luminous} halo of the MW and other galaxies.  
N-body simulations  \citep{Oh95,Piatek95,Johnston95,Johnston99,
Johnston02,Munoz2007} show how a dwarf galaxy can experience tidal 
disruption in its outer parts during close encounters with the central 
potential well of MW-like, parent systems.  Stars that escape the 
system will form tidal tails, like those observed in the Sagittarius 
(Sgr) system \citep[e.g.,][]{Ibata95,Newberg02,Majewski03b,Belokurov2006d}.  
Such extratidal stars can create  ``breaks'' in the projected radial 
star-count profiles of the satellites as the density law transitions from 
a steeply-declining central density law for bound stars to a much more 
gradual decline at radii where unbound stars start to contribute significantly.  
Such ``breaks'' have been observed in several MW dwarf spheroidal 
(dSph) satellites \citep{Eskridge88a,Eskridge88b,Irwin95,Majewski00a,
Majewski03b,Palma03,Wilkinson04,Westfall06,Siegel05,Munoz06}.  
For example, the radial density profiles of nearly every one of the
then-known dSphs of the Milky Way in the thorough
\citet[][hereafter IH95]{Irwin95} study 
(see their Figure 2) show an excess of stars with respect 
to the outer parts of their best-fitting model King profiles.  
In the particular case of Leo~I, because of its extreme 
distance, the number of dSph stars relative to the background level 
in the outermost regions was too low to judge conclusively the 
significance of its apparent break and limited the ability of IH95 to 
explore the Leo~I profile significantly past the King limiting radius 
with their photographic star counts.  Undoubtedly, a radial surface 
density profile with better signal-to-noise and spatial coverage will 
help verify whether this extremely distant Galactic companion 
shows the break profile trait seen in other MW dSphs.  Determining
the structure of Leo I to large radii is a primary goal of the present study.  

A second goal is to increase our understanding of the internal dynamics
of Leo I.  If dSph galaxies are the visible counterparts of the largest 
DM lumps \citep[e.g.][]{Stoehr02}, one might expect the process of tidal 
disruption to be inhibited or lessened in these massive subhalos. 
On the other hand, if dSphs can be proven to have tidal tails, the latter 
would place constraints on the dark matter components within these objects 
\citep{Moore1996}.  The Sgr dSph provides a vivid example of a tidally 
disrupting system for which the degree of disruption, as measured by the 
nature of the tails, has been used to constrain its dark matter content 
\citep{Law05}; on the other hand, the Sgr system is sometimes considered 
to be an exception to the norm for dSph galaxies \citep{Mateo98, MateoARAA98}.  
That Sgr may {\it not} be an exceptional case is suggested by studies of 
possible tidal disruption in other MW dSphs \citep{Gould92,Kuhn96,
Majewski00a,Majewski06,Gomez-Flechoso03,Palma03,Walcher03,Westfall06,
Munoz05,Munoz06,Munoz2007}.

The effects of tidal disruption can be inferred not only by the existence of 
``break'' in the density profiles of these other MW dSphs, but also, more 
recently, by the velocity characteristics of the dSphs at large radii. 
The past decade has seen substantial progress in measuring radial velocities 
for large numbers of stars in at least the more nearby MW dSph galaxies
\citep{Armandroff1995, Tolstoy2004, Wilkinson04, Munoz05, Munoz06, Walker2006}.
These studies not only imply large central mass-to-light ratios in many of the
satellites, but generally reveal relatively flat velocity dispersion profiles 
reaching into the break profile regions \citep{Munoz05,Munoz06}, a phenomenon
reproducible in N-body simulations of disrupting, {\it mass-follows-light} 
satellites \citep{Munoz07}.  

However, a tidal disruption interpretation of these velocity dispersion profiles 
to large radii is not unique based on the extant data, because flat velocity 
dispersion profiles  may also be accommodated to arbitrary radii by extending 
the DM halos in which the luminous dSphs are embedded.  These ``equilibrium models'' 
imply mass-to-light ratios that increase with radius and boost the global ratio 
--- in some cases quite substantially \citep[e.g.,][]{Kleyna02}.  Knowing whether 
mass follows light in dSphs or if their luminous components lie within extended 
dark halos is critical to establishing the regulatory mechanisms that have 
inhibited the formation of galaxies in all subhalos and the extent to which 
luminous satellites are vulnerable to disruption and populate the stellar halos 
of $L*$ galaxies.  

Leo~I provides an interesting contrasting case to other dSphs in that (1) this 
particular satellite clearly has an unusual orbit that has protected it from but 
a few potential tidal encounters, and (2) unlike the apparently high $M/L$ systems 
(like Ursa Minor, Draco and Carina), the previously measured mass-to-light ratio 
of Leo~I is only $\sim 6 M_{\sun}/L_{\sun, V}$, i.e. at the low end of the dSph 
$M/L$ scale.  Of course, it is of interest to know whether and how these two 
traits of this satellite may relate, and whether the properties of Leo~I, if 
explored more extensively, may lend new insights into the issues raised above.  
For example, as we shall show in this paper, new signatures found in the velocity 
distribution of Leo~I stars at large radii (namely, an asymmetry) may provide a 
way to break the degeneracy in interpretations of the velocity dispersion 
profile for at least some dSphs.

The heart of our study is a new photometric survey covering 4.5 deg$^{2}$ of 
the sky centered on Leo~I, and new Keck spectroscopy of 105 more Leo~I giant 
stars than available in the literature.  As with other contributions in this 
series, we adopt a technique based on multicolor Washington filter imaging, 
including the use of the $DDO51$ filter centered on the gravity-sensitive 
MgH+Mgb triplet spectral feature near 5150 \AA\ to identify giant stars 
associated with the Leo~I dSph.  We use bright K giants of Leo~I as a 
tracer population of the structure of the dSph because (1) they are the most 
easily detected type of stars over large areas with the use of a mid-size 
telescope, (2) the Washington+$DDO51$ separation technique has already been 
proven successful in the discrimination of metal-poor dSph giant stars from 
foreground metal-rich field dwarf stars and thereby to significantly increase 
the signal-to-noise of the faint, diffuse outer parts of Local Group dSphs 
\citep[e.g.,][]{Majewski00a,Palma03,Munoz05,Munoz06}, and (3) these stars 
are accessible to spectroscopic membership and dynamical follow-up with 
the currently largest telescopes.

Because we are able to explore the Leo~I dSph to large radii with much better 
signal-to-noise than previous studies, we can rederive the structural parameters 
for Leo~I (\S 4.1), which are important to assessing the $M/L$ using the 
traditional, \citet{King62} methodology based on the internal dynamics 
of the system.  A newly derived $M/L$ for Leo~I in this way is 
discussed in \S 4.2. We also explore the possible mass distribution for Leo I
within the context of a tidal disruption scenario using $N$-body simulations of 
Leo I--like satellites orbiting a MW--like galaxy in \S 6.2. 

The unusually high radial velocity of Leo~I at its extreme distance 
naturally leads to interesting questions about its specific orbit.  There have 
been two studies that are in disagreement about potential orbits for 
Leo~I.  On one hand, \citet{Byrd94} conclude that Leo~I was once 
loosely bound to M31 and now is in an unbound, hyperbolic orbit about 
the MW.  They estimate Leo~I's MW perigalactic distance to be 
$\sim 70$ kpc and for this nearest approach to have occurred 2-4 Gyr ago. 
On the other hand, in an earlier study, Z89 concluded that Leo~I 
probably did not originate in the M31 system and that {\it the most 
reasonable assumption is that Leo~I is bound to the MW}.  An unexpected
benefit of our study is that we have been able to derive 
new constraints on the Leo~I orbit from detailed 
study of its structure and dynamics (\S6.2), and we find reasonable 
agreement of these results to inferences about Leo~I's  orbit based 
on its star formation history.

Dwarf spheroidal galaxies, with other halo objects such as globular 
clusters and field stars, are also useful test particles for probing 
the large scale mass distribution and total mass of our Galaxy. 
Despite much work, the spatial extent and total mass of the MW remain 
among the more poorly established Galactic parameters.  
While traditionally thought to be a ``lesser sibling'' to the 
Andromeda galaxy (M31) in terms of mass, some recent work \citep{Cote00, 
Evans00, Evans00b, Geehan06, Seigar07} 
suggests that, in fact, the MW edges out M31 as the most massive 
galaxy in the Local Group.  With the traditional analyses, the mass 
estimate is sensitive to the inclusion or exclusion of one particular 
object, Leo~I because of its combined unusually large Galactocentric 
radial velocity ($+177 \pm 3$ km/s; \citealp[hereafter Z89]
{Zaritsky89}) and great distance ($257\pm 8$ kpc, see \S 1.2).  
Estimates of the Galactic mass by Z89 varied by a factor of 3--4 
depending on whether Leo~I is considered as bound to the MW or not.  
While recent studies by \citet{Wilkinson99}, \citet{Sakamoto03} have 
decreased the dependency of MW mass estimates on including Leo~I by 
using Bayesian likelihood methods and larger samples of halo objects, 
Leo~I is still considered to be a determining factor for fine tuning 
the results.

\subsection{Previous Photometric Studies of Leo~I}

The Leo~I dSph was discovered more than half a century ago by 
\citet{Harrington50} during the first Palomar sky survey.  
Due to its extreme distance and its angular proximity to the 1st-magnitude 
foreground star Regulus, photometric studies of Leo~I 
have been difficult.  Not until CCD arrays were developed were the 
first color-magnitude diagrams (CMD) of Leo~I constructed. 
\citet{Fox87}, \citet{Reid91}, \citet{Lee93}, and \citet{Demers94} 
presented early ground-based CMDs.  Later, \citet{Caputo99} and 
\citet{Gallart99a} presented Leo~I CMDs based on data taken with the 
{\it Hubble Space Telescope} ({\it HST}) and they reach the main-
sequence turnoff (MSTO) of the oldest ($> 10$ Gyr) Leo~I populations.
These deep CMDs allowed detailed studies of the multiple stellar 
populations and the complex star formation history of Leo~I 
\citep{Gallart99b}.  Leo~I was thought to be unique among the MW 
satellite dSphs for not having a conspicuous horizontal branch (HB) 
population until a 12\arcmin $\times$ 12\arcmin\ ground-based survey 
on Leo~I by \citet{Held00} revealed an extended HB structure in 
its CMD.  More recently, \citet{Held01} discovered more than 70 RR 
Lyrae variables with pulsational properties suggesting an 
intermediate Oosterhoff type similar to other dwarf galaxies in the 
Local Group \citep{Siegel00,Cseresnjes01,Pritzl02}.  The existence 
of both extended blue HB population and RR Lyrae stars suggest that 
Leo~I is in fact similar to other local dSph galaxies in having a 
$> 10$ Gyr population, likely formed in the initial collapse of 
the system.  The extended star formation history of Leo~I, which 
includes this initial starburst, followed by a quiescent phase and a 
new burst of star formation activity starting $\sim 7$ Gyr ago 
\citep{Gallart99b} may be intimately tied to its orbital dynamics, 
since close interactions between galaxies are known to be triggers 
of star formation.

The distance to Leo~I has been derived using various photometric 
methods.  \citet{Lee93} used the tip of red giant branch (TRGB) 
method to derive the Leo~I distance modulus $(m-M)_{0} = 22.18\pm 
0.11$, while \citet{Demers94} used both the apparent magnitudes of 
red clump and of the carbon stars to estimate $(m-M)_{0} = 21.56\pm 
0.25$.  \citet{Held01} used the mean magnitude of RR Lyrae variables 
in Leo~I and derived $(m-M)_{0} = 22.04\pm 0.14$, and more recently, 
\citet{Bellazzini04} provided a new estimate of $(m-M)_{0} = 22.02 
\pm 0.13$ using the TRGB method.  A weighted average of the distance 
modulus derived in the four studies gives $(m-M)_{0} = 22.05\pm 
0.07$, which converts to a distance of $257\pm 8$ kpc.  Throughout 
this study, we adopt these values.

The following sections include a presentation of the data from our 
photometric survey of Leo~I (\S 2), and a description of the 
photometric identification of the Leo~I giant star candidates (\S 3).
The two-dimensional distribution of Leo~I giant star candidates is 
discussed in \S 4, and new morphological parameters for the dSph are 
derived. In \S 5 we present new Keck Observatory spectroscopy of a 
subsample of our Leo~I giant candidates and in \S 6 we discuss the 
implications of our results, making use of new $N$-body simulations
of a tidally disrupting dSph satellite galaxy
that appear to generate similar Leo~I properties to those we have 
observed.  Finally, a summary of our work and conclusions are in \S 7.

\section{Photometric Observations and Data Reduction}

The images used in this study were obtained with the Mayall 4 meter 
Telescope at Kitt Peak National Observatory (KPNO) during the nights 
of UT 1998 November 17 and 2002 May 2-5.  We used the Mosaic I  
8K$\times$8K CCD, which has a pixel scale of $0\farcs26$ per pixel 
resulting in a 36$\times$36 arcmin$^2$~field of view.  The camera is 
an array of eight 2048$\times$4096 CCD chips.  We used the broad-band 
Washington $M$, Harris $I$, and intermediate-band $DDO51$ (hereafter, 
$D51$) filters.\footnote{Hereafter, we denote the $I$ filter as 
$T_{2}$ filter since their response curves are nearly identical 
(see, e.g., Fig. 1 of Lejeune \& Buser 1996 and discussion by Majewski 
et al. 2000a).} Our survey fields were selected to lie predominantly 
along the major axis of Leo~I.  The area to the north of Leo~I was 
sampled but a large part to the south of Leo~I was avoided due to 
Regulus.  All program fields were overlapped by 6\arcmin ~with 
adjacent fields as a check on consistency of the photometry. 
Table \ref{t:obs_log} summarizes the basic information of the CCD 
fields used in this study. 

The raw images were pre-processed using the CCDPROC task in the IRAF 
MSCRED package.\footnote{IRAF (the Image Reduction and Analysis 
Facility) is distributed by the National Optical Astronomy 
Observatories.}  The flat-fielding was done with special care since 
the KPNO $T_{2}$ and $D51$ passband images are affected by a pupil 
image that produces an artifact that increases the amount of 
background light near the center.  The $T_{2}$ and $D51$ images also 
suffer from fringing.  We carefully followed the procedures described 
in \citet{Valdes98} to correct for these effects. Once the 
pre-processing was done, we split each Mosaic image into its eight 
sub-images and performed stellar photometry via the 
DAOPHOT II/ALLSTAR \citep{Stetson87} package.  
A point-spread function (PSF) was constructed using 25-50 bright and 
isolated stars for each sub-image.  The quality of each PSF was 
improved by removing neighboring stars and reconstructing the PSF 
iteratively. PSF magnitudes were derived using UNIX shell scripts 
based on ALLSTAR.  The growth-curve analysis package DAOGROW 
\citep{Stetson90} was then used to correct for the missing light lying 
outside of the PSF tail (the aperture correction).

Measured instrumental magnitudes were calibrated against Geisler 
(1990, 1996) standards that were observed many times in different 
airmass ranges over each observing run.  We fit a transformation 
equation of the form:
\begin{eqnarray}
 MAG - mag = \alpha_{1} + \alpha_{2}X + \alpha_{3}C, 
\end{eqnarray}
where $MAG$ is the Geisler standard magnitude, $mag$ is the 
instrumental magnitude, $X$ is the observed airmass, and $C$ is the 
color index.  We tested for terms in $XC$ and these were found to be 
negligible.  The colors $M - T_{2}$ and $M - D51$ were used in the 
equations derived for the $M$ and $D51$ filters, respectively.  
A color term for the $T_{2}$ filter was also found to be unnecessary.
The RMS of the solutions for the transformation equations were less 
than 0.01 magnitude in all three filters.  For the 1998 observations, 
we used the existing transformations derived for other observations 
on these same nights by \citet{Ostheimer02}.  He notes that on the 
night of UT 1998 November 17 there were minor transparency variations, 
so this night's data are considered non-photometric.  We subsequently 
tied the 1998 photometry to that of May 2002 in the following manner: 
First, multiply-measured stars in the overlapping regions among the 
2002 observations were used to derive and apply frame to frame 
offsets, which in most cases were less than 0.01 magnitudes for all 
three filters. This ensures that all of the 2002 observations share 
the same photometric zero-point.  We then used stars in overlapping 
regions of the 1998 and 2002 observations to calculate the average 
magnitude offsets of each non-photometric frame relative to the 
photometric ones, and the corresponding offsets were applied to all 
objects in each non-photometric frame.  We iterated these steps until 
all of the average offsets among all frames were less than 
0.001 magnitudes.  

We note that images taken with mosaic CCDs such as those used in this 
study may suffer from chip-to-chip sensitivity differences that could make
the color terms be chip--dependent.  \citet{Ostheimer02} found that the 
chip-to-chip color terms of the KPNO Mosaic to be negligible by observing 
the standard fields in Washington and DDO51 filters  on every CCD chip 
in the Mosaic array and by cross-comparing the standard coefficients for 
each chip.

Astrometry of the detected objects was obtained by running the 
TFINDER task in the IRAF FINDER package and using USNO-A V2.0 catalog 
stars \citep{Monet98} as reference.  Since Leo~I is located at a high 
Galactic latitude ($b = 49$\arcdeg), the foreground reddening is not 
significant [according to Schlegel, Finkbeiner, \& Davis 1998, 
$E(B-V) = 0.037$].  It is, however, important to consider the 
variation of reddening within and across our program fields which may 
result in systematic differences in stellar magnitudes and colors.  
Each object in our data set has been corrected for reddening based on 
its Galactic coordinates (converted from the equatorial coordinates) 
and direct reference to the reddening map constructed by 
\citet{Schlegel98}.  The $E(B-V)$ of all objects in our fields range 
from 0.031 to 0.050.  

Our survey covers a large spatial area and our 
photometry is likely to contain a large number of galaxies. 
Bad columns, random cosmic rays, and photoelectron bleeding from 
saturated stars are other possible nonstellar contaminants.  
Nonstellar objects were eliminated from our photometry using the two 
DAOPHOT image quality diagnostic parameters SHARP and $\chi$.
All sources beyond the range $-0.3 <$ SHARP $< 0.3$ were considered 
as objects with extreme non-stellar morphology, and were rejected.
In case of $\chi$\footnote{The ratio of the observed pixel-to-pixel 
scatter from the image profile compared to the expected 
pixel-to-pixel scatter from a model stellar image profile.},
our experiments show that the acceptable range changes as a function 
of radial distance from the center of Leo~I because
crowding near the center affects the stellar profiles.
Therefore, we have elected to use a different $\chi$ selection criteria 
for sources inside a major axis radial distance\footnote{Until we 
derive new structural parameters of Leo I using our photometric 
data, we use those of IH95, {\it i.e.} $r_{t} = 12\farcm 6$, $PA$ = 
79\arcdeg, $e$ = 0.21, for defining the tidal boundary of Leo~I.} of 
8 arcmin ($\chi$$ < 2.5$) and outside this radius ($\chi$$< 1.3$).  

\begin{figure}
\epsscale{1.0}
\plotone{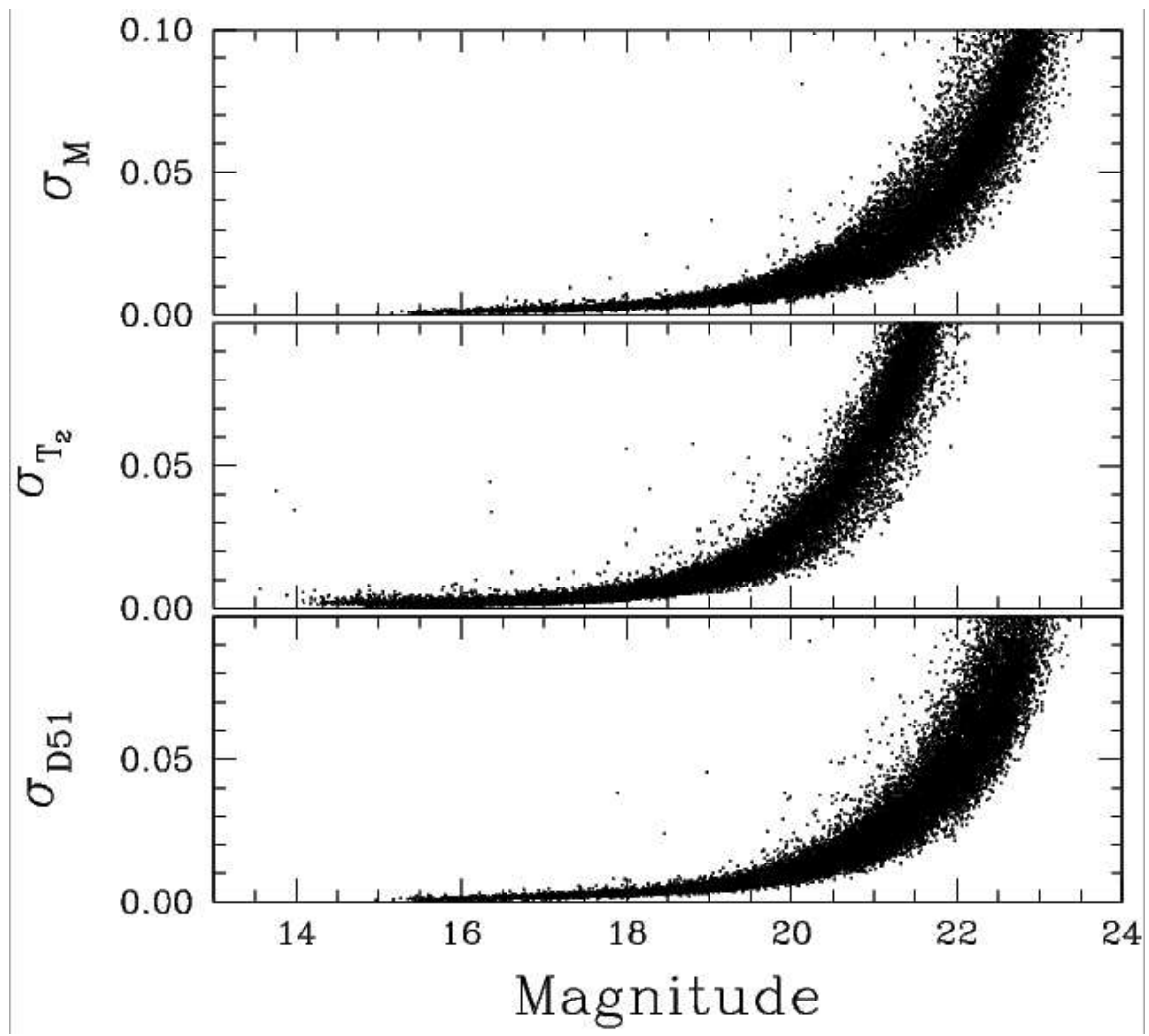}
\caption{ DAOPHOT internal errors for stellar objects in our program fields as 
          functions of magnitudes in each band.
         }
\label{f:magerr}
\end{figure}

In Figure~\ref{f:magerr}, we plot the DAOPHOT internal photometric 
error for the final catalog stars as functions of calibrated magnitudes 
in the three different bands.

\begin{figure}
\epsscale{0.8}
\plotone{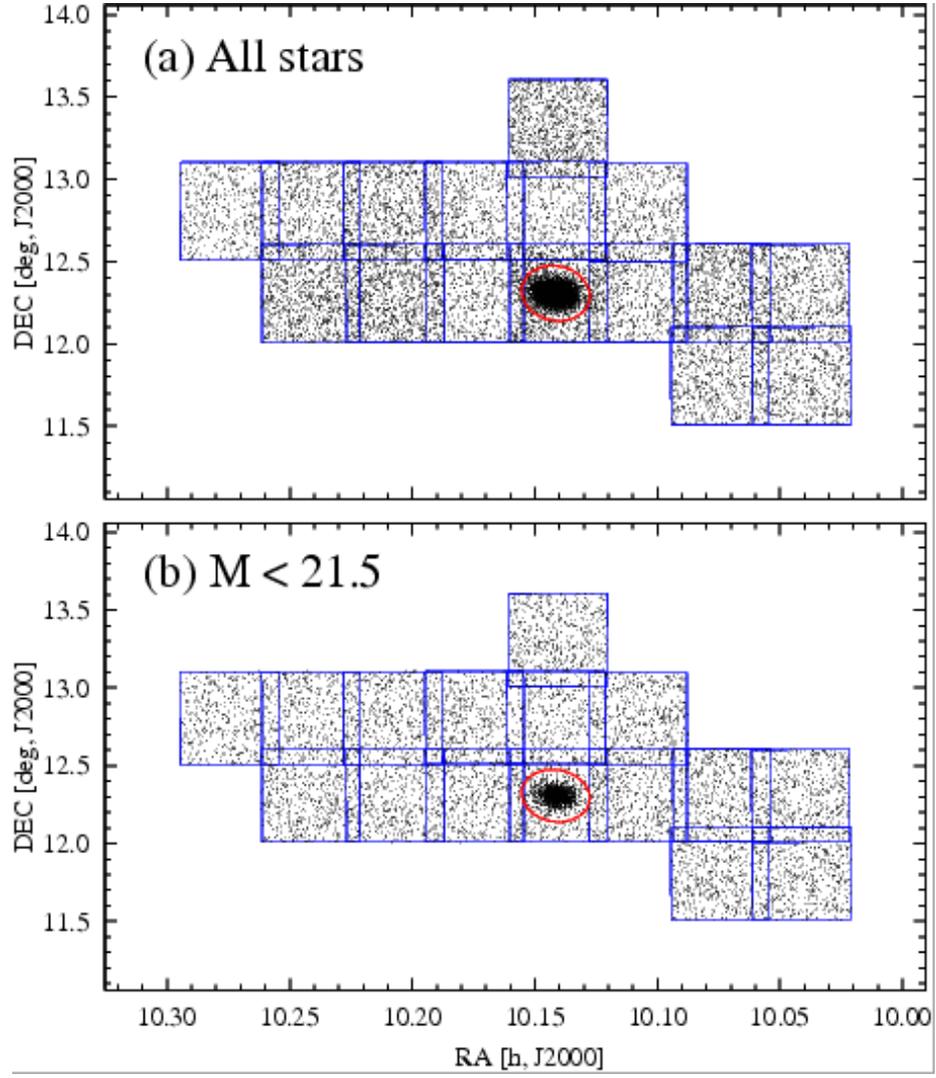}
\caption{ Maps of (a) all stars detected in our survey and (b) stars brighter 
          than $M = 21.5$.  The solid lines give rough indication of the 
          boundaries of each observed field.  The ellipses represent the Leo~I 
          tidal boundary derived by IH95 ($r = 12\farcm 6; PA = 79$\arcdeg; 
          $e = 0.21$) and the asterisk symbols mark the positions of the 1st 
          magnitude star Regulus.  The inhomogeneities in density among the 
          fields of (a) is a reflection of variation in limiting magnitude 
          across our survey area due to different observational conditions. 
          Note that the inhomogeneities are absent in (b).
         }
\label{f:entiremap}
\end{figure}

\begin{figure}
\epsscale{1.0}
\plotone{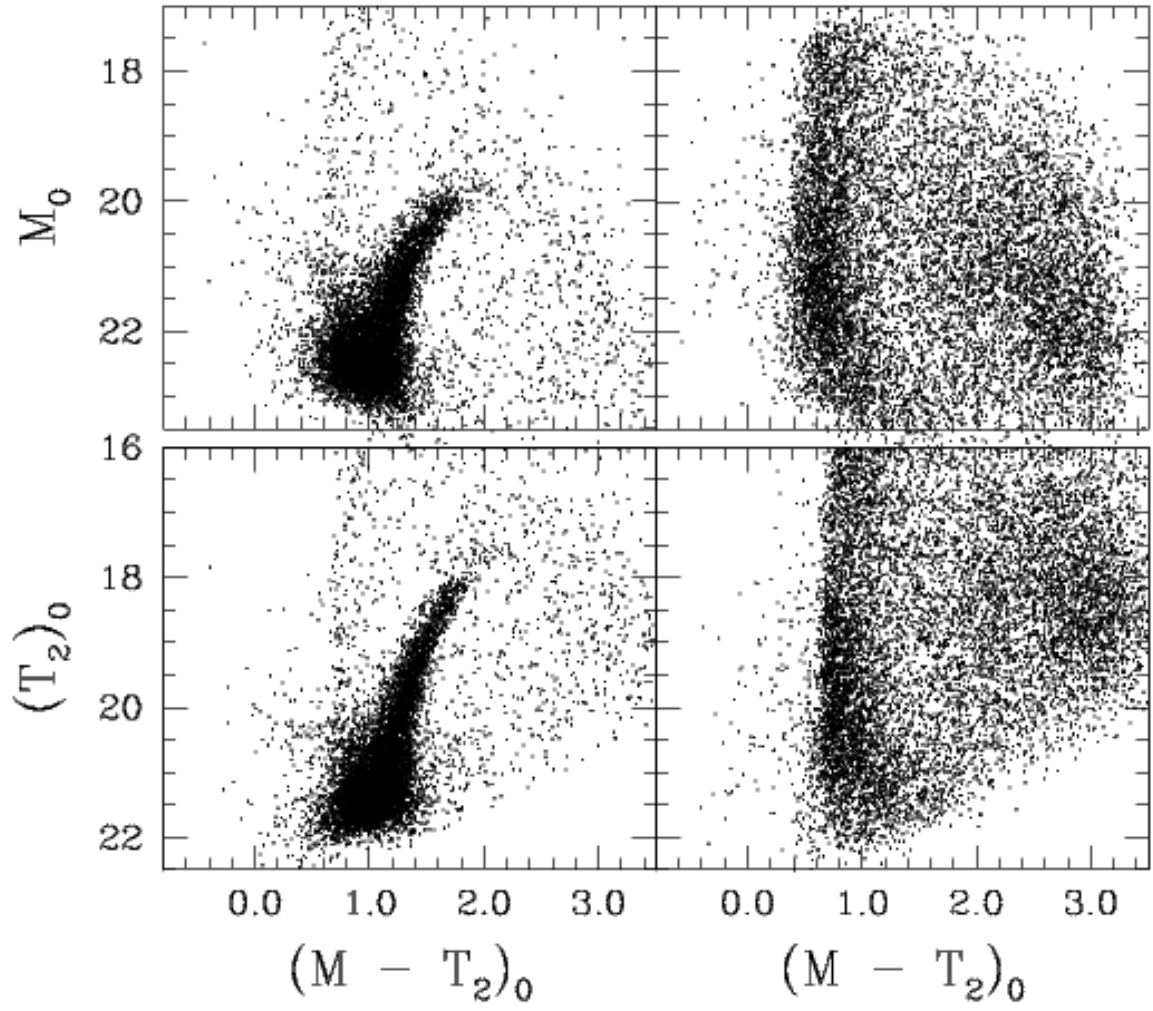}
\caption{ Dereddened $(M - T_{2}, M)_{0}$ and $(M - T_{2}, T_{2})_{0}$ color-magnitude 
          diagrams for stars in C frame ({\it left panels}) and all other 
          frames ({\it right panels}).
         }
\label{f:cmd}
\end{figure}

Figure~\ref{f:entiremap}a shows the sky distribution of all stars 
detected by DAOPHOT in celestial coordinates.  The varying density of 
sources reflects differences in limiting magnitude and seeing among 
our different fields.  Figure~\ref{f:entiremap}b shows the 
distribution of sources to a uniform depth.  In Figure~\ref{f:cmd}, 
we show the $(M-T_{2}, M)_{0}$ and $(M-T_{2}, T_{2})_{0}$ CMDs for 
the stars in Figure~\ref{f:entiremap}a. The left and right panels of 
Figure~\ref{f:cmd} show the CMDs for stars in the central field of 
Leo~I (C field) and in all other fields, respectively.  The dominant 
CMD structure seen in the left panels is the upper part of Leo~I red 
giant branch (RGB).  The prominent Leo~I red giant clump (RGC) is 
present at $21.5 < M_{0} < 23.0$ ($20.0 < T_{2,0} < 22.0$) and $0.6 
< (M-T_{2})_{0} < 1.6$.  A small clump of asymptotic giant branch 
(AGB) stars at $M_{0} \sim 19.5$ ($T_{2,0} \sim 17.5$) and 
$(M-T_{2})_{0} \sim 2.0$ is also apparent.  One other noticeable 
feature in the left panels of Figure~\ref{f:cmd} is the group of 
stars at $0.0 < (M-T_{2})_{0} < 0.7$ and $19.0 < M_{0} < 21.5$ 
($19.5 < T_{2,0} < 21.0$).  These stars are thought to be anomalous 
Cepheids or short period Cepheids of a few hundred Myr age 
\citep{Gallart99a}.  The right panels of Figure 3 are more or less
typical CMDs for a high 
Galactic latitude field containing a mixture of Galactic disk stars, 
halo giants, field HB and blue straggler stars.  The sharp edge 
at $(M-T_{2})_{0} \simeq 0.65$ is due to the main-sequence 
turnoff (MSTO) of field stars smeared in apparent magnitude by the 
range of distances along the line of sight 
\citep[see e.g.,][]{Reid93,Chen01,Siegel02}.  The many stars bluer than 
$(M-T_{2})_{0} = 0.65$ are likely field HB stars and blue 
stragglers.   However, some low density of Leo~I stars of all types may also
lie in the outer fields, swamped by the Milky Way foreground.

\section{Identification of Leo~I Giant Star Candidates}

The methodology we use in this study to select clean samples of Leo~I 
giant star candidates is adopted from that used by \cite{Majewski00b} and
subsequent papers in this series.  
In summary, we photometrically select stars that have (1) magnesium 
(MgH + Mg$b$) band/line strengths consistent with those of giant 
stars, and (2) combinations of surface temperature and apparent 
magnitude consistent with the RGB of Leo~I.  In the following 
sections, we demonstrate the applications of these two criteria using 
our $M$, $T_{2}$, and $D51$ photometry.  To improve the reliability 
of our Leo~I giant star catalog, we restrict our sample to the 
stars that have photometric errors less than 0.04 mag in each band.

\subsection{Giant Star Discrimination in the Color-Color Diagram}

The $D51$ filter measures the strength of the MgH+Mg$b$ spectral 
feature which is a good indicator of stellar surface gravity.  
Since dwarf and giant stars are differentiated by their surface 
gravities, we use the $(M-D51)$ colors  --- where $M$ filter acts 
as a {\it continuum} measure against $D51$ --- combined with the 
surface temperature sensitive $(M-T_{2})$ color\footnote{
\citet{Majewski00a} have shown that the $(M - T_{2})$ has a linear 
relationship with $V - I$ , which is a good surface temperature 
indicator for late type stars.} to separate the foreground dwarf 
stars and distant metal-poor giant stars.  The use of the 
$(M - T_{2}, M - D51)$ color-color diagram as a dwarf/giant 
separating tool has been utilized in several studies 
\citep{Majewski00a,Majewski00b,Morrison00,Palma03,Westfall06,
Munoz05,Munoz06}.

\begin{figure}
\epsscale{1.0}
\plotone{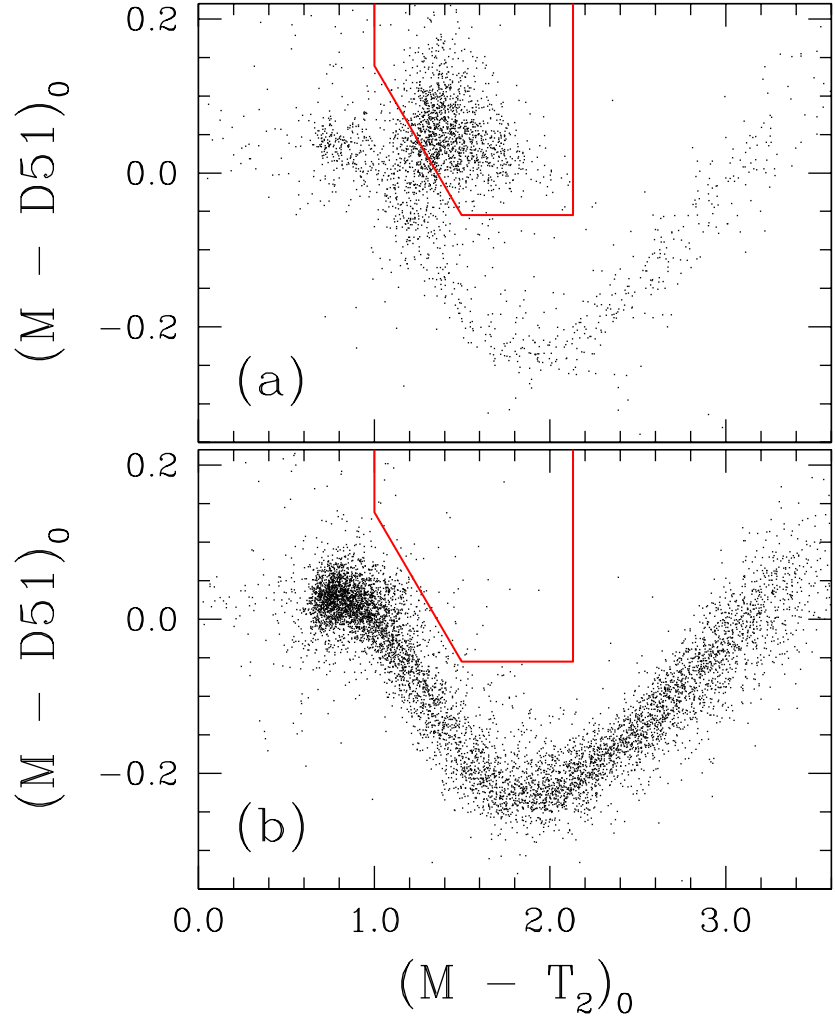}
\caption{ $(M - T_{2}, M - D51)_{0}$ diagrams for stars in (a) C frame and (b) 
          all other frames.  The box drawn with solid lines shows the bounding 
          region we have employed to select metal-poor giant star candidates.
         }
\label{f:ccdselect}
\end{figure}

Figure~\ref{f:ccdselect} shows the color-color diagrams for all 
stars that survived the magnitude error, SHARP, and $\chi$ rejection. 
To better show the difference of the $(M-T_{2}, M-D51)$ distribution 
between two distinct components, {\it i.e.} Leo~I giants and 
foreground dwarf stars, we divide our plot into stars in (a) the C 
field alone and (b) all other fields.  The central part of the top 
panel of Figure~\ref{f:ccdselect} is dominated by Leo~I giant stars 
whereas the lower panel shows mainly the elbow-shaped, high-metallicity 
dwarf star locus and only a few stars in the classic 
``giant'' region of the diagram (see Majewski et al. 2000b).  
The solid bounding box is defined as follows.  The bottom and right 
boundaries were set by using the distribution of Leo~I giant stars 
in the top panel.  The lower-left boundary was drawn roughly 
parallel to the dwarf locus but offset by about +0.1 magnitude in 
$M-D51$ to account for the increased color errors at the faint 
end of our data set.  We note from Figure~\ref{f:ccdselect}a that 
this conservative limit does result in the loss of the bluest Leo~I 
giant stars, but ensures that few dwarf stars will scatter into our 
selection.  We truncate the  upper-left boundary of the box at 
$(M-T_{2})_{0} = 1.0$ since this corresponds to the blue boundary 
of our color-magnitude selection region discussed below.  Stars are 
considered to be giant candidates if they are inside the bounding 
limits in the two-color diagram.  Our color-color selection rejects 
all but the most metal-poor ([Fe/H] $\lesssim -2.0$) foreground dwarf 
stars, whereas most dwarf stars at higher metallicities lie mainly on 
the elbow-shaped locus \citep{Paltoglou94,Majewski00a}.
 
\subsection{Identifying Leo~I Giant Star Candidates in the Color-Magnitude Diagram}

\begin{figure}
\epsscale{1.0}
\plotone{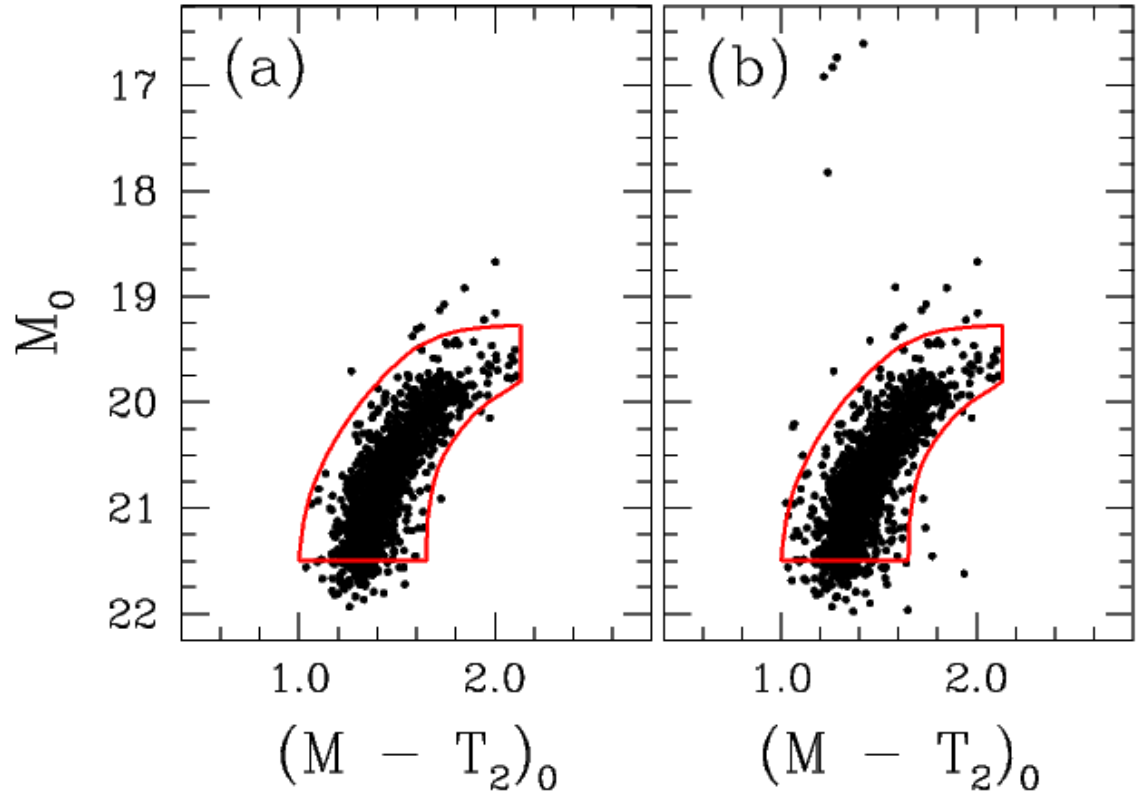}
\caption{ $(M-T_{2}, M)_{0}$ color-magnitude diagram for stars selected as 
          metal-poor giants in Figure~\ref{f:ccdselect} (a) within the tidal 
          boundary derived by IH95 and (b) within our entire survey area.  
          The bounding boxes shown by the solid lines are our CMD selection 
          criteria.
         }
\label{f:cmdselect}
\end{figure}

To select a purer sample of Leo~I giants, we now apply a second, 
color-magnitude criterion to our giant candidate list.  Regardless 
of the angular separation from the core of Leo~I, any giant star in 
Figure~\ref{f:entiremap} associated with Leo~I should have a 
combination of $(M-T_{2})$ and $M$ magnitude that places them within 
the giant branch of Leo~I.  The left panel of Figure~\ref{f:cmdselect} 
shows the CMD for giant candidate stars within the IH95 tidal boundary 
of Leo~I, while the right panel is for all of the giant candidate 
stars.  We use the 
stars in the left panel to delineate the RGB structure in the CMD. 
The boundaries were drawn to roughly follow the overall structure of 
the giant branch.  The box is slightly extended to a point brighter 
and redder than the tip of RGB to include some Leo~I stars on the 
tip of the asymptotic giant branch (AGB).  The lower boundary was set 
at $M_{0} = 21.5$ where the luminosity function of our color-color 
selected stars turns over due to incompleteness.  We note that the 
nine stars brighter than $M_{0} = 21.5$ lying outside of the RGB 
selection region in Figure~\ref{f:cmdselect}a are likely other (asymptotic) 
giant members of Leo~I, but we elect to exclude these few stars in the 
interests of maintaining a conservative selection that excludes 
interloping non-members.  This CMD selection criterion is finally 
applied to the entire sample of giant candidates as shown in 
Figure~\ref{f:cmdselect}b. In this way, a total of 1282 stars are 
selected as Leo~I giant candidates. 

\subsection{Leo~I Giant Stars in Earlier Spectroscopic and Photometric Studies}

It is worthwhile to compare the results of our giant star selection 
criteria against known spectroscopically and photometrically confirmed 
Leo~I members.  M98 obtained spectra of 32 stars within the core radius 
($r_{c}$) of Leo~I and another one just outside of $r_{c}$.  The results 
from their derived RVs show that all 33 of these stars in this densest 
part of Leo~I  are members of the dSph.  There have also been carbon star 
surveys of Leo~I using spectroscopic \citep{Azzopardi85,Azzopardi86} 
and photometric data \citep{Demers02}.  Stars 21 and 24 of M98 sample 
are the same stars listed as carbon stars Azzopardi18 and Azzopardi1, 
respectively in \citet{Azzopardi85,Azzopardi86}.  We have cross-
identified 28 red giant stars of M98, 14 carbon stars of 
\citet{Azzopardi85,Azzopardi86}, and 2 carbon stars of 
\citet{Demers02} with our photometry.  Other stars in the past 
literature that were not cross-identified in our survey are missing 
because they are situated within the physical gaps of the Mosaic CCD 
chips.

\begin{figure}
\epsscale{1.0}
\plotone{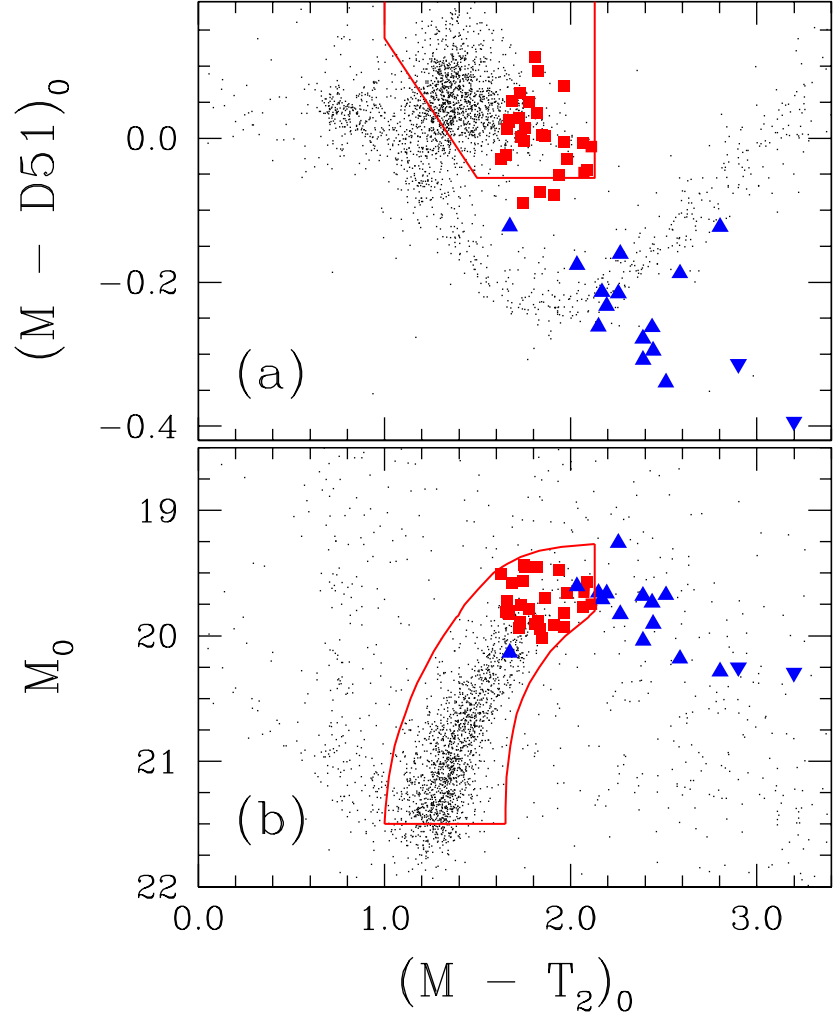}
\caption{ Positions of the previously identified Leo~I giant stars shown along 
          with our catalog of stars with photometric errors less than 0.04 
          in the C field: (a) in the color-color diagram, and (b) in the 
          color-magnitude diagram.  The {\it filled squares} are for red giants 
          and asymptotic giants spectroscopically observed by \citet{Mateo98}.
          The {\it filled triangles} are for carbon stars spectroscopically 
          confirmed by \citet{Azzopardi85,Azzopardi86}, and the {\it inverted 
          filled triangles} are for carbon stars photometrically identified 
          by \citet{Demers02}.
         }
\label{f:crossid}
\end{figure}

Figure~\ref{f:crossid} shows the identified giant ({\it filled 
squares}) and carbon stars ({\it filled triangles}) in the color-
color diagram (top panel) and the CMD (lower panel).  The two reddest 
carbon stars ({\it inverted filled triangles}) are from 
\citet{Demers02}.  It is immediately apparent that the carbon stars 
are not situated within the giant star selection region in the 
color-color diagram.  Instead, these stars extend to the lower right 
part of the diagram indicating that the carbon stars have strong 
absorption features within the $D51$ filter bandwidth.  In fact, 
carbon stars are well known to have a strong C$_2$ Swan band 
absorption feature at a bandhead of $\lambda$ 5165\AA\   
\citep[see e.g., Fig. 1 of ][for a representative spectrum of an 
R-type carbon star]{Christlieb01}, well within the bandwidth of the 
$D51$ filter.  Despite the weak surface gravity of these cool giant 
stars, the strong C$_2$ absorption lines dominate the $M-D51$ color 
and place the carbon stars away from the giant selection regions 
shown in Figure~\ref{f:crossid}.  On the other hand, all but three 
of the normal giant stars in the M98 sample are within the color-
color selection box (Figure~\ref{f:crossid}a).  We could include 
these three stars by extending the lower boundary of our color-color 
selection box, but because it might introduce more contaminants 
(such as metal-poor subdwarfs from the thick disk or halo), we opted 
not to do so.  Our RGB selection in the lower panel includes all M98 
giant stars.  The placement of {\it filled squares} in Figure 
\ref{f:crossid} thus substantiates the reliability of our selection 
technique, though we note that all cross-identified non-carbon giant 
stars of Leo~I lie in the bright end of the giant branch where 
photometric errors are on average only 0.01 magnitude in all three 
bands.  However, in \S 5 we present new spectroscopy showing that 
the catalog maintains its reliability to significantly fainter 
magnitudes. 

\subsection{Spatial Distribution of Leo~I Giant Star Candidates}

Figure~\ref{f:spatialplot2} shows the sky distribution of all stars 
selected as Leo~I RGB stars by our color-color and color-magnitude 
selection methods.  Of particular interest are the 52 stars that lie 
outside the limiting radius found by IH95 and 26 stars outside 
twice the IH95 limiting radius.  The inner part of the 
spatial distribution ({\it lower panel}) shows more clearly the fall 
off in the concentration of Leo~I giant candidates.  Interestingly, 
we find that stars {\it spill over} the IH95 boundary from the main 
body of Leo~I predominantly along the semi-major axis to both sides. 
We discuss the more extended distribution of ``Leo~I giant candidates''
below.

\subsection{Evaluation of Background Level}

The two-criteria Leo~I giant star selection method described above is 
designed to maximize the reliability of our candidate selection.  
However, our final sample may still contain contaminants.  Any halo 
giant or extreme ([Fe/H]$<-2.0$) subdwarf with an observed $(M-T_{2}, M)_{0}$ 
combination similar to our Leo~I RGB stars can be caught in our 
color-color and color-magnitude selection net.  So too can normal 
metallicity dwarfs artificially scattered into the selection from 
photometric errors (these will be main sequence turnoff dwarfs or
subgiants with $M_V \lesssim 6$, so that these will also be 
predominantly from the Milky Way halo).  The number of these 
contaminants can be estimated by a simple technique described in 
\citet{Majewski00b} and \citet{Palma03}: offset the RGB selection 
box to brighter magnitudes and count the stars that fall in the offset box.  
Assuming that the density of halo stars is approximately represented 
by a $R^{-3}$ power law \citep[e.g.,][]{Siegel02}, the number of halo 
stars per unit solid angle is flat with respect to magnitude to first 
order.  Therefore, the 
number of stars within the offset box should remain roughly constant 
as a function of magnitude offset.  The Carina and Ursa Minor fields 
were in compliance with such predictions \citep{Majewski00a,Palma03}.  
However, inspection of Figure~\ref{f:cmdselect}b already suggests 
that the amount of contamination in our case is very low because 
there are very few stars picked as giant candidates that are brighter 
than the RGB selection box.  Nevertheless, we repeated the exercise 
and the results are presented in Table \ref{t:halogiantcount}.  
The four faintest offset bins are still counting a number of Leo~I 
giants including the AGB stars left out by our color-magnitude 
selection box.  At the largest magnitude offsets, the sampling of 
the CMD is incomplete due to the saturation of bright stars in our 
fields.  Therefore, the bins to be considered are $-5.0 < \Delta 
M \le -2.0$.  We take the average of these six bins and estimate 
the background level in our case to be $3.5 \pm 1.9$ for the entire 
field and $3.4 \pm 1.8$ for the field outside an ellipse drawn 
from the IH95 structural parameters, which has an area of 
$\sim 0.11$ deg$^{2}$ (errors have been calculated assuming 
Poissonian statistics).  This converts to a background density of 
$0.8 \pm 0.4$ deg$^{-2}$.  The number of Leo~I giant candidates 
found outside the King limiting radius is twenty times this 
background level.  While our derived background level is considerably 
smaller than was found in the previous studies of the Carina and 
Ursa Minor fields, Leo~I is at a higher Galactic latitude 
and, moreover, we have employed a very conservative 
color-color selection.  For example, if we use a different 
color-color selection box by bringing the lower-left boundary closer 
to the metal-rich dwarf locus than the one used before by $\Delta 
(M-D51) = 0.05$, the background density becomes $12.5 \pm 3.5$ 
deg$^{-2}$.  Typically, the photometric errors are larger at fainter 
magnitudes.  Since fainter Leo~I RGB stars are bluer and closer to 
the dividing line between the giant and dwarf locus in the color-
color diagram, the likelihood of dwarfs contaminating our giant 
selection is greater at fainter magnitudes.  Our very conservative 
selection of Leo~I giant candidates gives us confidence that residual 
contamination will be very low.

The low contamination level of our Leo~I giant candidate list 
is illustrated by the upper panel of 
Figure~\ref{f:spatialplot2}: no Leo~I giant candidates exist in the 
large area east of RA $=$ 10$\fh$22, which suggests that the zero-level 
background is reached within our survey region.  But this begs
the question of the origin of the numerous, very widely separated Leo~I 
giant candidates in the {\it western half} of our survey area.  There
are 26 candidates outside of twice the King limiting radius in
our survey area (all but four of them to the west of Leo~I's center), 
but the background level derived above suggests
that only about 3-4 of these stars should be contaminants.
Until we can obtain spectroscopy of these stars it is difficult
to assess whether they are part of Leo I (as tidal debris), 
they are Milky Way contaminants, or some combination of both.  
Tidal disruption models (e.g. \S6.1.2.) predict that stars released 
during the latest perigalactic passage are mostly spread out along the 
major axis of the satellite, and this could account for at least half
of these spatial outliers.  Moreover, stars that became unbound 
during older perigalactic passages tend to be more irregularly distributed 
around the satellite (see Fig.\ 22).  It is therefore possible that many 
of the Leo I giant candidates located far out from the main body could 
very well be unbound stars of the latter type.  Certainly the large scale 
inhomogeneity of the distribution is highly suggestive of substructure 
of some kind.  Nevertheless, because of the present ambiguous nature of 
these widely separated stars, we have deliberately and conservatively 
elected to confine the remaining analyses in this paper to the central 
square degree around Leo I.

\section{Structure of Leo~I}

In this section, we present the results from profile 
fitting the two-dimensional distribution of Leo~I giant candidates.  
A mass-to-light ratio of the bound population is calculated via the 
new set of structural parameters using the standard techniques of 
\citet{King66} and \citet{Illingworth76}.  We comment on newly 
identified, potential extratidal Leo~I stars and also present 
isodensity contour plots of the Leo~I system.

\subsection{Profile Fitting and Structural Parameters}

Structural parameters for Leo~I were first derived by 
\citet{Hodge63,Hodge71}, and later by IH95.  All three studies used 
King profiles to fit their photographic observations.  We use the 
positions on the sky of our giant candidates to explore the radial 
density profiles.  To do so, we fit the surface density distribution 
of our Leo~I giant candidates using two different models: the 
single-component King model and a power law + core (PLC).  
The fittings were done using a combination of Bayesian and Maximum 
Likelihood techniques similar to that employed by \citet{Kleyna98}.  
Errors for each parameter were estimated using a Bayesian 
approach in conjunction with a Markov Chain technique.  The details 
of the fitting and error-estimating algorithms are described in 
\citet{Ostheimer02}.  A single-component King model has been widely 
used to fit many dSph galaxies ({\it e.g.}, IH95).  The PLC model was 
adopted by \citet{Kleyna98} to fit their Ursa Minor surface density 
profile.  We describe our fitting method below in detail.

\begin{figure}
\epsscale{1.0}
\plotone{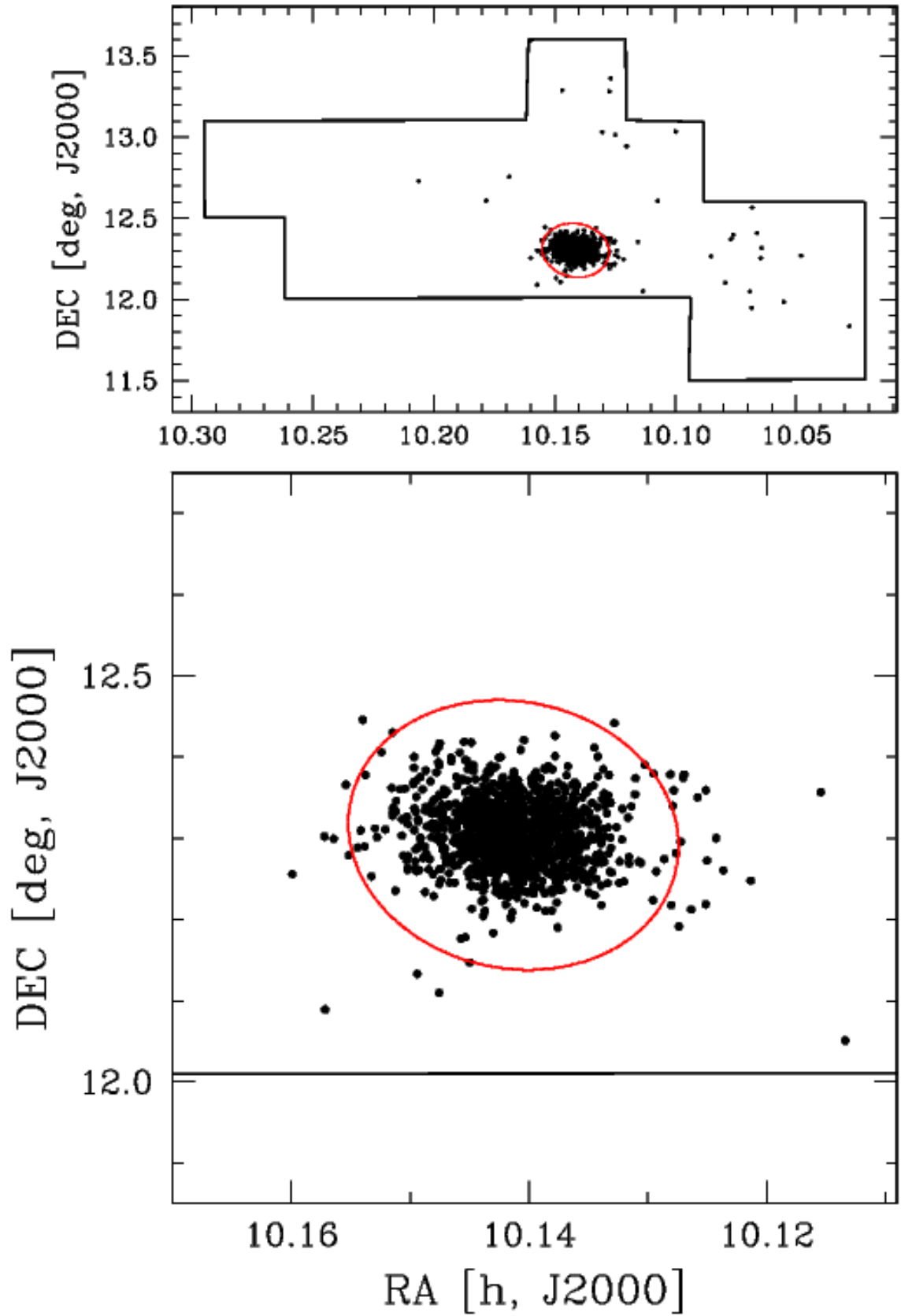}
\caption{ Distribution on the sky of stars selected as ``Leo~I-like'' giants by 
          our color-color and color-magnitude selection methods.  The 
          {\it lower panel} is a blow-up of the central part in the upper panel. 
          The region enclosed by the solid lines is our survey area.  
          The ellipse shows the tidal boundary derived by IH95 and the hatched 
          region in the bottom panel is outside of our survey area.
         }
\label{f:spatialplot2}
\end{figure}

Several dSphs are found to have outer profile ``breaks'' (i.e. slope 
changes) from their central King profile density distributions: 
Fornax and Sculptor \citep{Eskridge88a,Eskridge88b,Westfall06}, 
Carina \citep{Majewski00b,Majewski05,Munoz06}, Ursa Minor \citep{Palma03}, 
and Sgr \citep{Majewski03b}.  Although statistically insignificant, 
IH95 found an excess of stars with respect to their King fit in the 
outer region of Leo~I.  An excess of stars should show up in our 
radial density profile based on the appearance of stars spilled 
over from the main body of Leo~I along the east/west in Figure 
\ref{f:spatialplot2}.  Since a fit to a distribution of stars 
{\it including} the excess of stars in the outer regions will likely 
inflate the derived core and limiting radii, we have taken the 
following steps to find a new King profile fit: First, we limit the 
stars used to derive the structural parameters to those within 
$\sim 13\arcmin$ from the center of Leo~I in the east/west 
direction.  Next, six structural parameters (RA and Dec of the dSph 
center, position angle, core radius, ellipticity, and limiting 
radius) are derived from this stellar sample.  Finally, the background 
level derived in \S 3.5 is combined with the six parameters to 
represent a newly fitted single-component King model.  A similar 
approach was taken in the study of the Sgr profile by 
\citep{Majewski03b}.  The PLC model on the other hand was fit 
using the entire data set in one pass with the background level 
set to the same value used for fitting the King model.

Table \ref{t:structuralparams} lists the best-fit parameters and 
errors for each profile.  The first two lines in the table are for 
the new King and PLC profile fits.  For comparison, we also list 
the King profile parameters derived in earlier Leo~I studies 
by \citet{Hodge63}, \citet{Hodge71}, and IH95.  The newly derived 
center of Leo~I is at $(\alpha,\delta)_{2000} = $ (10$^{\rm h}$ 
08$^{\rm m}$ 28$^{\rm s}$.68, +12\arcdeg ~18\arcmin ~19\farcs 7).

\begin{figure}
\epsscale{1.0}
\plotone{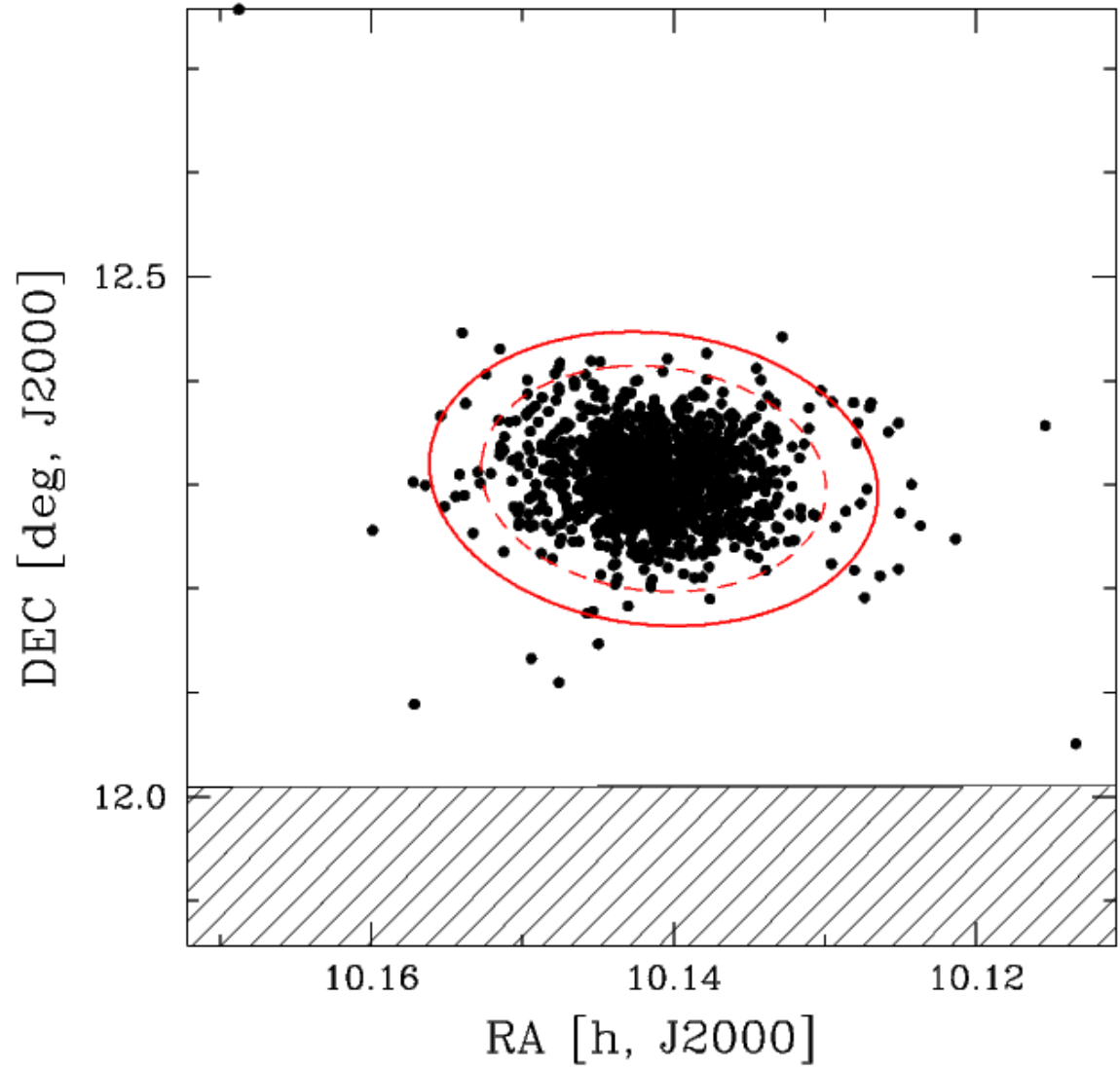}
\caption{ Sky distribution of Leo~I giant candidates with the new King limiting 
          radius ({\it solid line}) derived in this study.  The inner ellipse 
          drawn with a {\it dashed line} corresponds to the point where the 
          radial density profile starts to deviate from a single-component 
          King model (\S 4.1). 
          The hatched region is outside of our survey area.
         }
\label{f:spatialplot_center}
\end{figure}

Our derived core radius and ellipticity for either the PLC or King 
profile fit are larger than those derived by IH95, but are much 
closer to those of \citet{Hodge71}.  Similarly, our 
limiting radius is in better agreement with that of \citet{Hodge71}, 
although IH95's limiting radius is also consistent within the errors. 
To illustrate how our new structural parameters better fit the sky 
distribution of Leo~I giant candidates, the new King limiting radius 
is plotted over the sky distribution of Leo~I giant candidates in 
Figure~\ref{f:spatialplot_center} (compare to IH95 limiting radius 
size and shape in Figure \ref{f:spatialplot2}).  
Figure~\ref{f:spatialplot_center} also shows the ellipse 
corresponding to the break radius ($r_{break}$; see below 
for the determination of the break radius). 

\begin{figure}
\epsscale{0.8}
\plotone{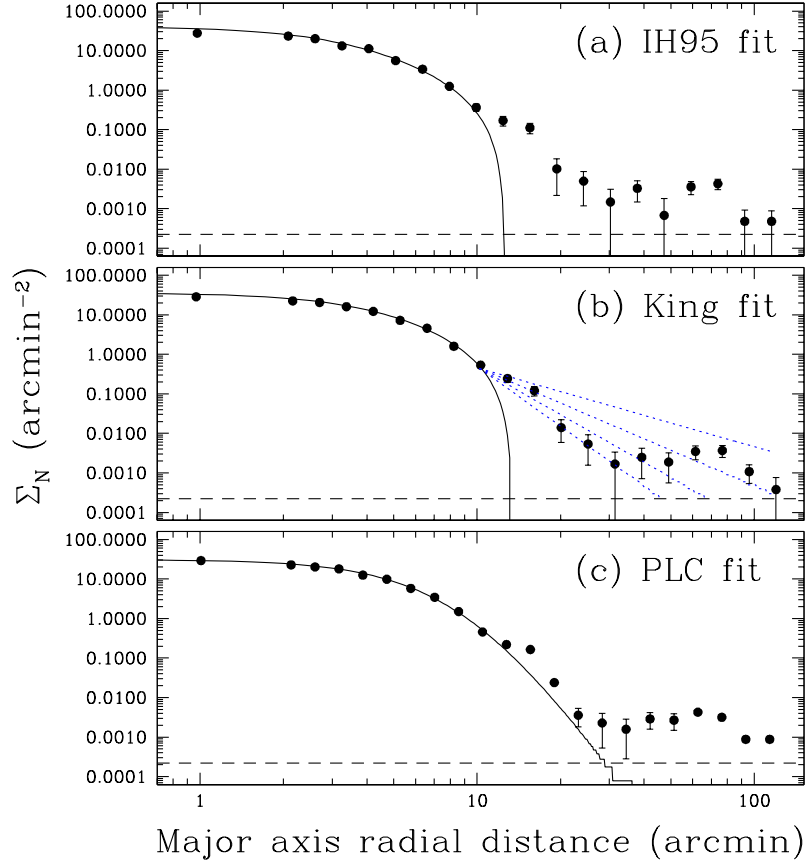}
\caption{ Fits of the stellar distribution of Leo~I using (a) a King model with 
          parameters derived by IH95, (b) our new King model with one power-law, 
          (c) new King model with two power-laws, and (d) a power-law with core 
          (PLC) model.  The horizontal dashed lines are the derived levels of 
          the backgrounds, which were subtracted from the data and the model 
          curves shown.  The data points in each panel 
          have slightly different positions because the ellipticities and 
          position angles derived from the fits require different binning of 
          the data points.  The dotted lines in the two mid panels show power 
          laws of the form $\Sigma_{N} \propto r^{-\gamma}$, where $\gamma = 4.4$ 
          for (b), and $\gamma = 1.7$ (shallower slope) and 10.5 (steeper slope) 
          for (c).  The errors for each point includes both 1$\sigma$ Poissonian 
          and background estimation errors.
         }
\label{f:profiles}
\end{figure}

With the newly derived Leo~I structural parameters, we construct 
radial density profiles by calculating the number density of Leo~I 
giant candidates in elliptical annuli of varying major axis radial 
size.  Although our coverage has small gaps from because of spaces between
the Mosaic CCD chips, our profile 
fitting program interpolates over them when doing any relevant calculation.  
Figure~\ref{f:profiles}a shows the King model with structural 
parameters derived by IH95 overlaid on the giant candidate radial 
profile.  Figures \ref{f:profiles}b show our new King model and 
Figure~\ref{f:profiles}c shows the PLC profile fit. 

We now concentrate on the newly derived King model 
(Figure~\ref{f:profiles}b).  The observed density profile 
clearly departs from the model beyond $r_{maj} \sim 10$ and this 
``break population'' dominates the density out to the entire survey 
range even when uncertainties in both counting 
and background estimation are taken into account.  Such profile 
breaks are also seen in N-body models of tidally disrupting satellite 
galaxies \citep{Johnston99,Johnston02}.  In these models, the density 
profiles of the break populations typically assume power law shapes, 
$\Sigma_{N} \propto r^{-\gamma}$.  In Figure~\ref{f:profiles}b, 
we overplot power laws with $\gamma = $ 2, 3, 4, and 5 in the range 
$r_{maj} > 10$\arcmin.  As seen from the figure, our observed profile 
beyond the break point is not fitted well by a single power law, 
but instead the first three radial bins starting at and including the 
break radius follow a power law with $\gamma = 3$, while the next 
three radial bins follow a power law density trend with $\gamma = 5$.  
Overall, a power law with $\gamma \sim 4$ best describes the surface 
density from the break radius to $r_{maj} \sim 35$\arcmin.  
It is worth noting that even for the PLC fit 
(Figure~\ref{f:profiles}c), the observed density profile starts to 
depart from the model at $r_{maj} \sim 10$\arcmin.  This indicates 
that the existence of the break population does not strongly depend 
on the choice of fitting models.

The surface density profile of Figure~\ref{f:profiles}b roughly 
remains flat in the range 35\arcmin $\la r_{maj} \la$ 80\arcmin, 
and then drops to the background level beyond that.  Such a flat 
density distribution at large radii has not seen in the density profiles 
of other dSph galaxies.  In \S6.1.1. we hypothesize that the complex 
form of the density profile might be attributable to specific patterns 
in the Leo I mass loss history.

\begin{figure}
\epsscale{1.0}
\plotone{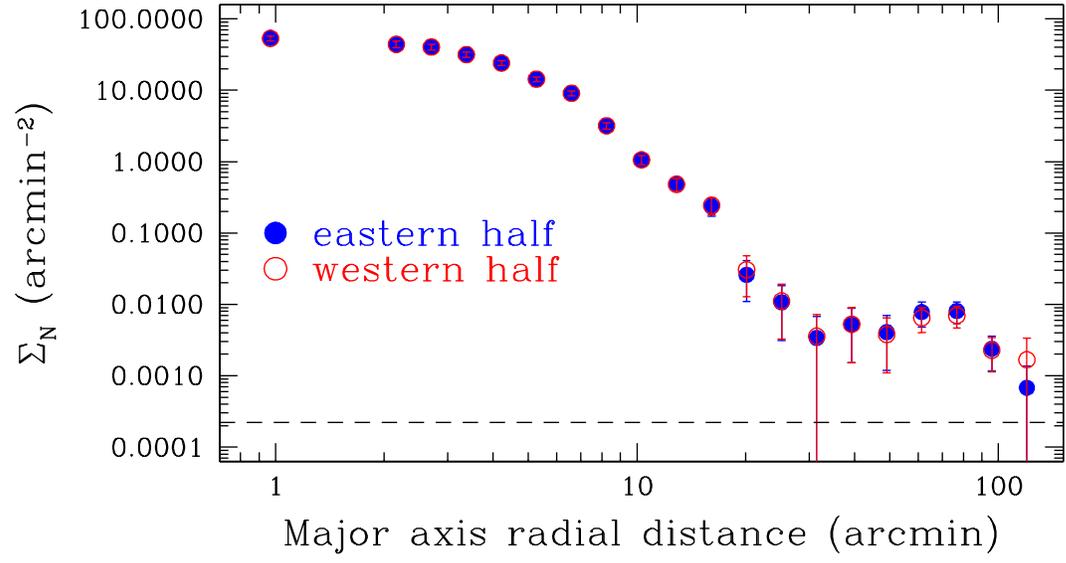}
\caption{ Radial surface density profiles for the eastern and western halves 
           of Leo~I.  The {\it closed} and {\it open circles} are for eastern 
           and western bins, respectively. 
         }
\label{f:profiles_eastwest}
\end{figure}

IH95 found that Fornax, Sculptor, and Ursa Minor show asymmetric 
residual structure after a best-fitting, smooth elliptical profile 
has been subtracted from their two-dimensional distribution.  
Figure~\ref{f:spatialplot_center} gives some impression of an 
East-West inhomogeneity in the distribution of Leo~I stars.  
To check this we plot in Figure~\ref{f:profiles_eastwest} the 
radial density profiles for the east and west halves of the galaxy, 
separately.  In the inner five bins, the western half has slightly 
higher densities, but at low statistical significance.  
The densities are then nearly identical out to the second break.  
For the next four bins after the second break, the western half 
is again higher in density and this corresponds to the excess stars 
to the west, just outside the limiting radius.  Nevertheless, the 
differences are within the Poissonian errors and we conclude that 
the structural difference between the eastern and western halves of 
Leo~I is of low significance.  However, the fact that the breaks 
are seen independently in both east and west samples reinforces 
the reality of these structural features.

\subsection{Mass-To-Light Ratio from Core-Fitting}

If we use the King profile parameters in Table 
\ref{t:structuralparams} to represent the {\it bound}, part of Leo~I, 
the bound mass can be estimated under the usual assumptions of virial 
equilibrium and the standard equation given by \citet{Illingworth76}:
\begin{eqnarray}
M_{\rm tot} = {{166.5 R_{c,g} \mu} \over {\beta}}
\end{eqnarray}
where $R_{c,g}$ is the geometric mean of the King core radius 
($= r_{c}\sqrt{1 - \epsilon}$) in parsecs ($271\pm 26$ pc for Leo~I), 
$\mu$ is the \citet{King66} mass parameter, and $\beta$ is a velocity 
parameter which is related to the observed velocity dispersion. 
Both $\mu$ and $\beta$ are strongly dependent on the concentration of 
a system \citep{King62}.  From the results in Table 
\ref{t:structuralparams}, the concentration of Leo~I is 
$\log(r_{t}/r_{c}) = 0.39\pm 0.04$.  The scaling parameter $\mu$ was 
taken from an extrapolation of Table II of \citet{King66} and $\beta$ 
from Figure 4-11 of \citet{Binney87}.  The central velocity 
dispersion of $\sigma_{0} = 8.8\pm 1.3$ km/s (M98) yields 
$M_{\rm tot} = 4.0(\pm 1.2) \times 10^{7} M_{\sun}$\footnote{Our own 
derived central velocity dispersion of 8.2 km s$^{-1}$ (see \S5.5) 
only lowers the estimated Leo~I mass by 14\%.}.  To obtain the 
corresponding luminosity, we integrate our best-fit model King 
profile for giant star candidates and scale it to the total 
luminosity using a central surface brightness of $\mu_{0, V} = 22.4$ 
mag/arcsec$^{2}$ taken from Table 4 of M98; this yields 
$L_{\rm tot} = 7.6\times 10^{6} L_{\sun}$.  The $M/L$ using this 
technique thus becomes $(M/L)_{tot, V} = 5.3\pm 1.6 
M_{\sun}/L_{\sun ,V}$, where the error is from the uncertainty in 
the mass derived above.  Though we have used different structural 
parameters, we obtain nearly the same $M/L$ as M98.  We note that 
this $M/L$ is actually not that large compared to those typically 
found for dSphs, and is fairly similar to that of dE galaxies and 
globular clusters of similar luminosity.  As has been previously 
found (M98), Leo~I has one of the lowest $M/L$ among the Galactic 
satellites, though one that, by this method, still suggests a 
significant amount of DM.

\subsection{Isodensity Contours}

\begin{figure}
\epsscale{1.0}
\plotone{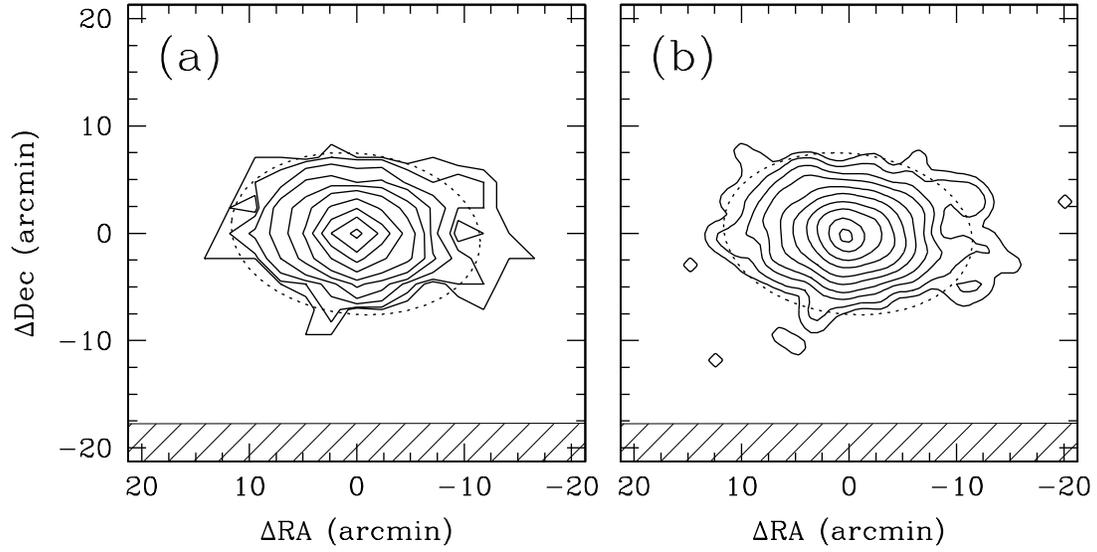}
\caption{ (a) Isodensity contour plots of Leo~I constructed using Leo~I giant 
          candidates.  This figure shows the $43\times 43$ arcmin$^{2}$ region 
          centered on Leo~I.  The newly derived center is at 
          $(\Delta {\rm RA}, \Delta {\rm Dec}) = (0,0)$.  The hatched region 
          is outside of our survey region. The contour levels are 1, 2, 5, 10, 
       	  20, 35, 60, 90, 120, and 160 stars per pixel.
          (b) Smoothed isodensity contour plot of Leo~I using the same 
          technique for making (a) but with each star represented by a  
          6\farcm0 wide two-dimensional Gaussian (see \S 4.3).  
          Each pixel has a dimension of roughly 0\farcm 7$\times$ 0\farcm 7.
          The contour levels are 0.8, 2, 5, 10, 20, 35, 60, 90, 130, and 160
          per pixel.
         }
\label{f:contourplots} 
\end{figure}

In Figure~\ref{f:contourplots}, we present the isodensity contour 
map of Leo~I.  This map was constructed using the Leo~I giant 
candidates in the following manner: The equatorial coordinates of 
each star were first converted to a flat Cartesian system via 
tangential projection centered at the newly derived system center.  
The Cartesian space was divided into a large grid at intervals of
2\farcm 7 in each dimension.  We then counted 
the number of stars in each grid interval to obtain the density for that
part of the map. The gaps between the chips of Mosaic CCD camera are 50 
CCD pixels in the rows and 35 CCD pixels in the columns, corresponding 
to $\sim 13\arcsec$ and $\sim 9\arcsec$ in the sky, respectively. 
In order to check whether the absence of stars in the CCD gaps produce 
any artifacts in Figure \ref{f:contourplots}a, we performed checks on 
our data as follows:  First, we divided the CCD gaps into segments of 
6\arcmin\ bins, and filled each bin with randomly placed artificial 
stars.  The number of stars that went into each bin was calculated by 
taking the average of the number of Leo~I stars on either side of the 
gaps in similar RA/Dec ranges.  Once the gaps were filled, an isodensity 
contour plot was constructed using the exact same procedure as described 
above.  We have also placed artificial gaps with same size as the real ones 
in random places and constructed several contour plots.  Only slight 
differences were found between the isodensity contours constructed in
these ways and Figure~\ref{f:contourplots}a.  This ensures us that 
the influence of the CCD gaps on Figure~\ref{f:contourplots}a is 
negligible.

In order to reduce the granularity from the griding process in our 
isodensity contour, we constructed a grid four times finer than the 
one used above, and replaced each star with a $6.0\times 6.0$ 
arcmin$^{2}$ grid filled with a two-dimensional Gaussian template 
($\sigma = 1.1$ arcmin; amplitude = 1).  We assigned density to each 
pixel by counting up the numbers in a similar manner discussed above. 
The resulting smoothed image is shown in Figure \ref{f:contourplots}b.

Overall, the inner contours of Leo~I are similar to the isopleth map
shown in IH95 (see their Fig.\ 1d).  However, the spill-over of stars 
at the profile break radius is evident in the outer contours of 
Figure~\ref{f:contourplots} as contour extensions
stretching out from the main body along the semi-major axis.  
It is important to note that the spatial density for the outermost 
contour in Figure~\ref{f:contourplots}a is still higher than 600 times 
the  background level derived in Section 3.5.

\section{Spectroscopy of Leo~I Giant Candidates}

\subsection{Observations and Basic Reductions}

We obtained spectra of a total of 135 stars in the Leo~I field; 
49 stars lying just inside and outside the King limiting radius to 
the west of the Leo~I center (Field KD1), and 86 stars near the 
center and along the east major axis of Leo~I (Field KD2).  
The goals of our spectroscopic observations are (1) to check further 
the reliability of our color-color and color-magnitude selections 
discussed in \S 3, (2) to verify the reality of the break population 
seen in the radial density profiles of Leo~I (see Figure 
\ref{f:profiles}), and (3) to study the dynamics of both the inner 
and outer parts of the Leo~I dwarf.

\begin{figure}
\epsscale{1.0}
\plotone{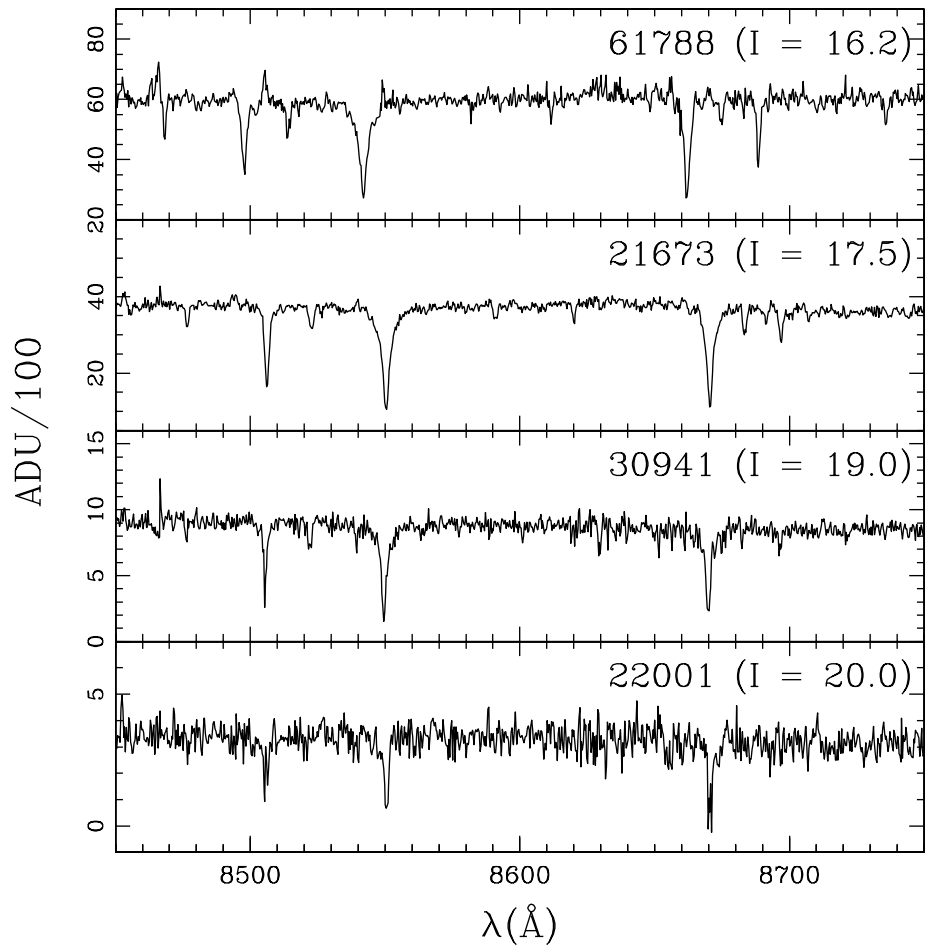}
\caption{ Spectra of four stars observed with the Keck DEIMOS spectrograph.  The
          wavelength region 8450--8750\AA\ shown highlights the Ca triplet lines.
	  The top spectrum is that of our chosen radial velocity template star 
	  (see text), and the bottom three stars are Leo~I giants of different 
	  brightnesses.
        }
\label{f:specsample} 
\end{figure}

The spectra were obtained with observations of two multislit masks 
with the Deep Imaging Multi-Object Spectrograph 
\citep[DEIMOS;][]{Faber03} on 
the nights of UT 2003 October 29 (Field KD1) and 2004 October 15--18 
(Field KD2) using the Keck II 10 m telescope (see Figure 
\ref{f:keckspec_spatialplot} for the placement of our masks on the 
sky with respect to the Leo~I giant candidates).  The masks were 
designed using coordinates from our photometric catalog, which have 
been locked to the astrometric system of the USNO-A V2.0 catalog 
(see \S 2).  Priority for slit selection was given to stars selected 
to be Leo~I giant candidates by the criteria described in \S 3.  
Even for this priority sample, all stars falling within the 
5\arcmin $\times$ 16\arcmin\ slitmask area of Field KD1 could not be 
observed because of slit overlap.  On the other hand, occasional gaps 
in mask area remained and these were filled with slits for, in 
priority order: (1) stars selected as giants, but not as Leo~I 
giants, and (2) any other star in the field without regard to 
classification.  Observations of stars classified as dwarfs prove 
useful as checks on the reliability of the dwarf/giant separation.  
The spectrograph was configured with the 1200 lines/mm grating 
and 1\arcsec slits with the central wavelength set at 7800\AA.  
This instrumental set up provides 0.33\AA /pixel dispersion, a 
spectral resolution FWHM of 1.95\AA\ (68 km s$^{-1}$ at the Ca 
infrared triplet) after accounting for the 0.7$\times$ 
anamorphic magnification factor, and a spectral coverage of 
6500--9100\AA ~(which varies a little from slit to slit 
depending on their particular placement within the mask field of 
view).  The total integration times were 80 minutes for Field KD1 
and 115 minutes for Field KD2, divided into four and five separate 
exposures, respectively, for cosmic ray removal.  
Typical DEIMOS spectra for different magnitude stars are shown in 
Figure~\ref{f:specsample}.
The typical seeing during the Leo~I spectroscopic observations were 
0\farcs8 -- 1\farcs0.  All steps of the data reduction to 
wavelength-calibrated, one-dimensional spectra (including bias 
subtraction, flat fielding, sky subtraction, wavelength solutions, 
and extraction of 1-d spectra) were done using the spec2d reduction 
pipeline (Cooper et al., in prep) developed at the University of 
California at Berkeley for the DEEP2 galaxy redshift survey project.

\subsection{Radial Velocity Determination}

Derivation of the radial velocity (RV) measurements uses a sequence of steps, 
each devised to address some aspect of noise suppression or bias compensation.
Combined, the resulting scatter in the achieved RV measurements
can be shown to be less than 1/40 of the FWHM of a resolution element 
for the best exposed Leo I stars.

The first phase of the RV reduction follows the 
masked, Fourier-filtered cross-correlation process outlined in 
\citet{Majewski04} and used in a number of our RV studies over more
than a decade (e.g., \citealt{Kunkel97}).  Briefly, each star is 
first Fourier-filtered to attenuate frequency components lower than those 
given by the typical absorption line and higher than permitted by the 
intrinsic line resolution (as might be introduced by cosmic rays, for example).  
Masking multiplies a filtered spectrum in wavelength space
with zero to suppress, or with unity to admit portions of a spectrum.  
Mask wavelengths are selected according to several line lists that each
serve specific functions.   For example, an OH night sky emission mask 
multiplies all candidate data spectra prior to the correlation process 
to suppress possible residual noise due to imperfect night sky subtraction.
Another mask similarly multiplies filtered spectra prior to
correlation to suppress telluric water and O$_2$ bands when their strength 
at the instrumental resolution exceeds a few percent. 
Finally, a third line list is used to mask the cross-correlation ``master 
template'', which is set to zero everywhere except at low-ionization or 
low-excitation metallic or Balmer features that, at the instrumental 
resolution and at mean target star signal levels, contribute a useful 
equivalent width component for a metallicity representative of the anticipated 
population (in this case, Leo~I).  The suitability of features selected for 
inclusion in this last mask line list is decided from visual inspection of 
the Arcturus Atlas \citep{Hinkle00}, which permits identification of 
satisfactory features as well as untrustworthy blends (usually from line 
pairs of dissimilar elements) for exclusion, and with reference to 
representative target star spectra to determine in a practical sense the 
visibility of the selected features.

In the specific case of our Leo~I reductions, our master cross-correlation 
template was based on the star 61788 in the KD1 mask, which was very 
well-exposed ($\sim10,000$ ADU pixel$^{-1}$ --- roughly twenty times
stronger than the weaker Leo~I spectra; see Fig~\ref{f:specsample}).
This source was used as an RV template because no exposure of a more 
standard RV reference star was obtained, since the Keck observations of 
Leo~I were obtained only as a backup program for the DEEP project.  
The benefit of using this particular star as a cross-correlation template 
is its close match in spectral type and metallicity with the Leo~I spectra.  
The disadvantage is that the true RV of this local reference is unknown, 
so the final RVs derived in the initial cross-correlation phase are subject 
to zero-point uncertainties, although their random errors are low.  
Since the key dynamical descriptors of the Leo~I population of relevance 
to our later analysis are based on {\it relative distributions} of RVs, 
any systemic RV offset present in the data does not explicitly affect our 
interpretations.  The correction of systematic velocity offsets to a known 
velocity system can be achieved by accounting for offsets against telluric 
line references, as described below.  In addition, we describe how we 
ultimately adjust to the zero-point of the M98 RV system in \S 5.3, a 
necessary additional step if we wish to combine the two data sets.

In preparing star 61788 to serve as a cross-correlation template, the 
masking procedure left predominantly the lines of the Ca infrared triplet,
as well as the stronger features of Mg, Fe, Ti, Ni, Si, Na, and H$\alpha$.  
These lines visible in the atmosphere of the star 61788 were preserved 
with the non-zero mask ``slats'' represented by a Gaussian of unit 
amplitude and a width roughly 1.2 times the instrument profile.  
A total of 85 lines were used in cross-correlations using this masked 
template, with each slat having a FWHM of 3.0\AA ~for a total used 
spectral range of 255\AA\ over the full spectral range.  Experiments 
using slats of 2.5\AA\ and 4.2\AA\ yielded no significant variations in 
the quality of the cross-correlation results.  As described above, before 
cross-correlation the other Leo~I stars have been similarly Fourier-filtered 
to the 61788 spectrum.  To describe the quality of each cross-correlation, 
we have adopted the RV quality index $Q$, which is a descriptor of 
the shape of the central cross-correlation peak with respect to 
sideband peaks.\footnote{See \citet{Majewski04} and \citet{Kunkel97} 
for descriptions on how $Q$ values are assigned.}
We note that although the meaning of the RV quality values are 
identical to those given in \citet{Majewski04}, the actual values 
of the cross-correlation peak (CCP) levels are on a relative scale 
that is only meaningful within the context of the present instrument 
set up.  In general, only those RVs with $Q \ge 4$ can be trusted as 
reliable and we discard any star that does not meet this criterion in 
subsequent analysis.

A primary limitation on the precision of RVs as just obtained is the 
non-uniform manner in which a spectrometer slit is illuminated, 
even when the star is perfectly centered in the
slit.  The presumption that a spectrometer slit is uniformly
illuminated is true only the when illuminating source is uniformly 
spread over the portion of sky that the telescope is sampling.  
Even in poor seeing a stellar point source, through spread over a
angle with a roughly Gaussian profile on the sky tends
to illuminate the slit aperture (along dispersion) differently
compared to the illumination from a lamp used to impose pieces of 
monochromatic light for wavelength calibration.  To the extent that this 
difference appears perpendicular to the slit aperture, there
is a displacement in the centroid of (monochromatic) stellar
light along the dispersion compared with that of light from a uniformly
illuminated source (such as the sky or a calibrating lamp). 
The effects are obviously greatly amplified in the case that the star is
{\it not} centered in the slit, whether due to random errors in the 
astrometry, imperfect tooling of the mask slits, and/or mask misalignment
at the time of observation.  The dispersion displacement varies with the 
circumstances of an exposure and can be determined only from signatures 
carried by the stellar signal itself.  Fortunately, the signature of this 
displacement is also imposed on absorption features originating from 
molecules of telluric O$_2$ and H$_2$O in the light path, and 
whose profiles and positions accurately track the mean integrated 
slit function of each particular star as passed through its slit.  
Thus, the mean wavelength offsets in telluric features can be used to 
derive slit displacements on a star-by-star basis.  To derive the telluric 
offsets we design a new cross-correlation mask in which all spectral 
features are blocked except the regions with the strongest telluric 
absorption features.  The wavelength ranges upon which the telluric 
offset determination is based are 
6866-6912\AA\  (the Fraunhofer B band), 
7167-7320\AA, 
7593-7690\AA\  (the Fraunhofer A band), 
8110-8320\AA, and 
8925-9120\AA, but all telluric 
features weaker than 20\% were not used.  A ``master'' template of 
telluric features was made from a fairly strong Keck spectrum of a 
star with weak stellar features.  Unlike stellar features, the 
telluric features will look identical in all stars (modulo the $S/N$ 
of their spectra), and so the dominant peak profile in 
cross-correlations with our masked telluric spectrum should appear 
the same.  This was confirmed visually, and the lowest ``$Q$'' values 
we would assign to any of {\it these} cross-correlation peaks would 
be a ``6''; even stellar spectra with bad $Q$ from the 
cross-correlation of {\it stellar} features still provide excellent 
telluric correlation profiles.  We note that application of the 
offsets derived from cross-correlation of telluric features also 
corrects the original RVs for errors in the pixel-to-wavelength 
calibration and, in addition, places the RVs to an absolute 
reference. 

The heliocentric RVs corrected for these offsets (with the additional 
zero-point correction to the Leo~I systemic velocity as derived in 
\S 5.3) are given in column (8) of Table \ref{t:keckspec}, along with 
their RA and Dec coordinates, $M_{0}$ magnitudes, $(M-T_{2})_{0}$ and 
$(M-D51)_{0}$ colors in columns (3) through (7).  We also list 
Galactocentric standard of rest (GSR) RVs\footnote{We convert to 
Galactocentric standard of rest by assuming a local standard of rest 
velocity of 220 km s$^{-1}$ and a solar peculiar velocity of 
$(u,v,w)=(-9, 12, 7)$ km s$^{-1}$.}, telluric offsets, 
cross-correlation peaks ($CCP$), and quality index ($Q$) in columns 
(9) through (12) of Table \ref{t:keckspec}.

\begin{figure}
\epsscale{1.0}
\plotone{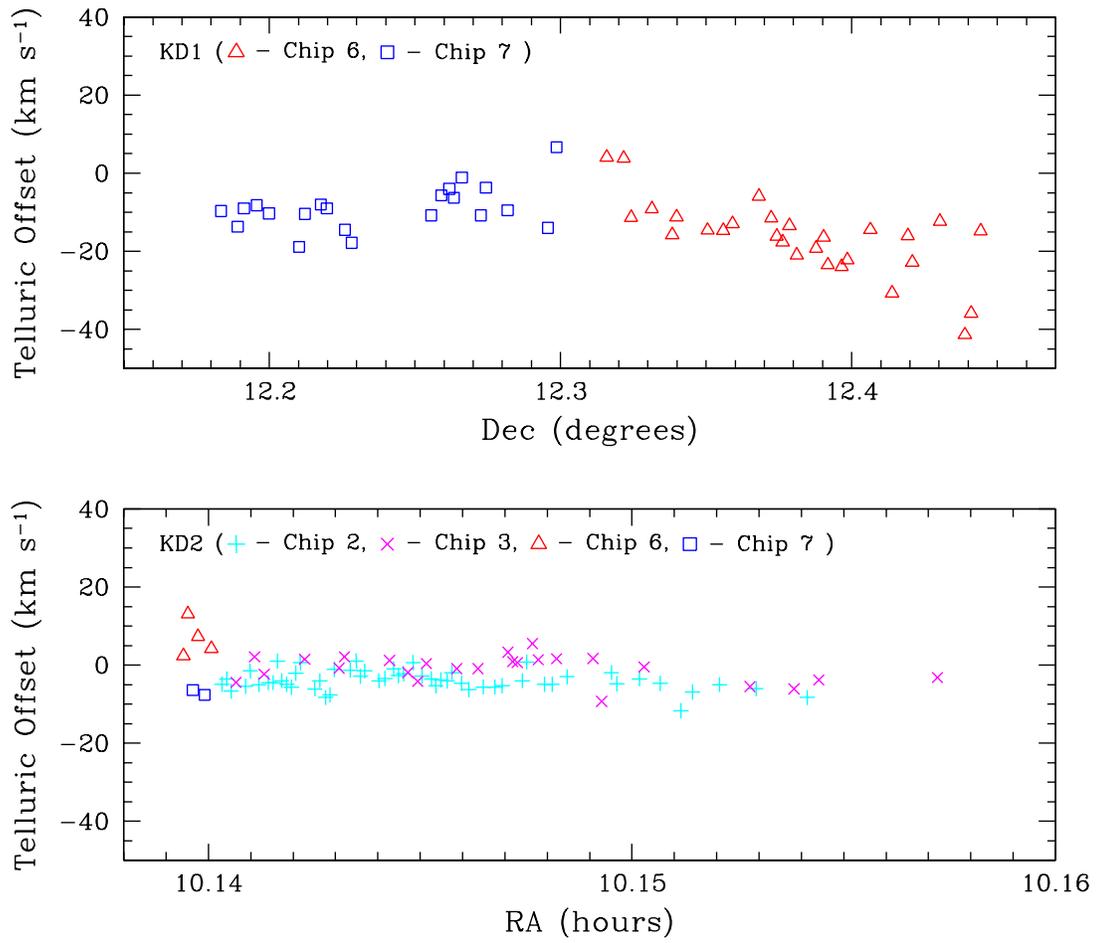}
\caption{ Telluric offsets for each stellar RV measurement as a function of 
          the position in the mask for (a) KD1 and (b) KD2.  
          The variation in astrometric solution from chip to chip in the original Mosaic
          camera data is revealed 
          by the trends in telluric offset by chip number.    
        }
\label{f:teloff_position}
\end{figure}

For those stars observed in 1\arcsec\ wide slits\footnote{Seven stars 
listed in Table \ref{t:keckspec} were actually alignment stars 
observed in 4\arcsec wide slits.  However, with the telluric offset 
corrections we can actually derive good RVs for these spectra with 
larger equivalent slit functions.  The larger telluric offset 
corrections needed for these stars are evident in the tabulated 
values.}, the mean telluric offset is $-11.5$ km s$^{-1}$ with a 
dispersion of $7.7$ km s$^{-1}$ for those stars in mask KD1 and 
$-2.8$ km s$^{-1}$ with a dispersion of 3.5 km s$^{-1}$ for those 
stars in mask KD2.  
That both mean offsets are non-zero is a reflection of systematics 
in the original template derived from star 61788; that the mean 
offsets are different for each mask reflects the differences in 
positioning of each mask relative to their respective star fields.  
The 7.7 and 3.5 km s$^{-1}$ dispersions are respectively equivalent 
to $\sim 0.2$ and $\sim 0.1$ arcsec variations in the slit centering, 
and reflects the quality of the original astrometry (see \S2), the 
slit manufacture and the {\it rotational} alignment of the slit mask 
on the sky.  Figure~\ref{f:teloff_position}, which shows the telluric 
offset correction for both masks as a function of position along the 
mask (approximately declination for mask KD1 and right ascension for 
mask KD2), demonstrates however that the initial astrometric 
reduction contributes significantly to the star-by-star slit miscentering, 
given the clear correlations of the offset trends with 
the original CCD frame on which the star was originally 
photometered (each CCD frame has a unique astrometric solution).  
The dispersions in the telluric offsets represent the actual 
precisions limits for multi-slit RVs in the case when the telluric 
offsets are {\it not} accounted for.  After applying the telluric 
offset the velocity precision is therefore well better than this.

\subsection{Determination of Sampling Errors}

Based on the relative strengths of the cross-correlations against 
the stellar and telluric templates, the dominant contribution to the 
RV errors are in the stellar absorption cross-correlation; 
the telluric cross-correlation peaks are almost always much
stronger than the kinematic correlation peaks, because the equivalent width of
telluric absorption features sum to an order of magnitude larger than those
intrinsic to the stellar atmosphere.
Errors in the stellar absorption cross-correlation are
partly a reflection of differences in stellar line strengths between 
the template and target spectra.  

An assessment of random sampling errors may be approached by
a variety of techniques; however, the method of \citet{TD79} is 
not one of them.  The adopted cross-correlation methodology 
employed here differs from that of Tonry \& Davis, having evolved over years 
of experience and motivated by the goal to cover as broad a range of (generally
unknown) spectral types among the target stars as possible.  The Tonry \& Davis 
method relies explicitly on the characteristic that a template and a target 
spectrum must be of identical shape, or spectral type, so that the resulting 
cross-correlation ideally becomes an even function with null imaginary terms.  
Then the imaginary terms that appear for real data may be utilized for the 
estimation of sampling errors.  In contrast, our methodology, by using 
slat masking, maximizes the tolerance of the cross-correlation to widely 
variable spectral types, so that F through M stars have in the past been measured 
equally successfully to full precision in the same correlation process with a 
single template design.  But such dissimilarities invalidate the reliance on the 
even character of the correlation function, and imaginary terms are generally 
non-zero even when sampling errors are rigorously zero.  Consequently alternate 
methods for estimating errors with our methodology are required.

To obtain a first estimate of the errors in our RVs, we repeated the 
stellar absorption cross-correlations twice with other RV templates 
manufactured from other stars in the Leo~I masks with much 
stronger stellar absorption features than stars 61788, namely stars 
61736, 61782.  With the cross-correlations there was little 
difference in the final results for the best quality target spectra, 
and there were only minor changes in $Q$ values using different RV 
templates.  For $Q=7$ spectra the RMS scatter in RVs was 1.2 km 
s$^{-1}$.  This rises to 3.0 km s$^{-1}$ when any one of the 
cross-correlation peaks dropped to $Q=4$.  
This analysis suggested an uncertainty scale of 1.2, 1.5, 2.0, 
and 3.0 km s$^{-1}$ for stars with $Q$ = 7, 6, 5, and 4, respectively.
However, these uncertainties must be regarded as lower limits, since
the use of different templates on the {\it same} target spectra do not
lead to completely statistically independent measurements.

Because repeat observations of our target stars are not available, an 
alternative error estimation derives from creating {\it artificial} 
``repeat observations'' via the convolution of the raw, unfiltered 
candidate spectra with randomized Poisson noise.  Practically, this is 
accomplished by using the original spectrum as a probability distribution 
function, whereby, at any wavelength, the envelope of the original 
spectrum defines the ``expectation'' of received source counts with respect 
to neighboring wavelengths.  With each wavelength ``bin'' set to the 
resolution FWHM width, the spectrum can be repopulated with Poisson
distributed photon events, dropped in randomly with wavelength destinations 
weighted according to the expectation probabilities.  
To get a representative spectrum whose noise characteristics are like 
those of the raw data (presuming a negligible contribution from read noise) 
one keeps adding Poisson events until the count of events equals that of 
the raw target spectrum.  On average, each wavelength bin in the new 
pseudo-spectrum retains the same amplitude, compared to neighbors, as 
the original spectrum, with modulation only due to Poisson noise.   
When a number of pseudo-spectra are so rebuilt but arbitrarily scaled to a 
much stronger total photon count than the original spectrum signal strength, 
their cross-correlations with the template show zero velocity scatter, as 
one expects for strong signals.  When, on the other hand, Poisson distributed 
photons are dropped into each slat so that the overall mean exposure level of 
the pseudo-spectra are comparable to the observed spectrum, the scatter 
in the derived velocities is found to be identical to that obtained by 
re-observing the same star as many times.\footnote{The basis for this claim 
derives from tests of this methodology on other data sets available to the 
authors wherein multiple observations of the stars are available.  In all 
cases tested, equivalent RV scatters were found for multiple observations 
of the same star compared to multiple pseudo-spectra generated from 
one observation.}

The above error estimation method was applied to a variety of Leo~I spectra
with the generated pseudo-spectra cross-correlated against the same RV template
using the same masks. This procedure packages into one procedure a test of all 
aspects of the data processing stream, including all instrumental traits, as 
well as the design of the line-lists, template fabrication, etc.
If we adopt test spectra near the {\it low end} limit of signal levels, 
i.e. $\sim500$ ADU, we obtain dispersions of 1.55, 
1.77, 2.87 and 4.91 km s$^{-1}$ for stellar feature
cross-correlations in $Q$= 7, 6, 5 and 4 spectra, respectively.  
Obviously, better precisions are achieved for the large number of
spectra with higher signal levels; e.g.,  
for $Q=7$ spectra, dispersions of 0.44, 0.50, 1.46 km s$^{-1}$
were obtained for signal levels of about 4200, 2500, and 1000 ADU. 
Independent 
tests of the telluric cross-correlations at these signal strengths lead to 
scatters of about 1.2 km s$^{-1}$ (and significantly better for stronger spectra).  
Adding these contributions in quadrature, and conservatively adopting the
worst case (i.e. low signal strength) dispersions for each quality, leads to 
the following estimated uncertainties for stars with $Q$ = 7, 6, 5 and 4, 
respectively: 2.0, 2.1, 3.1, and 5.1 km s$^{-1}$.  We adopt these as our 
velocity uncertainties for subsequent analyses.  However, it is worth pointing 
out that by adopting conservative estimates of our uncertainties at every level, 
we do risk {\it underestimating} derived true velocity dispersions after we 
remove the contributions from observational uncertainties.\footnote{For example, 
adopting the lower limits to the velocity uncertainties derived from our
first estimate above (i.e. 1.2, 1.5, 2.0 and 3.0 km s$^{-1}$ for $Q=$ 7 to 4)
raises the central Leo~I velocity dispersion (see \S5) by about 5\% (from 
8.2 to 8.6 km s$^{-1}$) and the ``cold point'' velocity dispersion near the King
limiting radius (see \S5) by the same fraction (from 3.9 to 4.1 km s$^{-1}$).}

\subsection{Testing the Photometric Selection of Leo~I Giants}

\begin{figure}
\epsscale{1.0}
\plotone{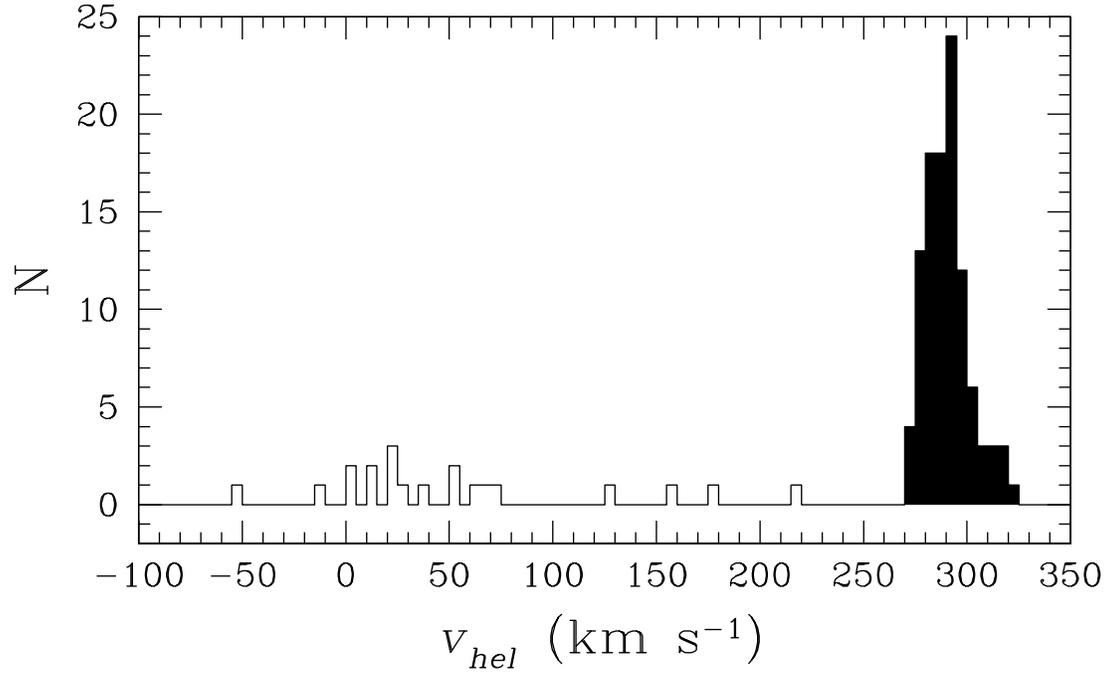}
\caption{ Histogram for heliocentric radial velocities of 125 stars observed 
          with Keck DEIMOS that have $Q \ge 4$.  Bin size is 5 km s$^{-1}$.
        }
\label{f:keckspec_hist}
\end{figure}

The extreme systemic RV of Leo~I, $v_{hel} = 287$ km s$^{-1}$ (M98) 
is a great advantage for distinguishing true Leo~I members from 
Galactic stars --- very few Galactic stars are expected to have such 
extreme velocities at this position in the Galaxy 
[$(l,b)=(226,+49)$\arcdeg].  Figure~\ref{f:keckspec_hist} shows a 
histogram of the observed RVs of stars listed in Table 
\ref{t:keckspec}; stars photometrically selected to be Leo~I giant 
candidates are shaded darkly.  Leo~I giants are easily identified by 
the clump at $v_{hel}/($km s$^{-1}) > 250$.  In KD1, 21 out of 42 
stars are identified as Leo~I members and in KD2 all 83 stars are 
Leo~I members, based on their RVs.  Stars we targeted simply as 
``mask fillers'' tend to have RVs clustering near 0 km s$^{-1}$ 
as expected for MW disk stars.

\begin{figure} 
\epsscale{1.0} 
\plotone{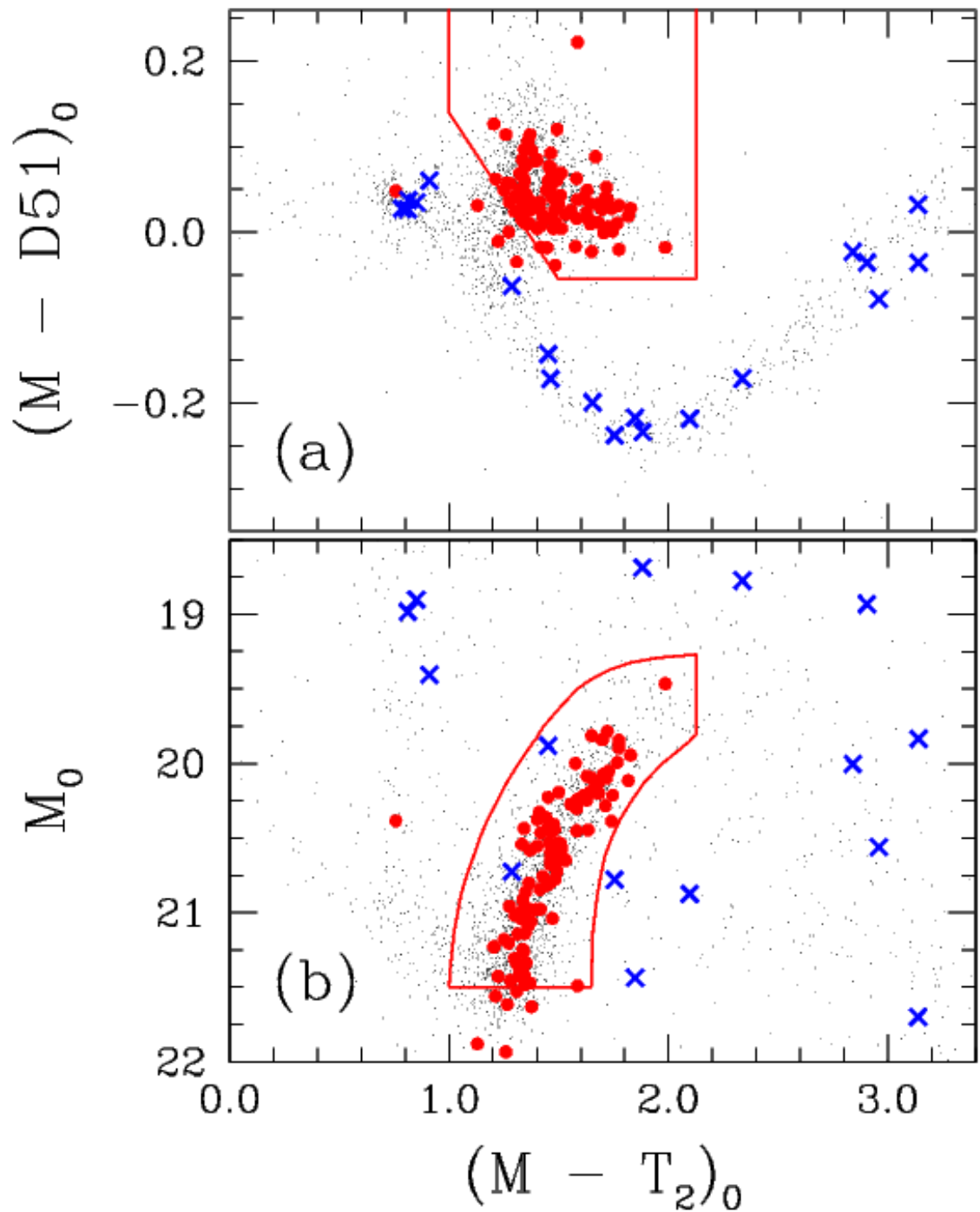} 
\caption{ Positions of the stars observed with the Keck DEIMOS: (a) in the 
          color-color diagram; (b) color-magnitude diagram.  The selection 
          boxes for the Leo~I giant candidates (from Figs. \ref{f:ccdselect} 
          and \ref{f:cmdselect}) are shown in both plots. The 
          {\it filled circles} are for stars that have been confirmed to be 
          members of Leo~I based on their heliocentric radial velocities, 
          while the {\it crosses} are for stars confirmed not be members.
        }
\label{f:keckspec_crossid}
\end{figure}

\begin{figure} 
\epsscale{1.0} 
\plotone{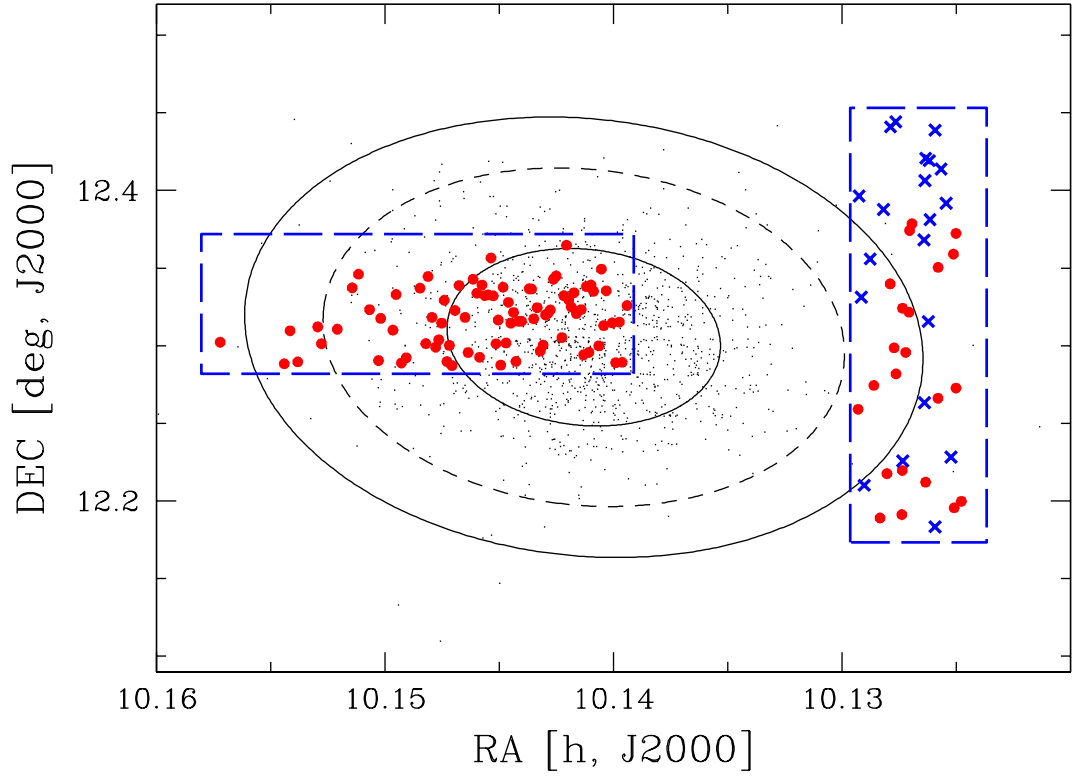} 
\caption{ Stars observed with the Keck DEIMOS plotted over the sky 
          distribution of Leo~I giant candidates.  The approximate field of 
          view for the Keck DEIMOS fields are shown with long dashed lines.
          The inner ellipse shown in solid line represent the King core radius, 
          and the outer two ellipses are same as those in Figure 
          \ref{f:spatialplot_center}.  The symbols are same as those 
          in Figure~\ref{f:keckspec_crossid}.
        }
\label{f:keckspec_spatialplot}
\end{figure}

In Figure~\ref{f:keckspec_crossid}, we plot the stars observed with 
the Keck DEIMOS over the color-color and color-magnitude diagrams 
with the selection regions shown.  In both the color-magnitude and 
color-color diagrams, the stars with measured velocity $v_{hel} > 
250$ km s$^{-1}$ are generally located within our primary giant 
selection region whereas the stars with lower velocities are not; in 
particular, in the two-color diagram the lower velocity stars all fall 
along the elbow-shaped, dwarf star locus, suggesting these stars to all 
be relatively nearby Milky Way dwarfs.  Even more gratifying is 
that for those 96 stars selected to be Leo~I giants using both the 
color-color and color-magnitude diagram and for which we obtained 
RVs, {\it all} of them are Leo~I members by the RV criterion; this 
suggests that the {\it reliability} of our selection is very good, 
and we emphasize that {\it a significant portion of our sample 
includes stars selected in the very low density outer parts of the 
Leo~I system} (see Figure~\ref{f:keckspec_spatialplot}).  We also 
observe that our Leo~I giant candidate selection criteria are on the 
conservative side, because nine additional stars that we did 
not pick as giant candidates are also Leo~I RV members.  In most cases a 
slight expansion of our selection criteria would have picked up all 
but one\footnote{This one Leo~I RV member well outside our 
color-magnitude selection box at $(M-T_{2})_{0} < 1.0$ (see Figure 
\ref{f:keckspec_crossid}) is likely to be an anomalous Cepheid.} 
of these other Leo~I stars, but this expansion would have brought 
in false positives.  In particular, it may be seen that our catalog 
can be trusted to even fainter magnitudes than the conservative 
$M_{0} < 21.5$ limit we adopted for the analyses presented in our 
photometric studies.  It may also be seen that use of the $M-D51$ 
color eliminates contaminant dwarf stars that fall within the 
color-magnitude locus of the Leo~I RGB, and thus improves the 
efficiency of our target selection.  This reliability analysis lends 
credibility to the structural features we have mapped using Leo~I giant 
candidates selected in this way.

Figure~\ref{f:keckspec_spatialplot} shows the sky distribution of 
the stars in Figure~\ref{f:keckspec_crossid}.  Out of the total 
105 Leo~I giants observed with Keck DEIMOS, 90 stars lie within the 
King limiting radius and 48 stars lie within the core radius, 
leaving 15 confirmed Leo~I members beyond the limiting radius -- 
i.e., in the domain of the ``break'' population we identified in our 
study of the Leo~I light profile.
The new Keck spectroscopy confirms that our selection 
of Leo~I giant candidates is both effective and reliable, and so 
lends confidence that the population of stars falling outside the 
King limiting radius (which may be the nominal tidal radius of the 
system -- see \S 6.1 below) and constituting our detected ``break 
population'' (e.g., Fig. \ref{f:profiles}) are actually stars 
associated with Leo~I.

\subsection{Velocity Dispersion}

The variation of velocity dispersion ($\sigma_{v}$) with projected 
radius of dSphs is important for testing dynamical models of dSphs 
\citep[e.g.,][]{Kroupa97,Lokas05,Strigari2007,Munoz07}.  For a bound 
stellar system where mass-follows-light $\sigma_{v}$ is expected to 
decrease with radius and approach zero at the tidal radius.  On the 
other hand, dSph models with extended DM halos predict a dispersion 
that falls off more slowly than mass-follows-light models 
\citep[e.g.,][]{Kleyna01,Wilkinson02}. Finally, tidal disruption 
models predict flat/rising $\sigma_{v}$ profiles 
(e.g., \citealt{Kroupa97}; Mu\~noz et al., in preparation) into the 
domain dominated by stars that have become unbound during/after tidal 
interactions with the host galaxy.  

\begin{figure}
\epsscale{1.0}
\plotone{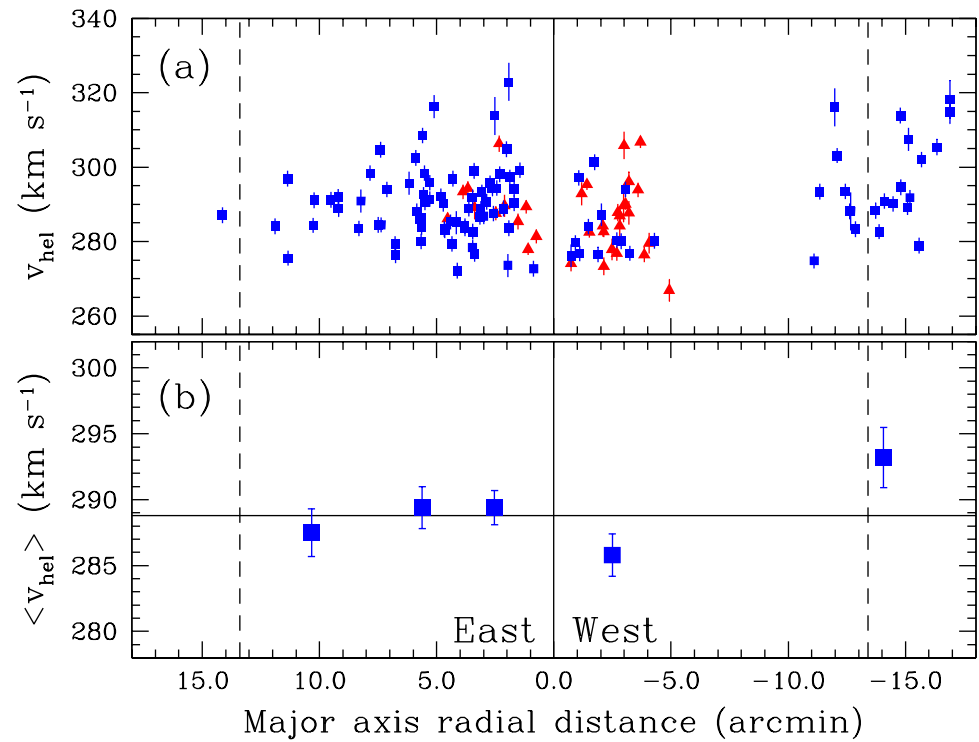}
\caption{ Distributions of (a) individual heliocentric radial velocities and 
          (b) mean radial velocities along the projected major-axis radial 
          distances.  {\it Filled squares} in the {\it upper panel} are stars 
          observed with the Keck DEIMOS, while {\it filled triangles} are 
          those from \citet{Mateo98}.
        }
\label{f:rvprofile}
\end{figure}

With our Keck DEIMOS data, we can explore the radial variation of 
$\sigma_{v}$ out to and beyond the King limiting radius.  
Before doing so, we adjust our RVs to the zero point of the M98 RV 
system so that both data sets may be considered together for 
improved statistics.  The weighted mean $\sigma_{v}$ of the 33 
stars observed by M98 is $8.6 \pm 1.2$ km s$^{-1}$.  These stars lie 
in the semi-major axis range $r_{maj}/r_{lim} = 0.055 - 0.367$ according 
to our newly derived structural parameters.  The $\sigma_{v}$ for 50 stars 
we observed in the same $r_{maj}$ range is $8.2\pm 0.9$ km s$^{-1}$, 
which is consistent with the M98's velocity dispersion within the 
error.\footnote{All velocity dispersion calculations throughout this 
paper follow the method of \citet{Armandroff86}.}  We take as the 
offset to our RV system the $+12.4$ km s$^{-1}$ difference between 
the error-weighted mean of our and the M98 samples, and apply this 
offset to the entire Keck DEIMOS data set (although it is not clear, 
in fact, which of the two surveys is closer to the true absolute 
RV system).  In the upper panel of Figure~\ref{f:rvprofile}, we plot 
the heliocentric RVs versus the projected major-axis radial distances 
for both Keck DEIMOS and M98 samples.  For comparison, we have 
plotted each sample with different symbols (and colors).  
The remarkable similarity of the RV distributions within 5 arcmin 
from the center of Leo~I assures us that both samples are on the 
same system.  We also show the mean RVs along projected radius in 
the lower panel of Figure~\ref{f:rvprofile}.  The mean RV trend shows 
that there is a hint of rotation in the inner $< 5$ arcmin, but with 
low statistical significance.  Spectroscopic observations for stars 
to the west in the range $r_{maj} = 5$ -- 10 arcmin will help reveal 
whether rotation is significant for Leo~I.  We use the combined 
sample of M98 and Keck spectroscopy in subsequent analyses.

\begin{figure}
\epsscale{0.8}
\plotone{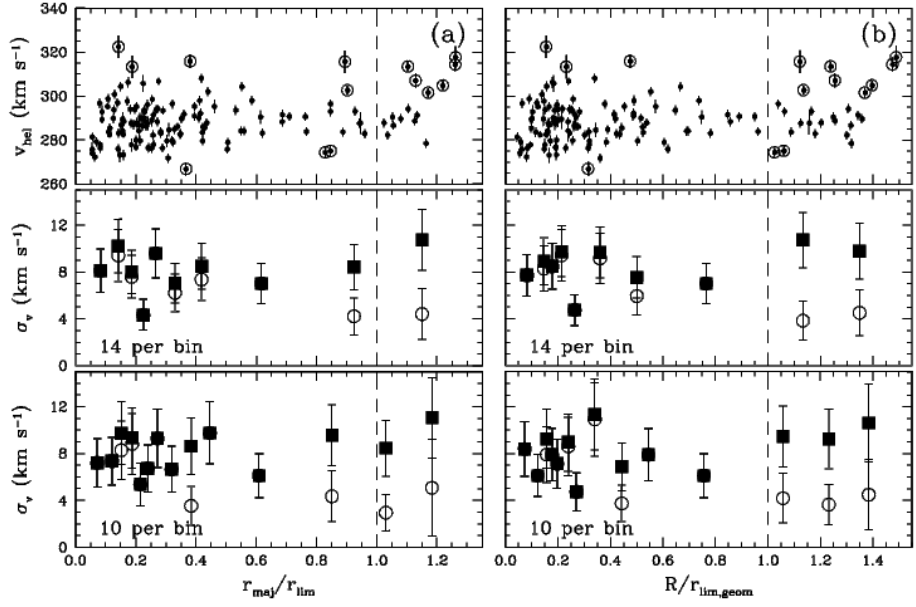}
\caption{ Radial variation of individual radial velocities and velocity 
          dispersions using (a) elliptical and (b) circular radii.
          The number of stars for calculating each data point ($\sigma_{v}$) 
          in the {\it mid panels} are 14, except for the last data point 
          where 11 stars were used.  Similarly, 10 stars were used for 
          calculating each data point in the {\it lower panels}, except 
          for the last data point where 7 stars were used.  The vertical 
          dashed lines denote the King limiting radius (geometric mean, 
          i.e., $r_{lim,geom} = r_{lim}\sqrt{1-e}$ for the right panels).  
          The radial velocity outliers (see text) are marked with large 
          circles in the {\it upper panels}, and velocity dispersions 
          computed without the outliers are shown in open circles in the 
          {\it mid} and {\it lower} panels.
        }
\label{f:vdprofile}
\end{figure}

In Figure~\ref{f:vdprofile}, we present $v_{hel}$ and $\sigma_{v}$ 
as a function of radial distance from the Leo~I center, calculated 
for elliptical ({\it left panels}) and circular radii ({\it right 
panels}).  We use different number of stars per bin for the 
middle and lower panels.  The overall trend of the 
$\sigma_{v}$ is an initial decline followed by a flat or rising 
profile.  Similar profile behaviors up to and beyond the nominal 
King radius are also seen in the velocity dispersion profiles of 
Ursa Minor \citep{Munoz05}, Draco \citep{Munoz05}, and Sculptor 
\citep{Westfall06}.  As discussed in these references, such behavior 
means either that we are seeing evidence for tidal disruption in 
these dSphs, or that these satellites have very extended DM components
dynamically traced by these stars at large radii.

\begin{figure}
\epsscale{1.0}
\plotone{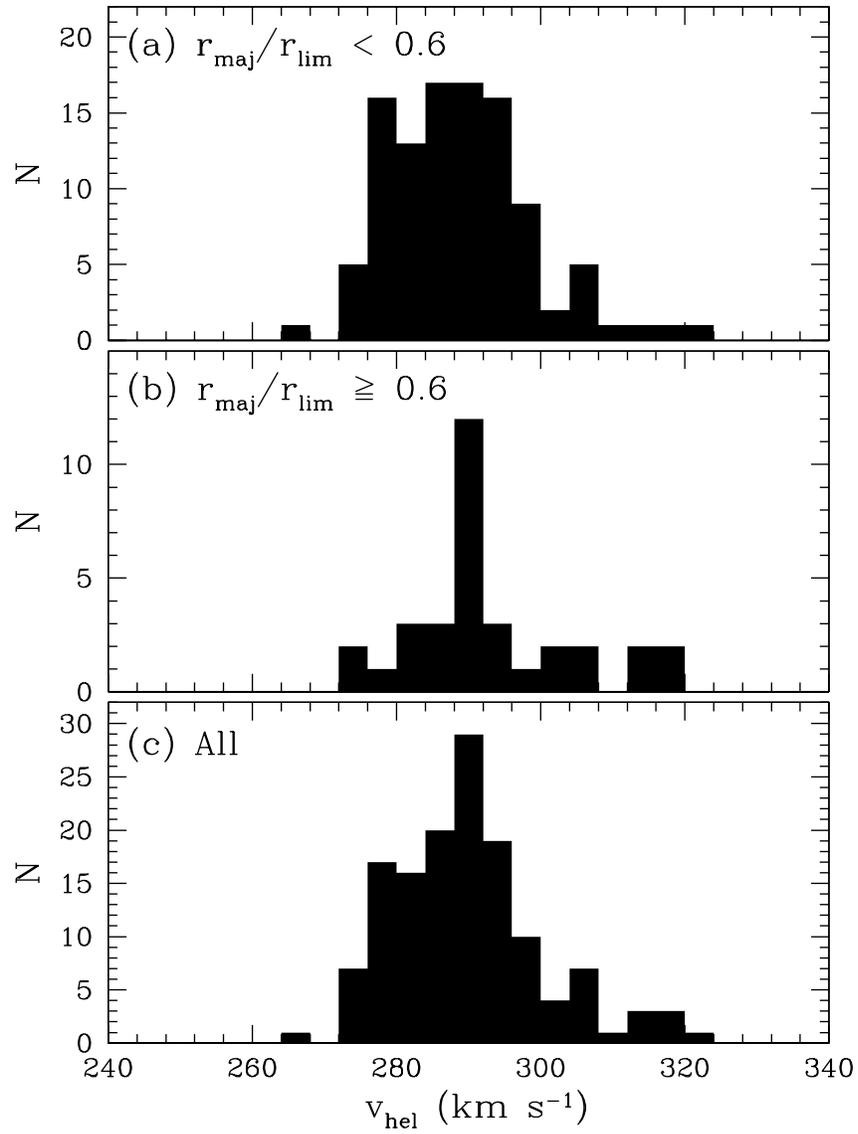}
\caption{ Histogram for heliocentric radial velocities of 
          (a) Leo~I giants with $r_{maj}/r_{lim} < 0.6$, 
          (b) Leo~I giants with $r_{maj}/r_{lim} \geq\ 0.6$, and 
          (c) all Leo~I giants.
        }
\label{f:keckspec_hist_leoi}
\end{figure}

Closer inspection of the upper panels of Figure~\ref{f:vdprofile} 
shows a number of stars (more at larger radius) that have an RV 
deviating significantly from the mean.  These 
{\it outliers}\footnote{Although we use the term outliers, these 
are obviously associated with Leo~I.} are predominantly 
toward the high RV side, and create an obvious asymmetric tail in the 
distribution of Leo~I velocities shown in Figure 
\ref{f:keckspec_hist}.  Indeed, as shown in Figure 
\ref{f:keckspec_hist_leoi}, it appears that these outliers create 
a secondary hump in the RV distribution about 20 km s$^{-1}$ 
higher than the Leo~I mean RV of 288.8 km s$^{-1}$, but in the least
there is an obvious asymmetric tail towards larger RVs.

Members of the outlier population we identify as those stars with RVs 
beyond 2.3-$\sigma$ (where $sigma$ refers to the velocity dispersions of the
entire sample, i.e. 8.9 km s$^{-1}$) for the Leo~I RV sample with 
$r_{maj}/r_{lim} < 0.6$, and those with RVs beyond 1.5-$\sigma$ for 
$r_{maj}/r_{lim} \geq\ 0.6$ Leo~I stars.  Our tighter criterion for the stars 
at larger radii is based on examination of the RV distribution shown in 
Figure~\ref{f:keckspec_hist_leoi}: the tight peak in the $r_{maj}/r_{lim} 
> 0.6$ distribution clearly stands out from a broader distribution of RVs 
and is well approximated by a Gaussian of dispersion of 3.9 km $s^{-1}$.  
Because this tight peak is centered on the mean velocity of the Leo~I core, 
it seems appropriate to associate it with the typical RV population at 
smaller radii (though the latter has a broader distribution).  But it is 
apparent that about half of the outer Leo~I stars are part of a second, 
much more broadly RV-distributed, ``outlier'' population in 
Figure~\ref{f:keckspec_hist_leoi}b.
Adopting a 3-$\sigma$ cut based on the rather cold, 3.9 km $s^{-1}$ central 
dispersion peak among the stars with $r_{maj}/r_{lim} > 0.6$ identifies the same 
set of RV outliers as a 1.5-$\sigma$ scheme where the entire Leo~I sample $\sigma$ 
is used.

\begin{figure}
\epsscale{1.0}
\plotone{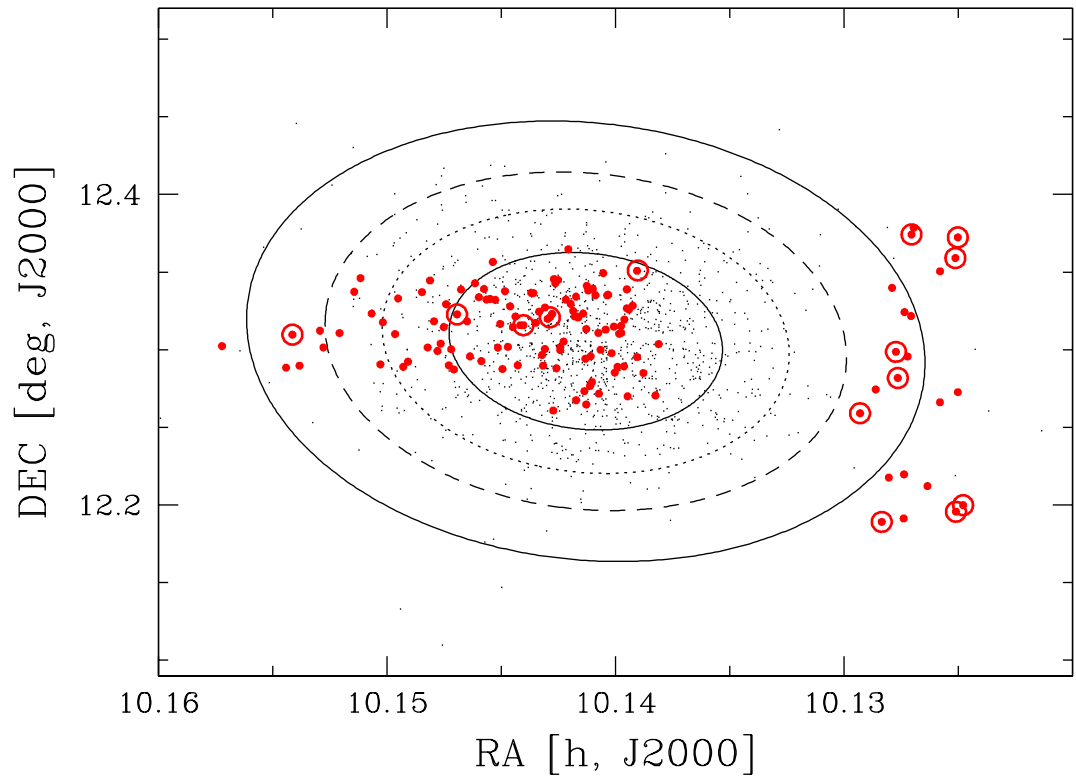}
\caption{ Same as Figure~\ref{f:keckspec_spatialplot}, but the radial velocity
           outliers in Figure~\ref{f:vdprofile} marked in open circles.
           The ellipse drawn in dotted line corresponds to $r_{maj}/r_{lim} = 0.6$.
          }
\label{f:keckspec_outliers}
\end{figure}

The outliers so identified are marked with {\it open circles} in the 
top panels of Figure~\ref{f:vdprofile}.  The spatial distribution of 
these stars is shown in Figure~\ref{f:keckspec_outliers} with 
respect to the entire sample of Leo~I giant candidates,  where it may be seen 
that not only do they tend to lie to the west side of Leo~I, they are 
predominantly found near  or beyond the nominal King radius.  
The latter point is not simply due to the combined facts that we have used a 
tighter ``outlier'' definition for stars at large radii and that {\it all} of 
the stars in the western DEIMOS mask are at large radii: In fact, there are 
still 11 stars to the east side of Leo~I center with $r_{maj}/r_{lim} \geq\ 0.6$, 
but only one of these is an outlier by the above definition, compared 
to 9 outliers among 22 $r_{maj}/r_{lim} \geq\ 0.6$ stars on the west side.  
Of the 9 outliers to the west, eight are outliers to the positive RV side.
 
Exclusion of these outliers from the calculation of $\sigma_{v}$ 
trends leads to different results, as shown by {\it open circles} 
in the middle and lower panels of Figure~\ref{f:vdprofile}:
the velocity dispersion tends to {\it fall} with radius, and to a 
relatively cold value at large radius.  For the stars beyond the King 
limiting radius, the dynamically cold population has a mean RV of 
$287.7 \pm 1.6$ km s$^{-1}$, close to the systemic velocity of the 
Leo~I core, and a velocity dispersion of $4.0 \pm 1.6$ km s$^{-1}$.  
These dynamically cold stars also appear to be more in line with the 
Leo~I major axis, as seen in Figure~\ref{f:keckspec_outliers}.
The RV outlier stars, when included, increase the dispersion 
outside the limiting radius to $10.3 \pm 2.2$ km s$^{-1}$.

Both the presence and meaning of velocity dispersion ``cold points'' 
near the nominal King radii of various dSphs has been debated 
recently \citep{Wilkinson04,Lokas05,Munoz05}.  The reality of the 
phenomenon turns on how one deals with (i.e., includes or excludes) 
apparent outliers and how one bins the data \citep{Lokas05,Munoz05}.  
In the present case, where the Leo~I systemic velocity is so 
extreme, it is hard to believe that the outliers are anything but 
Leo~I-associated stars, and so we are presented with a situation 
where the outliers and the dispersion characteristics demand an 
explanation within the context of a dSph structural model.

The lopsided nature of the RV outliers (i.e., that they tend to be 
outliers to the high RV side) provides a significant clue to the 
nature of the outer parts of the Leo~I dSph.  Were the RV outliers 
bound stars within a large Leo~I DM halo, it is hard to 
understand why they would not exhibit symmetrical dynamics --- i.e., 
a hot population of stars with members both approaching and receding 
relative to the Leo~I core.  The lack of any particular spatial 
concentration of the RV outliers (other than tending to the west side 
of Leo~I) does not support the idea that the RV asymmetry is caused 
by a star cluster or ``dynamical fossil'' like that recently 
suggested to be ``sloshing back and forth within the'' DM 
halo of the Ursa Minor dSph \citep{Kleyna03}.  The observation of 
significant spatial ``substructure'' within dSphs 
\citep[e.g.,][]{Olszewski85,Demers95,Eskridge01,Kleyna03,Palma03,
Coleman04,Coleman05,Olszewski05} has long been a concern for DM 
models, where high DM densities should quickly smooth such 
substructure out.  Here we are presented with an apparent
substructure evident both dynamically and spatially.

A more natural explanation of the RV distribution is that the RV 
outliers and many of the stars outside the nominal King limiting 
radius represent stars that have likely been tidally stripped from 
Leo~I, whereas the more dynamically cold component at the nominal 
Leo~I systemic velocity seen at large radii represents the outermost 
reaches of the bound population of Leo~I stars.  The decreasing RV 
dispersion trend when the outliers are excluded suggests that we are 
seeing the tapering gravitational field of a Leo~I apparently lacking an 
extended DM halo.  Moreover, the observed asymmetric RV distribution 
is naturally produced by tidal debris, as we show in \S 6.3.  
A spectroscopic survey over a larger area would of course help 
clarify the dynamics of the outer Leo~I system --- in particular, 
spectroscopic observations at large radius on the {\it eastern} side 
of Leo~I would provide particularly strong leverage on the tidal 
debris model, which would predict {\it negative} RV outliers 
to the east side of Leo~I.

\section{Discussion}

\subsection{Dark Matter, Extended Halos and Tidal Stripping}

The question of the origin of the apparently large $M/L$ in dSph 
galaxies remains one of the most vexing in Local Group research.  
While the presence of large amounts of DM is the most popular 
explanation (e.g., M98), other proposals, including Modified 
Newtonian Dynamics \citep[MOND; see][and references therein]
{Sanders02}, tidal heating and/or disruption, dynamical resonances 
\citep{Kuhn89}, and even the notion that dSphs are completely unbound 
\citep{Kroupa97,Klessen98} have been proposed to explain the large 
velocity dispersions observed in the cores of the Galactic dSphs.  
Leo~I provides an interesting opportunity to revisit some of these 
mechanisms, because its large Galactocentric distance and velocity 
implies that Leo~I could only have had a few encounters with the MW 
center.  Tidal heating/disruption or resonance effects, if they 
occur, will have been limited to those that could organize and be 
sustained over only one or two perigalactica.  Because of this, 
Leo~I offers the unique chance to gauge the importance of such 
effects on a per perigalacticon basis.

\subsubsection{The Case for Tidal Disruption}

In \S 4.2 we have rederived the $M/L$ of Leo~I using standard 
core-fitting prescriptions and the central velocity dispersion.  
The results are not much different than previous findings, which 
suggest a modest DM component in Leo~I compared to that in
other dSphs.  A primary point of this 
paper is to establish whether, independent of the net DM content 
of the dSph, it may have extended structure indicating that it is 
being tidally disrupted.

\begin{figure}
\epsscale{0.8}
\plotone{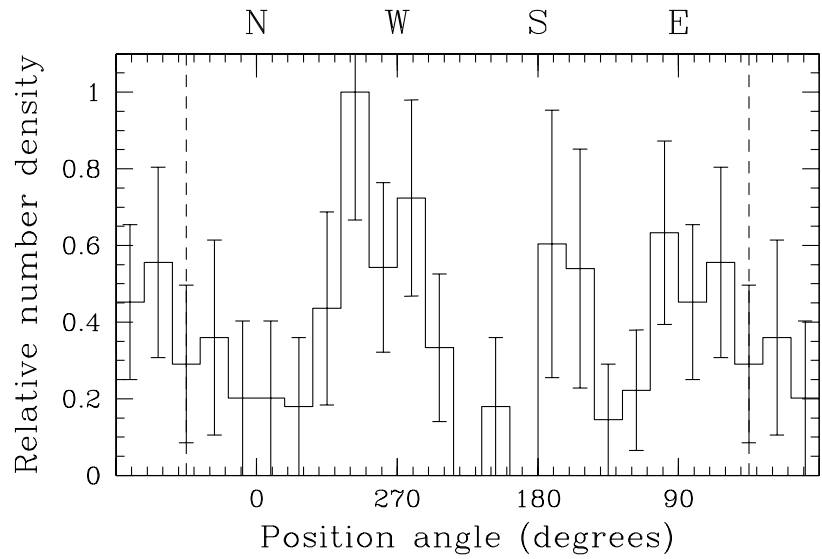}
\caption{ Normalized number densities of ``extratidal'' stars as a function of 
          azimuthal sectors (18\arcdeg wide) around Leo~I.  The histogram 
          represents the number density in each elliptical sector outside the 
          break radius (major axis radial distance of 10.2 arcmin) --- see 
          Figure~\ref{f:spatialplot_center}.  The final results were divided 
          by the maximum number density to scale the relative number densities 
          from 0 to 1.  The repeated dashed line at PA = 45\arcdeg ~is added 
          for guidance.  The broad peaks to the east and west reflect apparent 
          tidal arms stretching out from the main body.  The excess of density 
          at 180-120\arcdeg ~is due to the ``bridge'' of stars that extends 
          to the southeast, and is likely a statistical anomaly.
        }
\label{f:azcount}
\end{figure}

We believe the weight of evidence suggests compellingly that Leo~I 
has indeed been tidally disrupted and, moreover, that the derived 
King limiting radius approximates the true tidal radius in this particular
system: 

(1) We find a significant number of widely placed giant candidates 
associated with Leo~I in our survey area (see Figure 
\ref{f:spatialplot2}), and with an especially pronounced density 
near the King limiting radius.  While it might be argued that these 
stars represent a secondary, bound ``halo'' population around Leo~I 
--- perhaps tracing an extended DM halo --- as can be seen in Figure 
\ref{f:spatialplot_center} the ``break population'' stars are more 
spread out along the east-west direction than in north-south in a 
manner that {\it resembles} tidal arms.  In order to demonstrate the 
``bipolarity'' of these outer stars more clearly, we divide the 
region outside the break radius of Leo~I into azimuthal sectors of 
18\arcdeg\ in width and 17\arcmin\ in major axis radial length 
($r_{maj} = 10\farcm 4 - 27\farcm 5$), and count the giant candidates 
in each sector\footnote{Careful examination of the stellar 
distribution in model galaxies for tidal disruption models show that 
stars released during the most recent perigalactic pass form a tidal 
tail-like feature mostly aligned with the semi-major axes of the 
satellite, whereas those that become unbound during older perigalactic 
passes tend to be spreaded out irrespective of the semi-major axes.  
One demonstration can be found in Figure~\ref{f:nbody_orbit}, an orbital 
plot of our $N$-body simulations where we use different colors for 
stars that become unbound during different perigalactic passes.
Consequently, photometrically chosen dSph candidates located on
degree scales away from the satellite's main body will not necessarily 
show a clearly defined tidal tail-like configuration.  For this 
reason, we have limited 
our sample in $r_{maj}$ when constructing the azimuthal sector 
count plot.}.  The counts were then normalized by the surveyed 
area in each sector to obtain the mean densities as a function of 
azimuthal sector presented in Figure~\ref{f:azcount}.  As may be seen 
the ``arms'' are represented by two broad density peaks separated by 
$\sim 180$\arcdeg\ to the east and west.  A minimum in density 
is apparent to the north, but a corresponding minimum to the south is 
interrupted by the presence of a small peak at 145--180\arcdeg\ which 
originates from the bridge of seven stars extending to the southeast (visible
in Fig.~\ref{f:spatialplot_center}). 
Whether this feature is a chance alignment of Leo~I stars or a real 
dynamical structure may require high resolution velocity data to 
resolve, but we note that if these stars were a bit more spread out 
in PA, the ``minimum'' of ``beyond the break'' stars to the south 
would look similar to that of the north.  Overall, the stars beyond 
the break radius appear to be spread out more along the major axis 
than the minor axis, and have the appearance of a nascent tidal 
tail system.  It is difficult to understand how the observed 
increasing ellipticity of the system would originate and be sustained 
in a dSph with an extended DM halo; a very elongated luminous 
``halo'' population with anisotropic velocity ellipsoid structure 
overlapping and extending beyond a second, more concentrated 
population with a rounder distribution would be implied.  

(2) Our ``King+break'' radial density profile for Leo~I is similar to 
the profiles observed in previous models of disrupting satellites 
\citep[see also our new $N$-body simulation presented in \S 6.2]
{Johnston99,Johnston02} as well as in the Sagittarius dSph \citep{Majewski03b}, 
an acknowledged example of a tidally disrupting satellite.

(3) We have found a flat overall velocity dispersion profile for 
Leo~I, a profile that is expected in the case of a tidally disrupting 
dSph (\citealt{Kroupa97}; Mu\~noz et al., in preparation) and is actually 
observed in the case of the tidally disrupting Sgr dSph \citep{Majewski04}.  
While such profiles could also be formed in the presence of an extended DM 
halo \citep{Kleyna01}, the detailed velocity distribution at large radius 
is not what is expected for the latter model, whereas it
compellingly resembles (Fig.~\ref{f:keckspec_hist_leoi})
a dynamically cold (likely bound) stellar population at the systemic 
Leo~I velocity with a superposed population of unbound stars with larger
velocity dispersion.  

(4) The {\it asymmetry} of the RVs of the outliers would seem to 
challenge the bound, extended halo population hypothesis, since in this
model stars at a variety of orbital phases (i.e. both approaching and 
leaving the apocenters of their internal orbits) would be expected to yield 
a symmetrical velocity distribution.  On the other hand, such asymmetries 
are not only accommodated, but expected in tidal tail models (as we show 
in \S 6.2). 

We conclude that the most straightforward and natural interpretation 
for the observed Leo~I structure and dynamics is that (1) it is tidally 
limited near the observed King limiting radius with the dwindling bound 
population creating a dynamically cold signature at its outermost extent, 
and (2) increasing numbers of unbound stars are being observed at larger 
radius and forming nascent tidal tails that contribute the RV outliers at 
large radius.  If this proposed model for the structure of Leo~I is true, 
we may infer several things immediately about the mass loss history and 
orbit of this satellite galaxy, even without resorting to new $N$-body 
simulations.

\subsubsection{Inferred Mass Loss Rate}

Models of tidally disturbed satellites around the MW by 
\citet{Johnston99} and \citet{Johnston02} predict breaks in satellite 
radial density profiles similar to that shown in Figure 
\ref{f:profiles}b.  Under the assumption that Leo~I stars found past 
the radial density profile break are unbound, extratidal debris, we 
may estimate the mass-loss rate.  \citet{Johnston99} provide an 
algorithm for determining the approximate mass-loss rate of a 
satellite using the strength of the break population.  If the density 
profile of a satellite galaxy shows a break at $r_{\rm break}$ and 
the extra-break population is well defined out to a radius of 
$r_{\rm xt}$, the fractional mass-loss rate can be estimated from 
\begin{eqnarray}
\left ({{\rm d} f} \over {{\rm d} t} \right )_{1} = g(\theta){r_{\rm break} 
\over {r_{\rm xt} - r_{\rm break}}} {n_{\rm xt} \over n_{\rm break}} 
{\pi \over T_{\rm orb}}
\end{eqnarray}
where $\theta$ is the angle between the line of sight and the plane 
perpendicular to the satellite velocity, $g(\theta)$ is a geometric factor 
corresponding to the orbit (which can be approximated as 
$\cos{\theta}$), $n_{\rm xt}$ is the number of extratidal stars between 
$r_{\rm break}$ and $r_{\rm xt}$, $n_{\rm break}$ is the number of 
stars within $r_{\rm break}$, and $T_{\rm orb}$ is the period for a 
circular orbit at the present satellite Galactocentric distance.  
For now we adopt the values of $\theta$ and $T_{\rm orb}$ for Leo~I 
listed in Table 4 of \citet{Johnston99} (but provide a new model for
this orbit in \S6.2).  By examining the radial density profile in 
Figure~\ref{f:profiles}b, we determine $r_{\rm break} = 10.2$ arcmin.
For $r_{\rm xt}$, we have adopted the major axis radial distance of  
the farthest data point in Figure~\ref{f:profiles}b since the 
corresponding surface density is near the background level.  
Applying corrections to account for missing catalog giant stars 
falling in the Mosaic CCD chip gaps gives $n_{\rm break} = 1150$.  
For the extratidal count, we scale our counts of stars at each annulus
by the ratio of the total elliptical area to the amount of that 
annulus in our actual field coverage, and we find $n_{\rm xt} = 129$. 
Using these values we obtain a mass-loss rate of 
$({{\rm d} f}/{{\rm d} t})_{1} = 3.1\times 10^{-3}$ Gyr$^{-1}$, 
which implies an average of less than 1\% of the total mass of Leo~I 
is lost per gigayear.  We note that the mass-loss rate equation 
(3) above is technically for satellites with extratidal radial 
density profiles that follow $\Sigma_{\rm xt} \sim r^{-1}$.  
In our case, the measured outer density slopes are steeper than 
$\gamma = -1$ (see Figure~\ref{f:profiles}b).  
\citet{Johnston99} also note that this equation is only good to 
within a factor of two.  Therefore, the combined uncertainty in the 
mass-loss estimation is likely to be large.  Nevertheless, the small 
net mass-loss rate for Leo~I is consistent with the upper limit 
determined by \citet{Johnston99}, though our new density profile of 
Leo~I has allowed us to re-calculate this improved upper limit to the 
mass-loss rate.  

For satellites with a less well-defined extratidal 
population, \citet{Johnston99} derived an expression for estimating 
the upper limit of the fractional mass-loss rate, and using more 
realistic models, \citet{Johnston02} update the relation to:
\begin{eqnarray}
\left ({{\rm d} f} \over {{\rm d} t} \right )_{2} = 
\left ({2 \over 5} \right ){{\Sigma_{\rm xt}(r_{\rm break})} \over 
n_{\rm break}} {\pi \over T_{\rm circ}} 2\pi {r_{\rm break}}^{2},
\end{eqnarray}
where $\Sigma_{\rm xt}(r_{\rm break})$ is the number density at 
$r = r_{\rm break}$ and $T_{\rm circ}$ is same as $T_{\rm orb}$.  
Using the values for the repeated parameters from above, we obtain 
an upper limit of $({{\rm d} f}/{{\rm d} t})_{2} = 5.1\times 
10^{-2}$ Gyr$^{-1}$, similar to what \citet{Johnston99} found.

Our results confirm the results of \citet{Johnston99} that among the 
dSphs studied by IH95, Leo~I apparently has one of the smallest 
mass-loss rates.  The large orbit and few perigalactic passes of 
Leo~I may explain this low mass-loss rate.  For example, Leo~II which 
has a lower RV at a similar distance to Leo~I (which 
implies a smaller apogalacticon), has a higher estimated mass-loss 
rate.  Despite the relatively low fractional mass loss rate derived 
for Leo~I, that is has a {\it perceptible} one at all, and so clearly 
in the form of tidal tails, shows that no MW dSph with a smaller 
orbit and mass (i.e., virtually all other known Galactic dSphs) is 
likely to be immune from significant tidal effects \citep{Johnston02}.

\subsection{A Tidally Disrupting Satellite Model to Explain the Outer Leo~I Structure and Dynamics}

\subsubsection{Fitting a Leo~I Model}

\citet{Johnston02} have $N$-body simulated a number of examples of 
disrupting satellite galaxies in MW-like potentials, including one 
system with mass and orbital properties likely similar to Leo~I 
(their models 4 and 5), and show how tidal disruption can be expected even 
in extreme cases of a satellite in a large orbit like that of Leo~I. 
Here we undertake additional modeling specifically guided by the new 
Leo~I observations to see whether we can obtain a somewhat closer 
model match to the newly observed Leo~I structure and velocities 
found here.  Our primary goal is to understand the nature of the 
extended population observed in the Leo~I profile.  More 
specifically, we seek to test whether a tidal disruption scenario 
can explain the observed (1) elongated dSph structure at radii 
comparable to and larger than the limiting radius (i.e. the bipolar 
distribution of the break population), (2) shape of the overall 
density profile, (3) dispersion profile over all radii, and (4) 
the asymmetric RV distribution.  A disrupting model that 
simultaneously accounts for all or even some of these features 
would not only lend support to a tidal disruption scenario, but 
(as we shall show) also provide the first real constraints on the 
orbit of Leo~I.

The $N$-body simulation code we adopt has the same heritage as the 
\citet{Johnston02} mass-follows-light (i.e. single component) models. 
The static MW potential has a logarithmic, spherical ($q=1.0$) halo 
with circular speed 210 km s$^{-1}$.  The assumed solar distance to 
the Galactic center is 8.5 kpc and the total Galactic mass within 
50 kpc is $4.5 \times 10^{11}$ M$_{\sun}$.  Other aspects of the 
potential are as in \citet{Law05}.

The satellite is modeled by $10^5$ particles originally configured as 
a \citet{Plummer11} model, 
\begin{equation}
        \Phi=-{GM_{\rm satellite,0} \over \sqrt{r^2+r_{\rm 0}^2}},
\label{PlummerEqn}
\end{equation}
which has a physical scale length parameter,  $r_{\rm 0}$.  The model 
is constrained so that the present satellite position and RV match 
those observed for Leo~I: $(l,b)=(226,+49)$\arcdeg, a 259 kpc 
heliocentric distance, and Galactocentric velocity ($v_{GSR}$) of 180 
km s$^{-1}$.  However, because the proper motion of Leo~I is unknown, 
this is a free parameter that ultimately determines the shape of the 
model Leo~I orbit.  After assuming a given present proper motion, the 
Leo~I orbit (with the satellite as a point mass) is evolved back in 
time long enough to derive the phase space position of the satellite 
two apogalactica ago.  The point mass is then replaced by the Plummer 
model satellite at this phase space position (after the satellite has 
been allowed to evolve and relax at infinite distance), and a full 
$N$-body simulation is evolved forward to the present time.  
In general, for the models tested here the start of the simulations 
occurs about 11--12 Gyr ago.  In the course of the modeling efforts 
for Leo~I, some 100 differently configured models have been run.  

Initial models were run to fix a likely orientation of the orbit, 
under the assumption that the east-west position angle of the Leo~I 
ellipticity and the orientation of the 
``break population'' are caused by tidal effects and tidal stripping, 
respectively.  $N$-body simulations were run with the satellite 
having an orbital pole every 45\arcdeg\ along its allowed set of 
poles \citep[its ``great circle pole family''; see][]{Palma02}.  
It was found that a satellite with orbital pole near $(l,b)=(122,+13)
$\arcdeg\ yielded tails with the proper orientation, though this 
simulation cannot be discriminated (on the basis of the direction 
of Leo~I's stretching alone) from one with a satellite having an 
opposite orbital pole (i.e. [302, -13]\arcdeg).  We adopt either of 
these antipodal orbital poles for the remaining simulations, which 
thereby fixes the position angle of the present Leo~I proper motion 
to $\sim75$\arcdeg\ or $\sim255$\arcdeg\ in Galactic coordinates.

The remaining simulation variables we explore are the initial 
satellite mass and scale ($r_{\rm 0}$) --- which determine the size 
and density of the satellite and therefore the degree of satellite 
disruption --- and the magnitude of the proper motion --- which 
determines the properties (eccentricity, peri-and apogalacticon) of 
the orbit and also affects the degree of satellite disruption.  
Given the extreme distance and RV of Leo~I, it seems probable that 
Leo~I has a rather elliptical orbit \citep{Taylor04}.  Thus, our 
modeling efforts centered on orbits with perigalactica ranging from 
10 kpc to 50 kpc (and consistent with the observed RV of Leo~I); 
despite the large variation in perigalactica, this actually 
corresponds to orbital eccentricities ranging only from 0.80 to 0.96.
With such orbits Leo~I has an orbital period of about 6 Gyr.
It is found that after fixing the satellite mass to of order that 
found in \S4.2, a scale of $r_{\rm 0} = 0.3 \pm 0.1$ kpc yields a 
final Leo~I satellite with a tidal radius of order the observed 
Leo~I King limiting radius.  Thus, we adopt models with this general 
structure, and explore how varying the orbit shape is reflected in 
the resultant radial light profile and velocity distribution.  
For the latter, we ``observe'' radial velocities sampled from the 
model distribution in a spatial ``footprint'' mimicking that of the 
Keck DEIMOS spectra as shown in Figure~\ref{f:keckspec_spatialplot}; 
this gives the most direct and fair comparison to the observed RV 
distribution seen in Figures \ref{f:keckspec_hist} and 
\ref{f:vdprofile}.

\begin{figure}
\epsscale{0.9}
\plotone{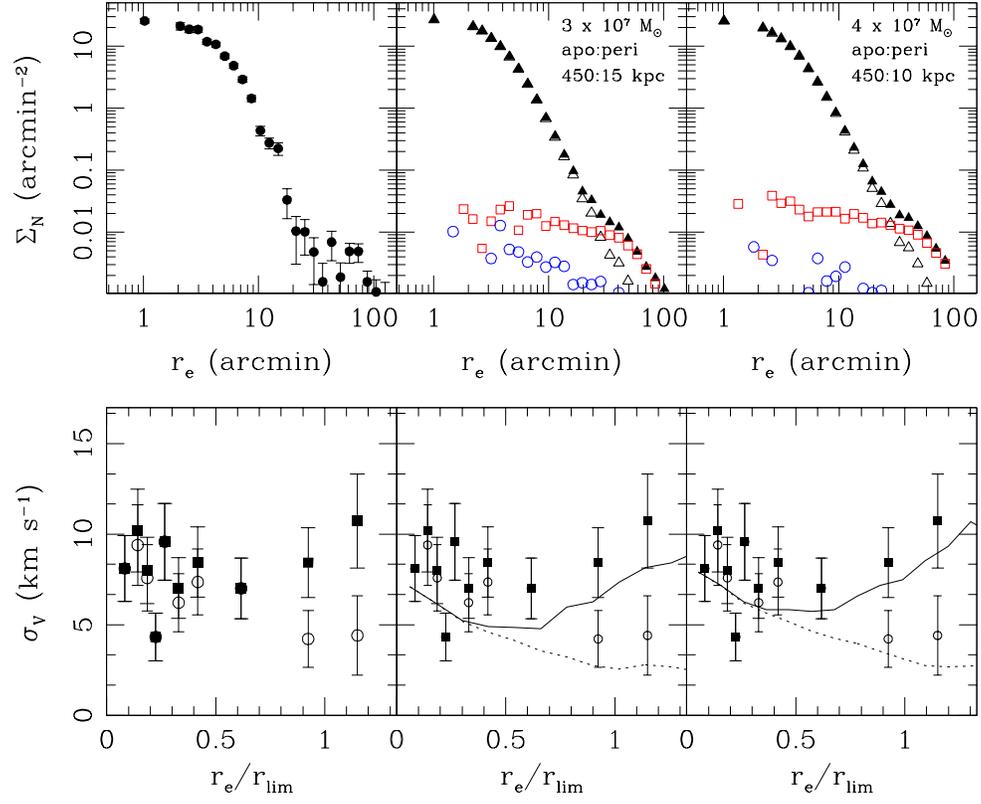}
\caption{ Comparisons of observed ({\it left panels}) surface radial density 
          profile ({\it upper panels}) and velocity dispersion profile 
          ({\it lower panels}) with those of $N$-body models.  The {\it closed 
          triangles} in the model surface density profiles are for the 
          entire sample, while {\it open triangles} are for the bound stars.
          The {\it open circles} and {\it open squares} in the {\it top panels} 
          are for unbound stars from the first and second mass loss events, respectively.
          The symbols used for the velocity dispersion profiles are same as those 
          in the {\it mid left panel} of Figure~\ref{f:vdprofile}, while the 
          {\it solid} and {\it broken lines} show the model velocity dispersion profiles 
          for the entire sample and the bound stars, respectively.
         }
\label{f:nbody_profiles}
\end{figure}

\begin{figure}
\epsscale{1.0}
\plotone{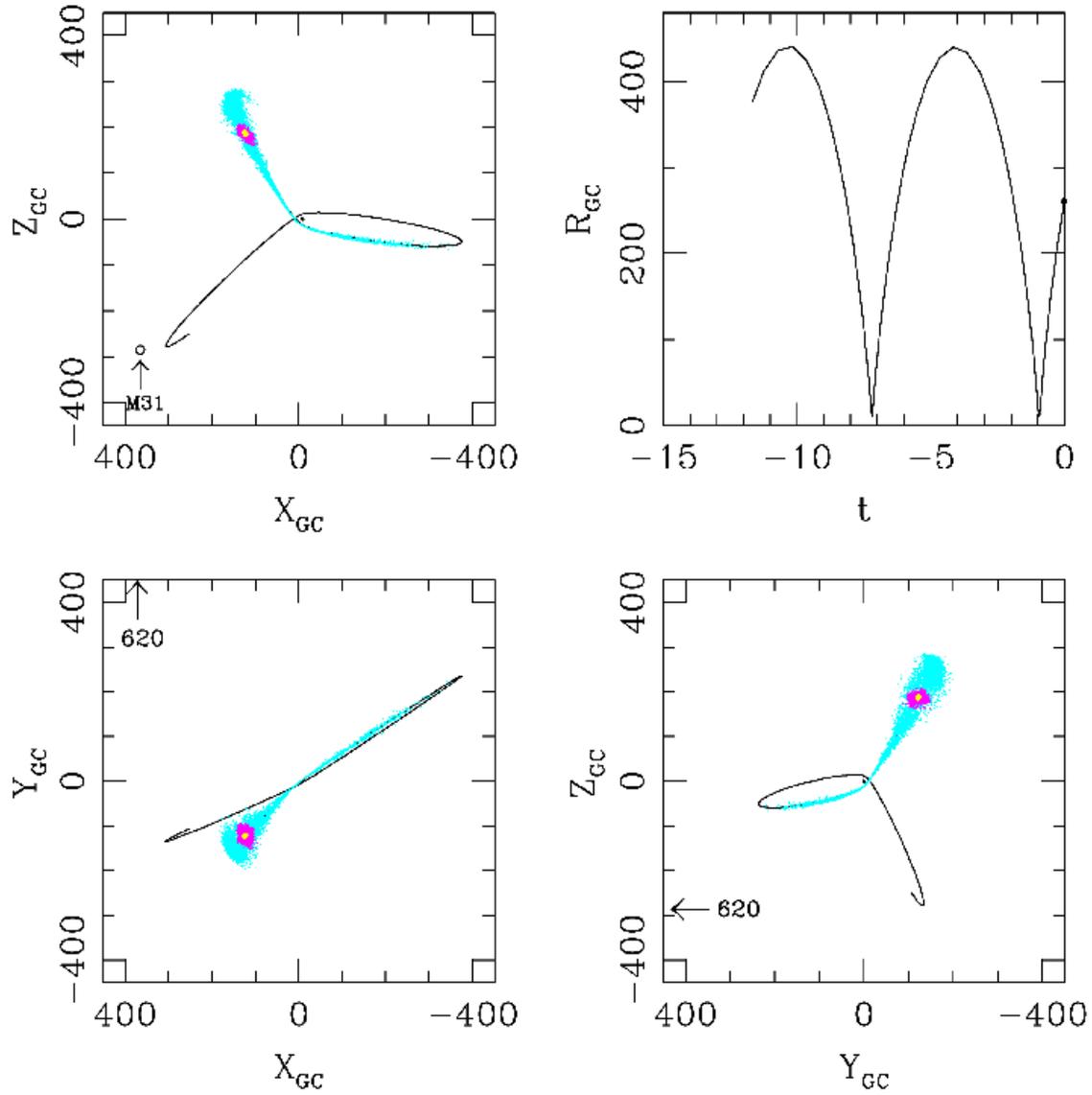}
\caption{ Orbital properties of model 117. This orbit is very similar to the one
          for model 111 since the primary difference between them is the closer perigalacticon
          distance in model 117. White dots (colored Yellow in electronic edition) represent 
          particles that are still bound.  Dark gray (magenta in electronic edition) and 
          light gray (cyan in electronic edition) mark particles that became unbound in the 
          first and second perigalacticon passages, respectively. The current position of M31 
          is indicated as arrows.
         } 
\label{f:nbody_orbit}
\end{figure}

It is found that a variety of simulations of Leo~I on an eccentric 
orbit can reproduce not only the overall ``King+break'' radial light 
profile that is characteristic of disrupting satellites 
\citep{Johnston98}, but also an asymmetric RV distribution and a more 
or less flat velocity dispersion profile like that observed.  
To fine tune the model, we took account of three general correlations 
in turn (see \citealt{Munoz07}): (1) The central velocity 
dispersion directly reflects the adopted satellite mass. (2) The size 
of the observed ``King profile'' part of the satellite is set by the 
initial Plummer model scale $r_{\rm 0}$.  (3) With mass and scale set 
by the previous conditions, the mass loss rate (hence the size of the 
break population) is then only driven by the orbital shape (i.e. 
impact parameter to the MW).  Following these general guides, a 
narrow set of mass, scale and orbital shapes were found to give 
reasonable matches to the observed Leo~I properties.  The density 
and RV properties of two of these ``matching'' models are compared 
against the data in Figure~\ref{f:nbody_profiles}, and their 
three-dimensional orbits are shown in Figure~\ref{f:nbody_orbit}.  

The parameters and results for adopted models are listed in Table 
\ref{t:modelparams}.  Models 111 and 117 have been run in the MW 
potential for 12.0 Gyr and 11.8 Gyr, respectively.  In both cases 
the net mass loss is modest, with only 16\% and 23\% of the initial 
mass being unbound after 12 Gyr.  We note that the implied mass loss 
rates for both models roughly match the estimates made in \S 6.1.2.  
As found in previously published simulations of satellite disruption 
\citep[e.g.,][]{Law05}, the break populations in the extreme orbit, 
Leo~I model satellite density profiles are also found to be due to 
tidal disruption; this is shown in Figure~\ref{f:nbody_profiles}, 
where stars that are still bound as well as those that have been lost 
and become unbound on each of the last two radial orbits are marked 
with separate symbols and colors.  Thus, our original supposition 
that the observed Leo~I break population may be due to unbound stars 
is supported by the models.  
However, the relative density at which these breaks occur does not 
match that of the Leo~I density profile if the beginning of the 
unbound debris is associated with the inner of the two radial 
profile breaks.  Said another way, these models provide a good 
match to all the Leo~I data if the first break in the radial 
density profile is considered as statistically insignificant or 
structurally unrelated to unbound tidal debris.  Even if the latter 
suppositions are not valid, we contend that these models still 
provide as good or a better simultaneous match to the overall observed 
structure {\it and} dynamics of any specific dSph than has been 
offered before.  

\begin{figure}
\epsscale{1.0}
\plotone{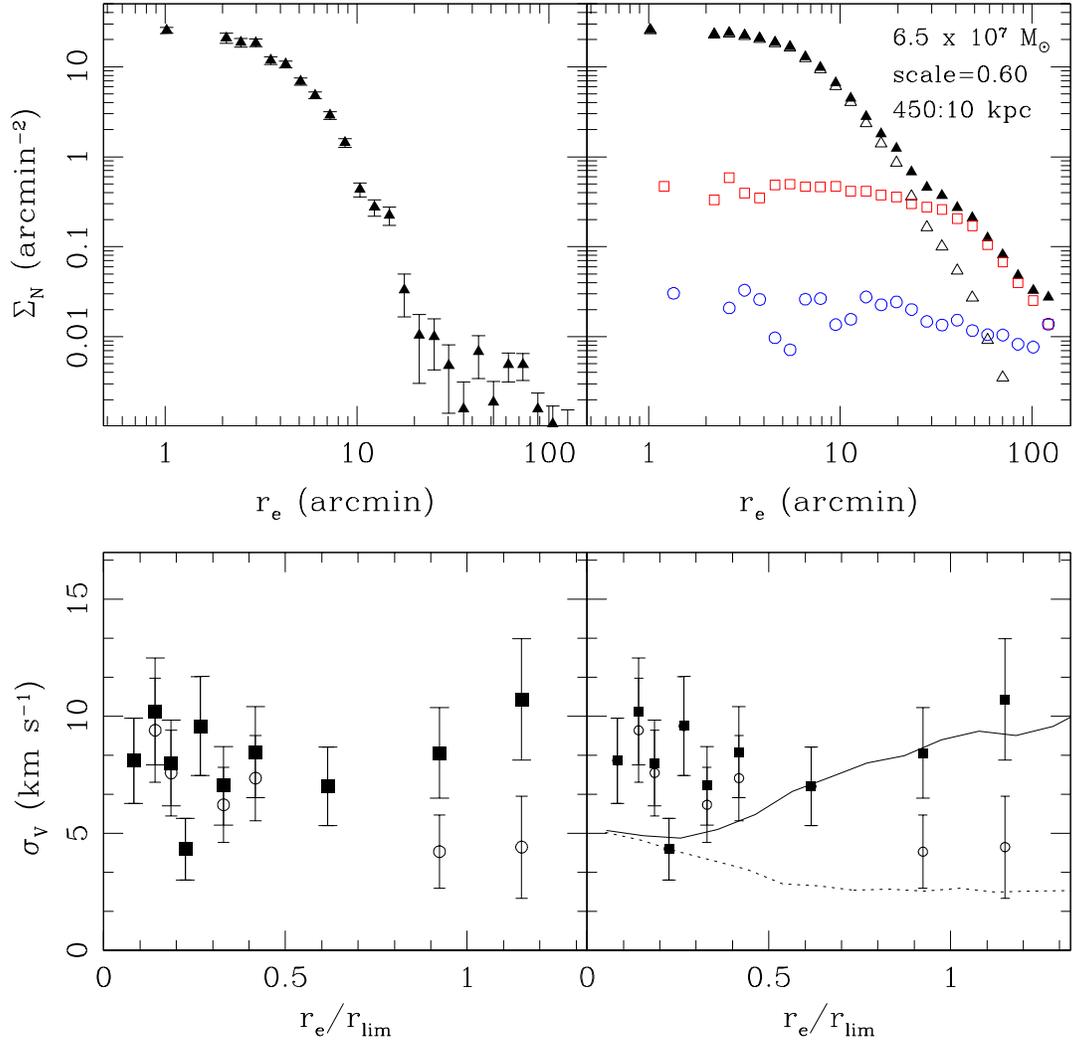}
\caption{ Comparisons of observed ({\it left panel}) surface radial density
          profile ({\it upper panels}) and velocity dispersion profile
          ({\it lower panels}) with those of model 122.  The symbols and lines are 
          same as those in Figure~\ref{f:nbody_profiles}
         }
\label{f:nbody_highbreak}
\end{figure}

We also attempted to find a model that could accommodate the inner 
density law break.  Since the relative density at which tidally 
induced break populations occur is directly related to the 
instantaneous rate at which debris is being generated, in order to 
match the relative position compared to the central density 
of the inner break at higher density we need to increase the mass 
loss rate of the satellite.  This can be done by decreasing the 
satellite density.  Figure~\ref{f:nbody_highbreak} shows the 
properties of model 122 which has an initial mass of $6.5 \times 
10^{7}$ M$_{\sun}$, a scale of 0.55 kpc and apo:perigalacticon of 
450:10 kpc.  Whereas the position of the model break is now matched 
to the inner of the two observed density breaks, the model satellite 
is found to be three times larger than the actual Leo~I, and the 
velocity dispersion as well as the overall density profile are no 
longer a good match.  We therefore find model 122 to provide a less 
satisfactory description of Leo~I than models 111 and 117.

\begin{figure}
\epsscale{1.0}
\plotone{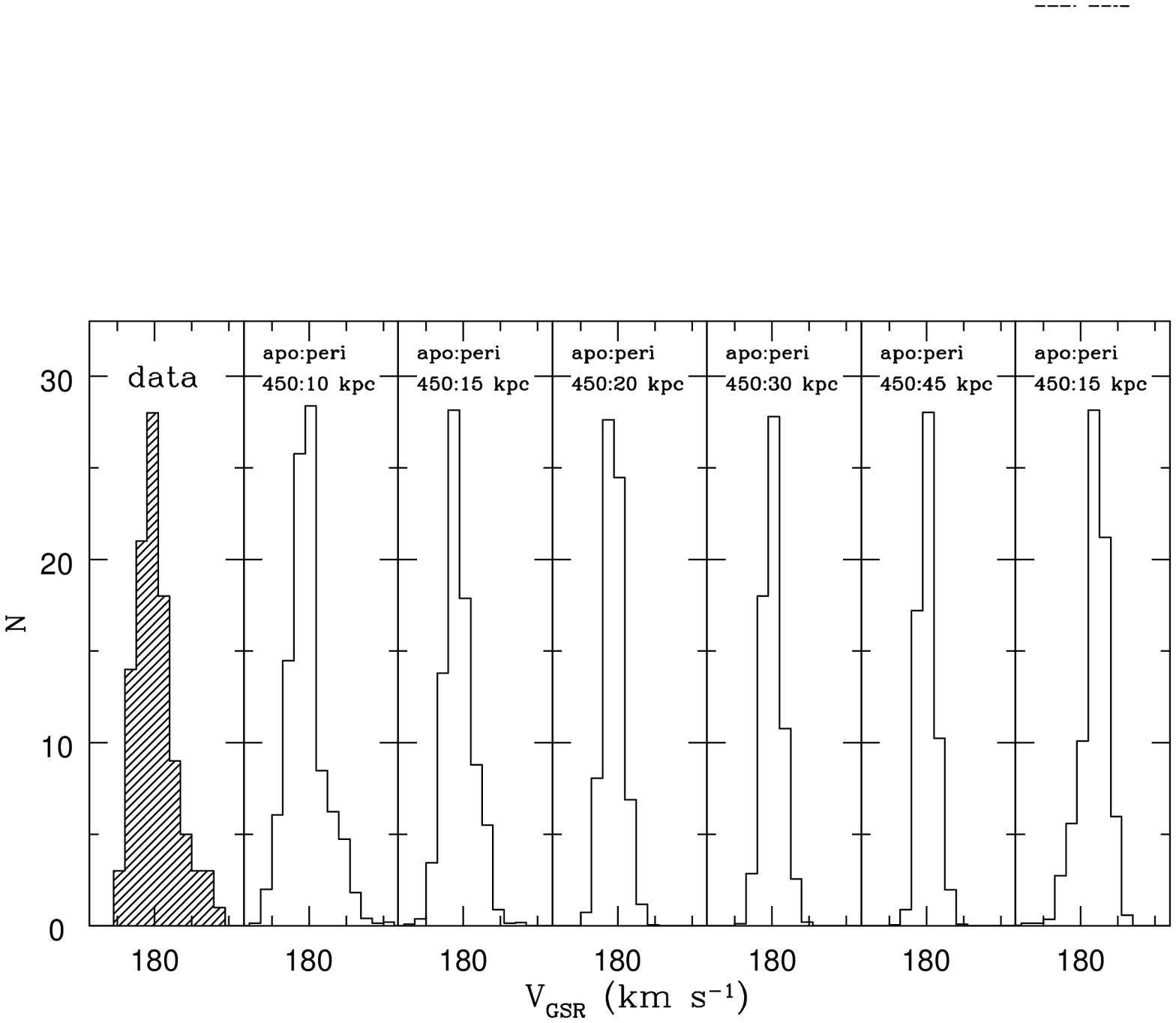}
\caption{ The observed Galactocentric radial velocity distribution ({\it leftmost 
          panel}) compared with the model velocity distributions using the DEIMOS footprint
          for Leo~I.  In each model 
          panel, the apogalacticon and perigalacticon distances are noted.  The 
          {\it rightmost panel} is for a model with opposite orbital pole.
         }
\label{f:nbody_hist}
\end{figure}

The final satellite masses for Models 111 and 117 (see Table 
\ref{t:modelparams}) are consistent with the Leo~I mass derived from 
core fitting in \S4.2 within the errors.  The implied current total Leo~I 
mass-to-light ratio from the models is $M/L = 3.1-4.5 (M/L)_{\sun}$.  
This lower $M/L$ than found in \S4.2 implies a relatively small DM 
content, and would bring Leo~I in line with the $M/L$ typical of dE 
galaxies and even globular clusters.  It is important to recognize 
that we have obtained a good match to the Leo~I RV asymmetry with 
both of the preferred models {\it even though this observed property 
was not used as a model constraint}.  That such a result comes 
naturally lends further confidence to the plausibility of our models.  
The RV asymmetry in the models arises from the long extension of the 
trailing tidal arms towards the inner Galaxy (and the observer) seen 
in Figure~\ref{f:nbody_orbit} and for which the innermost parts have 
a significant projection along the line of sight in Models 111 
and 117.  However, the degree of that projection, and therefore the 
velocity spread of the asymmetry is obviously tied with the 
eccentricity of the orbit.  This can be seen, for example, by the 
results (Fig. \ref{f:nbody_hist}) of a series of models run with the 
same mass and scale, but more circular orbits than shown in Figure 
\ref{f:nbody_orbit}.  As Figure~\ref{f:nbody_hist} shows, less 
eccentric orbits have less asymmetry in their observed RV 
distribution.  In addition, as might be expected, these models give 
poorer matches to the radial density and velocity dispersion 
profiles.  In this way, the observed RV distribution can apparently 
directly constrain the detailed {\it shape} of the Leo~I orbit, and 
by direct comparison to the observed RV distribution, we deduce that 
Leo~I has an orbital eccentricity similar to that shown in Figure 
\ref{f:nbody_orbit}.  We also deduce the general {\it direction} of 
the Model 111/117 orbits (i.e. the general direction of the proper 
motion) to be correct (i.e. to the east), since a satellite with a 
similar eccentricity orbit but opposite orbital pole yields the 
opposite RV asymmetry for our Keck spectroscopy mask footprint (Fig. 
\ref{f:keckspec_spatialplot}).  This is demonstrated by the model 
results shown in the right-most panel in Figure~\ref{f:nbody_hist}.

Finally, because we have shown by the models that stars in the break 
population are well described as nascent tidal tails, and by the analysis 
in \S 6.1.1 we have shown the Leo~I break population lies predominantly 
east-west, we conclude that the orbital pole of 
$(l,b)=(122,+13)$\arcdeg\ is approximately correct.  Thus, based on 
the apparent discriminatory power of the Figure 
\ref{f:nbody_profiles} parameters as well as the direction of the 
tidal arms, we conclude that the orbit shown in Figure 
\ref{f:nbody_orbit} to be a reasonable approximation to the true 
Leo~I orbit.  A check on the orbit will obviously be delivered by the 
measurement of the Leo~I proper motion, which can be expected in a 
decade or so from a key project (led by SRM) of NASA's Space 
Interferometry Mission.  The expected current proper motions for 
Leo~I are predicted to be $(\mu_l \cos b, \mu_b) = (0.0046, 0.0219)$ 
and $(0.0020, 0.0138)$ mas yr$^{-1}$, respectively for Models 111 and 
117 (including solar motion, which is assumed to have a 232 km 
s$^{-1}$ revolutionary speed about the MW, and additional peculiar 
motion of 9.0 and 7.0 km s$^{-1}$ in the Galactic radial and $Z$ 
directions, respectively).

In the meantime, verification of the overall picture painted by our model 
fitting here, including the orbit, can be made by obtaining more RVs 
on the {\it east} side of Leo~I.  Our models predict that at large 
radii on the east side of Leo~I an asymmetry of the RV distribution 
should be observed {\it opposite} (i.e., toward lower RVs) that we 
have found to the west.  The change in the sense of asymmetry arises 
from the sampling of the {\it leading arm} of the Leo~I tidal tails 
on the east side, whereas our western RVs have been sampling trailing 
arm debris.  Mapping the Leo~I tidal tails to larger angular 
separations would also provide significant leverage on the Leo~I 
orbit and mass loss history.  Figure~\ref{f:nbody_orbit} suggests 
that quite long tails should exist for two orbits of mass loss.  
However, the problem is quite observationally challenging because the 
tails are at fairly low surface brightness and the corresponding 
densities of Leo~I tidal tail giant stars will eventually become 
quite sparse.  A search for the main sequence turn-off CMD feature 
for the Leo~I tidal arms near the satellite will require reaching 
to $V > 28$ over large areas.  From the ground, such work is severely 
hampered by the difficulty of star/galaxy separation at these 
magnitudes (D. Martinez-Delgado, private communication).

\subsubsection{Implications for the Local Group Path of Leo~I and Mass of the Galaxy}

The provenance of Leo~I has important implications for the mass of 
the MW.  The high RV of Leo~I at 259 kpc translates to a large 
implied lower limit to the escape velocity of the MW if this 
satellite is bound.  If Leo~I has made two orbits about the MW, as 
assumed in our models, then it must be bound.  If it has made only 
one perigalactic passage, Leo~I could still be bound to the MW, 
having {\it become} so on the last orbit; however, in this case 
Leo~I may also be unbound and on a hyperbolic orbit.

\begin{figure}
\epsscale{1.0}
\plotone{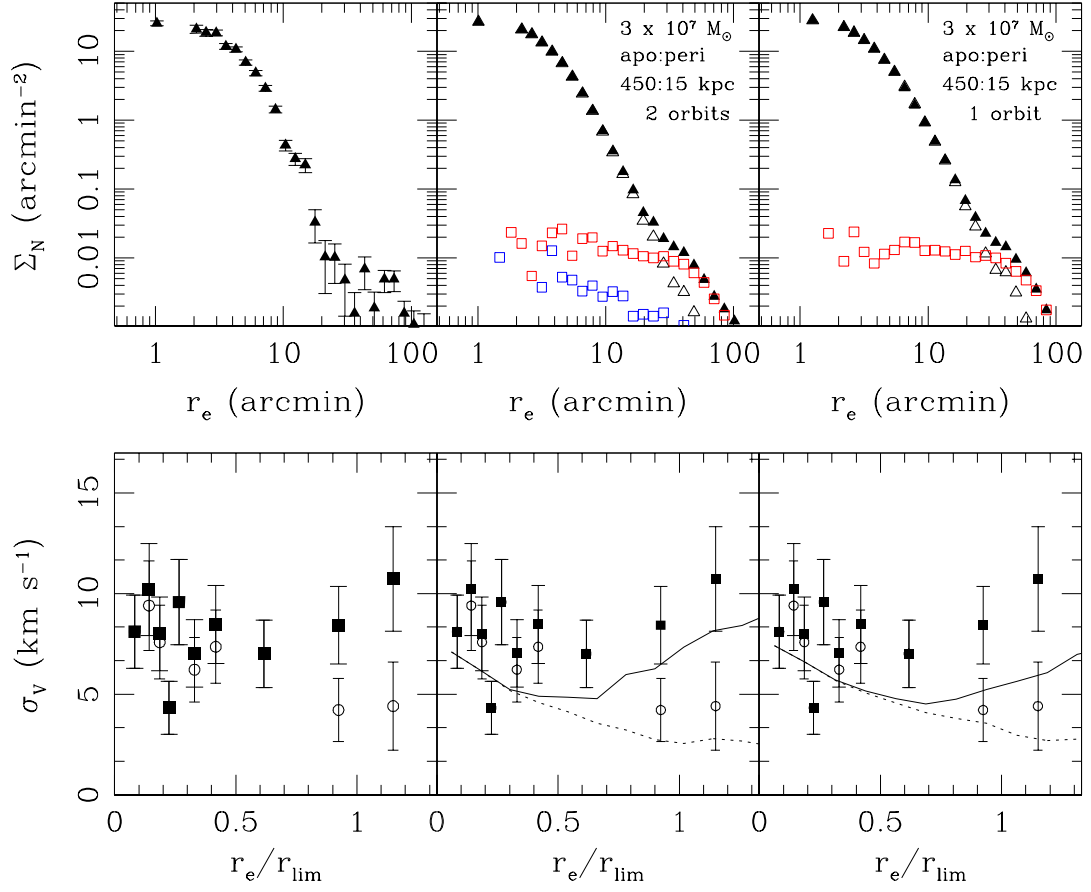}
\caption{ Comparisons of observed ({\it left panel}) surface density profile
          and velocity dispersion profile with those of models having the same
          orbital and structural properties, but where one has had 2 perigalacticon
          passage (mid panels) and the other only one (right panels).  
          The symbols and lines are same as those in Figure~\ref{f:nbody_profiles}.
         }
\label{f:nbody_1orbit}
\end{figure}

Unfortunately, it is not clear that we can, with certainty, establish 
whether two or only one orbits have occured.  According to the 
models, much of the unbound debris observed near Leo~I has detached 
in the last orbit (see Fig. \ref{f:nbody_orbit}).  To verify whether 
any differences can be discerned between a one and two orbit Leo~I, 
we ran a simulation (Model 118) with the same satellite and orbital 
properties as Model 111, but where only one radial orbit has occured 
(i.e. the model is started 4.3 Gyr ago).
Figure~\ref{f:nbody_1orbit} shows that the radial density profile 
from such a simulation is virtually indistinguishable from the two 
orbit Model 111.  This is because of the small contribution of older 
debris to the density of Leo~I {\it over the currently observed 
area}.  On the other hand, this small contribution does become more 
obvious in its influence on the observed {\it velocity dispersion}, 
as seen in the bottom panels of Figure~\ref{f:nbody_1orbit}: The one 
orbit model has a smaller velocity dispersion at large radii compared 
to that of the two orbit model, where older debris helps inflate the 
dispersion.  The latter, two orbit dispersion profile is a closer 
match to that observed for Leo~I, and so, based on this evidence 
alone can we tentatively suggest that a two orbit model is favored 
for Leo~I. 

The star formation history of Leo~I may lend circumstantial support 
to the two orbit scenario.  Models 111 and 117 suggest that Leo~I 
endured fairly substantial tidal shocking both $\sim1$ and $\sim7.5$ 
Gyr ago.  This general orbital picture seems to coincide with the 
star formation history of Leo~I studied by \citet{Gallart99b}, who 
finds that most of its star formation activity occurred between 
7 and 1 Gyr ago.  The oldest ($>10$ Gyr) Leo~I population, discovered 
by \citet[][blue HB stars]{Held00} and \citet[][RR Lyrae variables]
{Held01}, likely formed as a result of the initial collapse of gas 
that led to the formation of Leo~I.  The ``major'' star formation 
that started $\sim7$ Gyr ago is roughly timed with the previous 
perigalactic passage in our model (see also models by Mayer et al. 
2001).  We can only speculate as to the 
causes of the abrupt drop in star formation at $\sim1$ Gyr ago.  
Leo~I may have simply ran out of gas to fuel the star formation.  
The absence of gas in Leo~I\footnote{\citep{Blitz00} finds extended 
\ion{H}{1} emission around Leo~I but not toward the galaxy itself.} 
\citep{Knapp78,Bowen97} is consistent with this picture.  
The coincidence that Leo~I passed through the inner Galaxy about this 
time (and possibly with a rather small impact parameter; Fig. 
\ref{f:nbody_orbit}) may also suggest gas stripping by the MW 
disk or perhaps by a dense high velocity cloud.  A massive 
($3\times10^{6} M_{\sun}$) fragmented \ion{H}{1} cloud structure 
around Leo~I \citep{Blitz00} perhaps indicates that the gas has been 
perturbed by tidal shocking.  Nevertheless, something triggered star 
formation at a time that matches well the timing of the first 
perigalacticon in our two orbit model, and, together with the velocity 
dispersion data, our results support the hypothesis of Z89 that Leo~I 
is bound to the MW and has had two radial orbits in this state.

If Leo~I is bound, a large MW mass is implied.  The total mass of the 
MW within $260$ kpc from the center in our simulations is $1.8 \times 
10^{12}$ M$_{\sun}$, and it has a profile yielding a mass interior to 
50 kpc that is consistent with those found in the analyses by Z89, 
\citet{Wilkinson99}, and  \citet{Sakamoto03} when these analyses 
include Leo~I as a bound satellite.  Verifying the length of the 
Leo~I tidal tails would verify whether it has had multiple orbits 
around the MW and is thus bound.  Obviously, a definitively measured 
proper motion for Leo~I would also help determine the true orbit of 
Leo~I and refine the mass of our Galaxy.

What do our results say about the origin of Leo~I?  First, it is 
probable that a hyperbolic orbit for Leo~I would produce similar 
results (over the sky area we have surveyed here) to those seen for a 
single orbit, bound model.  Moreover, such an orbit will have an 
overall shape not dissimilar in overall orientation and general 
direction to the last orbit of Models 111 and 117, and thus we may 
determine from which direction Leo~I approached the MW in this case.
Interestingly, in either the one orbit (bound or hyperbolic) or two 
orbit case, our modeling seems to rule out any close association of 
Leo~I with M31 over relevant timescales to tidal stripping: (1) In 
the bound case, Figure~\ref{f:nbody_orbit} shows that Leo~I is not in 
the vicinity of M31 (which has current MW coordinates of 
$(X,Y,Z)_{GC} \sim (375,620,-285)$ kpc) since approximately two 
orbital periods ago, and even then the inferred distance between M31 
and Leo~I is $\sim 700$ kpc (and this ignores the motion of M31 over 
a Hubble time!).  (2) In the hyperbolic case, Leo~I would have 
apparently entered the Local Group in an orbital plane almost 
perpendicular to the direction of M31 (i.e., from the general 
direction of the last apogalactica shown in Figure 
\ref{f:nbody_orbit}).  Were the \citet{Byrd94} hypothesis that Leo~I 
was once bound to M31 to be true, it would have had to have been 
released from M31 at least $\sim10$ Gyr ago.  

Finally, we add a note of caution about interpreting the potential 
shape of the Leo~I orbit as inferred or extrapolated from Figure 
\ref{f:nbody_orbit}, which represents a model run in a {\it static} MW 
potential.  Obviously, currently favored hierarchical galaxy formation models 
imply a continuing growth of the Galactic potential with time, and this
will alter the orbits of satellites.  On the other hand, the MW formation
models of \citet{BJ05} suggest that destroyed satellites contributing mass
to the MW halo accrete predominantly before the last two postulated
Leo~I perigalactica (i.e., before 7 Gyr ago).  Thus, at least over this 
timescale, a static MW approximation, and the resulting Figure 
\ref{f:nbody_orbit} orbit, may be appropriate descriptors.  Moreover, 
as recently shown by \citet{Penarrubia2006}, tidal streams respond
adiabatically to evolving potentials, so that even if the MW potential
evolved over the past 7 Gyr or so, the Leo~I tidal stream visible today
would resemble that derived in a static MW potential with the present day
mass profile.

\section{Summary and Conclusions}

We have photometrically surveyed a 4.5 degree$^2$ region centered on 
Leo~I in the $M$, $T_{2}$, $DDO51$ filter system in order to explore 
the extended morphology of this dwarf spheroidal galaxy, which is 
currently regarded as the most distant of the known Galactic 
satellites (unless the more distant Phoenix system is bound to the 
MW).

The photometric data were used to select Leo~I giant candidates based 
on the two criteria: (1) the gravity sensitive ($M-DDO51$, $M-T_2$) 
diagram, which separates distant giant stars from contaminating, foreground, 
metal-rich disk dwarfs, and (2) CMD positions commensurate with the  
temperature--apparent magnitude combination of 
stars on the Leo~I giant branch.  The background level of our 
``Leo~I giant star'' sample is determined to be small, and a 100\% 
reliability in the identification of bona fide Leo I giants is found 
via testing with a total of 133 stars in the Leo I field with 
previously published or new Keck spectroscopy.
 
We derive a new set of Leo~I structural parameters by fitting a 
single-component King profile to the Leo~I giant candidates.  
Coupling this to the central velocity dispersion we have measured (\S 5.5), 
we use core-fitting techniques to derive a mass for the Leo I system of
$3.5 \times 10^{7} M_{\sun}$ and a total $(M/L)_{V}$ of 
4.6$M_{\sun}/L_{\sun ,V}$,\footnote{These values are 14\% smaller than those in
\S4.2, where the M98 central velocity dispersion was used to derive the mass.} 
values not too dissimilar from previous 
study using the same technique (M98).  The two-dimensional 
distribution of Leo~I giant candidates shows many giant stars outside 
the derived King limiting radius.  These are primarily along the 
major axis and spectroscopy of a subsample of these ``extratidal'' 
stars shows they are actually associated with the dSph.  
This population of stars shows up as a break at a major axis radial 
distance of $\sim 10$ arcmin in the radial density profile.  
Our new Keck spectroscopy confirms this excess of stars beyond the 
King limiting radius to be made of true Leo~I members.

Our Leo~I velocity dispersion profile shows a flat and then rising trend at large 
radii.  We also find that the Leo~I RV distribution, particularly 
for stars at large angular separations (which in our data are 
primarily to the west side of Leo I), includes a population of stars with a broad
and skewed, asymmetrical 
distribution toward positive RVs overlapping a second, very much colder population 
at the nominal Leo~I mean velocity.

We interpret these features as support for a picture wherein Leo~I has 
been tidally disrupted on at least one, but at most two, 
perigalactic passages by a massive Local Group member (most likely 
the MW for both), and find these phenomena naturally produced in 
mass-follows-light $N$-body simulations of a tidally disrupting Leo~I analog
in a Milky Way-like potential.  The best-matching $N$-body 
simulations to both the observed structural morphology and 
velocity distribution of Leo I are those where the tidally disrupting 
satellite is on a rather high eccentricity (0.93--0.96), small 
perigalacticon (10-15 kpc), bound orbit around the Milky Way, and has 
a present total $M/L \sim 3-4.5 (M/L)_{\sun}$.  These best fitting 
model masses are 58--85\% the $M/L$ derived from the central velocity 
dispersion and core fitting, but it is not unreasonable to presume 
that the latter method yields somewhat inflated masses because of 
systematic increases in the true dispersion due to 
astrophysical processes such as the 
presence of binaries and the atmospheric jitter common to giant star 
atmospheres.  Thus, the likely $M/L$ of the satellite is rather 
modest and not unlike those of other elliptical systems of approximately similar 
mass scale that are typically regarded as low in, or devoid of, DM 
(dE galaxies and globular clusters).  

Because of the rather close match between our model results and 
observations, and because disrupting satellites on highly radial 
orbits appear to yield great discriminatory power in this regard, 
we have been able to constrain the likely orbit of Leo I {\it without 
the measurement of its proper motion}.  The orientation of the 
current satellite orbital plane can be fixed by matching model tidal 
tails to the predominant direction of the observed break population 
(i.e., more or less along the major axis), whereas we find that the 
{\it direction of angular momentum} in this orbital plane is well 
constrained by the sense of the RV asymmetry we have observed to the 
west side of the satellite.  Our models demonstrate that a positive 
RV asymmetry in the models is produced by {\it trailing} tidal 
debris for the receding satellite.  Thus, we predict that an opposite 
RV asymmetry will be found on the east side of Leo~I from leading 
tidal debris.  Such a result would provide an important verification of the tidal 
disruption model we have put forward in this paper.  

Our observed Leo~I RV distribution is most consistent with a two MW 
orbit history for Leo~I, with both orbits around the MW, however we 
cannot yet definitively rule out a one orbit scenario.  However, 
whether Leo~I is bound to the MW or on an unbound, hyperbolic orbit 
around the MW, our results seem to rule out a Leo~I orbit that 
includes a previous association with M31 within the last 10 Gyr, 
in contradistinction to the \citet{Byrd94} scenario of a
relatively recent origin of Leo~I from M31.  

Leo I has long played a ``spoiler'' role in setting the mass of 
the MW because of its huge $rv^2$ lever arm in Jean's equation-based 
determinations using tracer particles of unknown proper motion 
(i.e. unknown orbit size and shape.  The large implied escape velocity at Leo~I's 
distance implies large MW masses if Leo~I is bound to 
our galaxy (Z89, \citealt{Wilkinson99}), although its influence 
on the mass determination is lessened when more complete samples of 
objects with complete phase space data are employed 
\citep[e.g.,][]{Sakamoto03}.  Our observed RV distribution for Leo~I 
slightly favors a two perigalacticon pass, bound MW orbit for Leo~I, 
thereby suggesting that higher mass estimates for the MW may be more 
correct.  

In general, our tidally disrupting mass-follows-light satellite 
models provide a quite satisfactory match to the observed properties 
of Leo~I, but a few details --- namely the apparent double break in 
the density profile and the density at which the inner of these 
breaks occur --- we have yet to account for in these initial 
modeling efforts.  Nevertheless, we contend that the scenario of a 
tidally disrupting, low $M/L$ system on a highly radial orbit 
provides a rather complete explanation for the observed properties 
of Leo~I.  While some properties of Leo~I and other dSphs (e.g., 
flat velocity dispersion profiles and break populations) have also 
been explained by postulating that these systems are embedded in 
extended DM halos, such an explanation in the case of Leo~I appears 
less compelling in that it cannot account for as many of the 
observed properties of the system (e.g., the asymmetry in the RV 
distribution and increasing ellipticity with radius).  In contrast, 
tidal tails by now provide a well-established observational 
paradigm for dSph satellites, with the Sagittarius dSph the most vivid example.  
We contend that the tidal disruption of Sagittarius is not unique and that 
Leo~I may be another example of the phenomenon, albeit at a much 
lower mass loss rate commensurate with a satellite on such an 
extreme orbit.  Tidal disruption observed among both the closest 
and farthest of the Milky Way satellites suggests that this process 
may be ubiquitous, and that similarly structured satellites with elliptical orbits and
distances between these two examples might also be expected to be 
experiencing tidally induced mass loss, likely with an intermediary 
range of mass loss rates. 

\acknowledgments

The authors wish to thank Chris Palma for providing several useful computer
programs and Alison Coil for aiding the Keck DEIMOS observations.  
The referee has provided a number of useful comments that have improved the paper.
Support for this work at Virginia was provided by NSF grants 
AST-9702521 and AST-0307851, a Cottrell Scholar Award from The 
Research Corporation, NASA/JPL contract 1228235, the David and 
Lucile Packard Foundation, and the generous support of The F. H. 
Levinson Fund of the Peninsula Community Foundation.  
KVJ's contribution was supported through  NASA grant NAG5-9064 and 
NSF CAREER award AST-0133617.  MHS received thesis travel support 
from KPNO that was used during the collection of the Mosaic data
and support from the STScI Director's Discretionary 
Research Fund.

\clearpage


\begin{deluxetable}{lccccc}
\tablewidth{0pt}
\tabletypesize{\footnotesize}
\tablecaption{ Leo~I observation log. \label{t:obs_log}}
\tablehead{ \colhead{}      & \colhead{RA}      & \colhead{Dec}     & \multicolumn{3}{c}{Exp time(sec) / Airmass / FWHM(arcsec)}   \\
            \cline{4-6}
            \colhead{Field} & \colhead{(J2000)} & \colhead{(J2000)} & \colhead{$M$} & \colhead{$T_{2}$} & \colhead{$DDO51$}
          }
\startdata
 C     & ~10:08:28.69~ & ~$+$12:18:17.2~ &  ~70\hspace{0.1cm}/\hspace{0.1cm}1.16\hspace{0.1cm}/\hspace{0.1cm}1.2~ &  ~70\hspace{0.1cm}/\hspace{0.1cm}1.15\hspace{0.1cm}/\hspace{0.1cm}0.8~ &  ~700\hspace{0.1cm}/\hspace{0.1cm}1.13\hspace{0.1cm}/\hspace{0.1cm}0.9~ \\
 NN    & ~10:08:29.27~ & ~$+$13:18:16.8~ &  ~70\hspace{0.1cm}/\hspace{0.1cm}1.20\hspace{0.1cm}/\hspace{0.1cm}1.2~ &  ~70\hspace{0.1cm}/\hspace{0.1cm}1.19\hspace{0.1cm}/\hspace{0.1cm}0.9~ &  ~700\hspace{0.1cm}/\hspace{0.1cm}1.17\hspace{0.1cm}/\hspace{0.1cm}1.0~ \\
 EE    & ~10:12:28.46~ & ~$+$12:18:08.7~ &  ~70\hspace{0.1cm}/\hspace{0.1cm}1.29\hspace{0.1cm}/\hspace{0.1cm}1.1~ &  ~70\hspace{0.1cm}/\hspace{0.1cm}1.28\hspace{0.1cm}/\hspace{0.1cm}0.9~ &  ~700\hspace{0.1cm}/\hspace{0.1cm}1.25\hspace{0.1cm}/\hspace{0.1cm}1.0~ \\
 WW    & ~10:04:28.91~ & ~$+$12:18:25.2~ &  ~70\hspace{0.1cm}/\hspace{0.1cm}1.46\hspace{0.1cm}/\hspace{0.1cm}1.2~ &  ~70\hspace{0.1cm}/\hspace{0.1cm}1.44\hspace{0.1cm}/\hspace{0.1cm}1.0~ &  ~700\hspace{0.1cm}/\hspace{0.1cm}1.39\hspace{0.1cm}/\hspace{0.1cm}1.1~ \\
\hline
 N     & ~10:08:28.12~ & ~$+$12:48:23.4~ & ~100\hspace{0.1cm}/\hspace{0.1cm}1.08\hspace{0.1cm}/\hspace{0.1cm}2.2~ & ~200\hspace{0.1cm}/\hspace{0.1cm}1.09\hspace{0.1cm}/\hspace{0.1cm}1.8~ & ~1000\hspace{0.1cm}/\hspace{0.1cm}1.10\hspace{0.1cm}/\hspace{0.1cm}2.3~ \\
 E     & ~10:10:28.12~ & ~$+$12:18:23.4~ & ~100\hspace{0.1cm}/\hspace{0.1cm}1.21\hspace{0.1cm}/\hspace{0.1cm}2.0~ & ~200\hspace{0.1cm}/\hspace{0.1cm}1.23\hspace{0.1cm}/\hspace{0.1cm}1.6~ & ~1000\hspace{0.1cm}/\hspace{0.1cm}1.27\hspace{0.1cm}/\hspace{0.1cm}1.8~ \\
 W     & ~10:06:28.11~ & ~$+$12:18:23.4~ & ~100\hspace{0.1cm}/\hspace{0.1cm}1.15\hspace{0.1cm}/\hspace{0.1cm}1.8~ & ~200\hspace{0.1cm}/\hspace{0.1cm}1.16\hspace{0.1cm}/\hspace{0.1cm}2.0~ & ~1000\hspace{0.1cm}/\hspace{0.1cm}1.19\hspace{0.1cm}/\hspace{0.1cm}1.7~ \\
 NE    & ~10:10:28.12~ & ~$+$12:48:23.4~ & ~100\hspace{0.1cm}/\hspace{0.1cm}1.35\hspace{0.1cm}/\hspace{0.1cm}1.6~ & ~200\hspace{0.1cm}/\hspace{0.1cm}1.38\hspace{0.1cm}/\hspace{0.1cm}1.6~ & ~1000\hspace{0.1cm}/\hspace{0.1cm}1.44\hspace{0.1cm}/\hspace{0.1cm}1.7~ \\
 NW    & ~10:06:28.55~ & ~$+$12:47:57.1~ & ~100\hspace{0.1cm}/\hspace{0.1cm}1.13\hspace{0.1cm}/\hspace{0.1cm}1.9~ & ~200\hspace{0.1cm}/\hspace{0.1cm}1.14\hspace{0.1cm}/\hspace{0.1cm}2.1~ & ~1000\hspace{0.1cm}/\hspace{0.1cm}1.20\hspace{0.1cm}/\hspace{0.1cm}1.7~ \\
 EEE   & ~10:14:28.12~ & ~$+$12:18:23.4~ & ~100\hspace{0.1cm}/\hspace{0.1cm}1.09\hspace{0.1cm}/\hspace{0.1cm}1.1~ & ~200\hspace{0.1cm}/\hspace{0.1cm}1.09\hspace{0.1cm}/\hspace{0.1cm}1.2~ & ~1000\hspace{0.1cm}/\hspace{0.1cm}1.11\hspace{0.1cm}/\hspace{0.1cm}1.3~ \\
 NEE   & ~10:12:28.11~ & ~$+$12:48:23.4~ & ~100\hspace{0.1cm}/\hspace{0.1cm}1.13\hspace{0.1cm}/\hspace{0.1cm}1.4~ & ~200\hspace{0.1cm}/\hspace{0.1cm}1.14\hspace{0.1cm}/\hspace{0.1cm}1.4~ & ~1000\hspace{0.1cm}/\hspace{0.1cm}1.16\hspace{0.1cm}/\hspace{0.1cm}1.3~ \\
 NEEE  & ~10:14:28.12~ & ~$+$12:48:23.4~ & ~100\hspace{0.1cm}/\hspace{0.1cm}1.19\hspace{0.1cm}/\hspace{0.1cm}1.3~ & ~200\hspace{0.1cm}/\hspace{0.1cm}1.20\hspace{0.1cm}/\hspace{0.1cm}1.5~ & ~1000\hspace{0.1cm}/\hspace{0.1cm}1.24\hspace{0.1cm}/\hspace{0.1cm}1.8~ \\
 NEEEE & ~10:16:28.12~ & ~$+$12:48:23.4~ & ~100\hspace{0.1cm}/\hspace{0.1cm}1.27\hspace{0.1cm}/\hspace{0.1cm}2.5~ & ~200\hspace{0.1cm}/\hspace{0.1cm}1.29\hspace{0.1cm}/\hspace{0.1cm}2.2~ & ~1000\hspace{0.1cm}/\hspace{0.1cm}1.34\hspace{0.1cm}/\hspace{0.1cm}1.8~ \\
 WWW   & ~10:02:30.17~ & ~$+$12:18:23.4~ & ~100\hspace{0.1cm}/\hspace{0.1cm}1.76\hspace{0.1cm}/\hspace{0.1cm}1.5~ & ~200\hspace{0.1cm}/\hspace{0.1cm}1.81\hspace{0.1cm}/\hspace{0.1cm}1.5~ & ~1000\hspace{0.1cm}/\hspace{0.1cm}1.95\hspace{0.1cm}/\hspace{0.1cm}1.4~ \\
 SWW   & ~10:04:28.12~ & ~$+$11:48:23.4~ & ~100\hspace{0.1cm}/\hspace{0.1cm}1.26\hspace{0.1cm}/\hspace{0.1cm}1.6~ & ~200\hspace{0.1cm}/\hspace{0.1cm}1.27\hspace{0.1cm}/\hspace{0.1cm}1.9~ & ~1000\hspace{0.1cm}/\hspace{0.1cm}1.32\hspace{0.1cm}/\hspace{0.1cm}1.6~ \\
 SWWW  & ~10:02:28.12~ & ~$+$11:48:23.4~ & ~100\hspace{0.1cm}/\hspace{0.1cm}1.48\hspace{0.1cm}/\hspace{0.1cm}1.8~ & ~200\hspace{0.1cm}/\hspace{0.1cm}1.51\hspace{0.1cm}/\hspace{0.1cm}1.8~ & ~1000\hspace{0.1cm}/\hspace{0.1cm}1.59\hspace{0.1cm}/\hspace{0.1cm}1.4~ 
\enddata
\end{deluxetable}

\begin{deluxetable}{cc}
\tablewidth{0pt}
\tablecaption{ Number of color-color selected giants in RGB bounding boxes offset along brighter magnitudes. \label{t:halogiantcount}}
\tablehead{ \colhead{$\Delta M$} & \colhead{Number of Giants} }
\startdata
 \phantom{$-$}0.0 &          1196\tablenotemark{a}\\
 $-0.5$           &       \phn709\tablenotemark{a} \\
 $-1.0$           &       \phn253\tablenotemark{a} \\
 $-1.5$           &    \phn\phn27\tablenotemark{a} \\
\hline
 $-2.0$           & \phn\phn\phn5 \\
 $-2.5$           & \phn\phn\phn2 \\
 $-3.0$           & \phn\phn\phn1 \\
 $-3.5$           & \phn\phn\phn5 \\
 $-4.0$           & \phn\phn\phn4 \\
 $-4.5$           & \phn\phn\phn4 \\
\hline
 $-5.0$           & \phn\phn\phn0\tablenotemark{b} \\
 $-5.5$           & \phn\phn\phn0\tablenotemark{b} \\
\enddata
\tablenotetext{a}{Sample contains stars from the Leo~I system.}
\tablenotetext{b}{Sample has an incomplete sampling of the CMD at the bright end.}
\end{deluxetable}

\begin{deluxetable}{lccccc}
\tablewidth{0pt}
\tablecaption{ Fit Parameters for Leo~I.\label{t:structuralparams}}
\tablehead{ \colhead{}          & \colhead{$r_{c}$}  & \colhead{$r_{t}$}  & \colhead{$\epsilon$} & \colhead{P.A.\tablenotemark{a}} & \colhead{} \\
            \colhead{Reference} & \colhead{(arcmin)} & \colhead{(arcmin)} & \colhead{($=1-b/a$)} & \colhead{(deg)}                 & \colhead{$\nu$\tablenotemark{b}} }
\startdata
This study (King) & $5.4\pm 0.4$ & $13.4\pm 0.7$ & $0.37\pm 0.02$ & $84\pm 2$ & \nodata \\
This study (PLC)  & $7.3\pm 0.8$ & \nodata       & $0.37\pm 0.02$ & $85\pm 2$ & $4.8\pm 0.6$ \\
\tableline
Hodge 1963        & \nodata      & $14.3\pm 1.0$ & $0.31\pm 0.07$ & \nodata   & \nodata \\
Hodge 1971        & $4.5       $ & $13.9\pm 0.5$ & $0.31\pm 0.07$ & \nodata   & \nodata \\
IH95              & $3.3\pm 0.3$ & $12.6\pm 1.5$ & $0.21\pm 0.03$ & $79\pm 3$ & \nodata
\enddata
\tablenotetext{a}{Major-axis position angle measuring from North=0\arcdeg to East=90\arcdeg.}
\tablenotetext{b}{See equation (1) of \citet{Kleyna98}.}
\end{deluxetable}

\begin{deluxetable}{lccccccrrrcc}
\tablewidth{0pt}
\tabletypesize{\scriptsize}
\tablecaption{ Observed Radial Velocities of Stars\label{t:keckspec}}
\tablehead{ \colhead{Slit} & \colhead{}   & \colhead{RA}        & \colhead{Dec}       & \colhead{}        & \colhead{}                           & \colhead{}                            & \colhead{$v_{hel}$\tablenotemark{a}}    & \colhead{$v_{GSR}$\tablenotemark{b}}    & \colhead{Tell. Off.\tablenotemark{c}}   & \colhead{}                     & \colhead{} \\
            \colhead{No. } & \colhead{ID} & \colhead{(J2000.0)} & \colhead{(J2000.0)} & \colhead{$M_{0}$} & \colhead{$\!\!\!(M$-$T_{2})_{0}\!\!\!$} & \colhead{$\!\!\!(M$-$D51)_{0}\!\!\!$} & \colhead{$\!\!\!$(km s$^{-1}$)$\!\!\!$} & \colhead{$\!\!\!$(km s$^{-1}$)$\!\!\!$} & \colhead{$\!\!\!$(km s$^{-1}$)$\!\!\!$} & \colhead{CCP\tablenotemark{d}} & \colhead{$Q$\tablenotemark{e}}
          }
\startdata
KD1-00 & 61782\tablenotemark{f} & 10:07:33.33 & $+$12:26:20.1 & 18.69 & 1.88 &           $-0.23$ & $  0.5$ & $-106.6$ & $-41.4$ & 0.59 &  5 \\
KD1-01 & 61736\tablenotemark{f} & 10:07:40.39 & $+$12:26:27.7 & 18.17 & 1.65 &           $-0.20$ & $ 39.3$ & $ -67.8$ & $-35.9$ & 0.40 &  6 \\
KD1-02 & 72520\tablenotemark{f} & 10:07:30.84 & $+$12:13:41.8 & 17.99 & 0.79 & \phantom{$-$}0.03 & $158.6$ & $  50.7$ & $-17.8$ & 0.39 &  4 \\
KD1-03 & 72418\tablenotemark{f} & 10:07:44.50 & $+$12:12:36.8 & 16.44 & 0.81 & \phantom{$-$}0.04 & $-13.1$ & $-121.0$ & $-18.9$ & 0.50 &  5 \\
KD1-04 & 61774                  & 10:07:34.30 & $+$12:25:09.4 & 18.93 & 2.90 &  	 $-0.04$ & $ 62.5$ & $ -44.7$ & $-16.0$ & 0.46 &  5 \\
KD1-05 & 61788                  & 10:07:31.61 & $+$12:23:30.4 & 17.56 & 1.46 &  	 $-0.17$ & $ 26.2$ & $ -81.1$ & $-23.5$ & 0.59 &  5 \\
KD1-06 & 61770                  & 10:07:34.86 & $+$12:25:15.0 & 18.77 & 2.34 &  	 $-0.17$ & $ 12.4$ & $ -94.8$ & $-22.8$ & 0.57 &  6 \\
KD1-07 & 61785                  & 10:07:32.43 & $+$12:24:50.0 & 19.84 & 3.14 &  	 $-0.04$ & $ 10.9$ & $ -96.3$ & $-30.7$ & 0.34 &  4 \\
KD1-08 & 72466                  & 10:07:38.44 & $+$12:13:33.4 & 20.00 & 2.84 &  	 $-0.02$ & $ 24.0$ & $ -83.9$ & $-14.5$ & 0.53 &  6 \\
KD1-09 & 61692                  & 10:07:45.33 & $+$12:23:47.6 & 20.56 & 2.96 &  	 $-0.08$ & $  4.0$ & $-103.3$ & $-23.9$ & 0.50 &  6 \\
KD1-10 & 72449                  & 10:07:40.94 & $+$12:13:03.6 & 20.11 & 1.82 & \phantom{$-$}0.02 & $288.3$ & $ 180.4$ & $ -8.0$ & 0.91 &  6 \\
KD1-11 & 72472                  & 10:07:37.98 & $+$12:17:44.3 & 20.19 & 1.65 & \phantom{$-$}0.01 & $293.5$ & $ 185.9$ & $-14.0$ & 0.87 &  6 \\
KD1-12 & 72456                  & 10:07:39.52 & $+$12:16:54.6 & 20.24 & 1.58 & \phantom{$-$}0.04 & $303.1$ & $ 195.4$ & $ -9.5$ & 0.64 &  7 \\
KD1-13 & 61733                  & 10:07:40.43 & $+$12:20:23.4 & 20.19 & 1.50 & \phantom{$-$}0.03 & $288.1$ & $ 180.6$ & $-11.2$ & 0.80 &  6 \\
KD1-14 & 72526                  & 10:07:30.06 & $+$12:16:21.6 & 20.41 & 1.48 & \phantom{$-$}0.00 & $290.1$ & $ 182.4$ & $-10.8$ & 0.82 &  6 \\
KD1-15 & 72493                  & 10:07:34.86 & $+$12:12:43.9 & 20.57 & 1.51 & \phantom{$-$}0.00 & $291.8$ & $ 183.9$ & $-10.4$ & 0.75 &  6 \\
KD1-16 & 72463                  & 10:07:38.60 & $+$12:11:28.7 & 20.55 & 1.47 & \phantom{$-$}0.04 & $278.9$ & $ 170.9$ & $ -9.0$ & 0.74 &  6 \\
KD1-17 & 61790                  & 10:07:30.46 & $+$12:21:32.6 & 20.65 & 1.53 & \phantom{$-$}0.02 & $302.0$ & $ 194.6$ & $-12.9$ & 0.53 &  6 \\
KD1-18 & 61757                  & 10:07:37.01 & $+$12:22:42.9 & 20.77 & 1.48 & \phantom{$-$}0.01 & $289.1$ & $ 181.8$ & $-13.4$ & 0.65 &  7 \\
KD1-20 & 72407                  & 10:07:45.47 & $+$12:15:32.7 & 21.31 & 1.30 & \phantom{$-$}0.02 & $274.8$ & $ 167.1$ & $ -5.7$ & 0.49 &  7 \\
KD1-21 & 72432                  & 10:07:43.00 & $+$12:16:27.9 & 21.37 & 1.33 & \phantom{$-$}0.05 & $293.4$ & $ 185.7$ & $ -3.7$ & 0.49 &  6 \\
KD1-22 & 61783                  & 10:07:32.88 & $+$12:21:01.9 & 21.42 & 1.35 & \phantom{$-$}0.01 & $294.6$ & $ 187.1$ & $-14.6$ & 0.52 &  6 \\
KD1-23 & 61755                  & 10:07:37.37 & $+$12:22:27.3 & 21.41 & 1.32 & \phantom{$-$}0.07 & $313.9$ & $ 206.5$ & $-16.2$ & 0.36 &  6 \\
KD1-24 & 61753                  & 10:07:37.44 & $+$12:19:18.2 & 21.49 & 1.59 & \phantom{$-$}0.22 & $283.3$ & $ 175.8$ & $ +3.8$ & 0.35 &  6 \\
KD1-25 & 61746                  & 10:07:38.47 & $+$12:19:27.3 & 21.93 & 1.26 & \phantom{$-$}0.11 & $288.1$ & $ 180.6$ & $-11.3$ & 0.28 &  4 \\
KD1-26 & 72464                  & 10:07:38.53 & $+$12:13:11.0 & 21.56 & 1.21 & \phantom{$-$}0.06 & $290.9$ & $ 183.0$ & $ -9.0$ & 0.34 &  6 \\
KD1-27 & 72525                  & 10:07:30.36 & $+$12:11:44.5 & 21.62 & 1.27 & \phantom{$-$}0.06 & $314.8$ & $ 206.8$ & $ -8.2$ & 0.25 &  5 \\
KD1-28 & 61769                  & 10:07:34.97 & $+$12:24:23.2 & 21.40 & 3.42 & \phantom{$-$}0.07 & $ 53.2$ & $ -54.1$ & $-14.5$ & 0.42 &  6 \\
KD1-29 & 72505                  & 10:07:33.39 & $+$12:11:00.4 & 18.90 & 0.85 & \phantom{$-$}0.03 & $178.3$ & $  70.3$ & $ -9.7$ & 0.42 &  5 \\
KD1-30 & 61743                  & 10:07:39.55 & $+$12:26:39.4 & 18.98 & 0.81 & \phantom{$-$}0.03 & $129.7$ & $  22.6$ & $-14.8$ & 0.28 &  4 \\
KD1-32 & 72490                  & 10:07:35.03 & $+$12:15:48.2 & 19.88 & 1.45 &           $-0.14$ & $ 52.7$ & $ -55.0$ & $ -6.3$ & 0.57 &  6 \\
KD1-33 & 61707                  & 10:07:43.58 & $+$12:21:21.2 & 19.41 & 0.91 & \phantom{$-$}0.06 & $219.1$ & $ 111.7$ & $-14.7$ & 0.18 &  4 \\
KD1-34 & 61694                  & 10:07:44.99 & $+$12:19:52.8 & 21.70 & 3.14 & \phantom{$-$}0.03 & $ 24.2$ & $ -83.3$ & $ -9.1$ & 0.47 &  6 \\
KD1-36 & 61766                  & 10:07:35.12 & $+$12:22:05.2 & 20.88 & 2.10 &           $-0.22$ & $ 68.6$ & $ -38.8$ & $ -5.9$ & 0.57 &  6 \\
KD1-37 & 61772                  & 10:07:34.39 & $+$12:18:57.0 & 20.78 & 1.76 &           $-0.24$ & $ 74.9$ & $ -32.7$ & $ +4.1$ & 0.48 &  6 \\
KD1-38 & 61775                  & 10:07:34.18 & $+$12:22:52.2 & 20.72 & 1.29 &           $-0.06$ & $-52.6$ & $-159.9$ & $-21.0$ & 0.48 &  4 \\
KD1-39 & 61724                  & 10:07:41.49 & $+$12:23:16.0 & 21.44 & 1.85 &           $-0.22$ & $ 24.9$ & $ -82.4$ & $-19.2$ & 0.44 &  5 \\
KD1-40 & 72531                  & 10:07:29.23 & $+$12:11:59.5 & 20.38 & 0.76 & \phantom{$-$}0.05 & $318.1$ & $ 210.1$ & $-10.3$ & 0.18 &  4 \\
KD1-42 & 72439                  & 10:07:42.05 & $+$12:11:20.8 & 21.21 & 1.27 & \phantom{$-$}0.00 & $307.5$ & $ 199.5$ & $-13.7$ & 0.35 &  5 \\
KD1-45 & 72508                  & 10:07:32.88 & $+$12:15:58.1 & 21.43 & 1.22 &           $-0.01$ & $282.7$ & $ 175.0$ & $ -1.1$ & 0.59 &  7 \\
KD1-46 & 72455                  & 10:07:39.82 & $+$12:17:55.1 & 21.53 & 1.31 &           $-0.04$ & $316.1$ & $ 208.5$ & $ +6.7$ & 0.31 &  4 \\
KD1-50 & 61792                  & 10:07:30.07 & $+$12:22:20.6 & 21.88 & 1.13 & \phantom{$-$}0.03 & $305.3$ & $ 197.9$ & $-11.5$ & 0.34 &  6 \\
\hline
KD2-03 & 21673                  & 10:08:36.93 & $+$12:20:11.5 & 19.47 & 1.99 &           $-0.02$ & $299.0$ & $ 191.7$ & $ -2.8$ & 1.08 &  7 \\
KD2-04 & 20868                  & 10:08:44.71 & $+$12:20:20.9 & 19.79 & 1.72 & \phantom{$-$}0.04 & $292.2$ & $ 185.0$ & $ -1.9$ & 0.93 &  6 \\
KD2-05 & 23429                  & 10:08:25.15 & $+$12:20:07.5 & 19.85 & 1.78 &           $-0.02$ & $294.0$ & $ 186.7$ & $ -4.9$ & 1.00 &  6 \\
KD2-06 & 31007                  & 10:08:40.95 & $+$12:18:06.7 & 19.94 & 1.83 & \phantom{$-$}0.03 & $293.3$ & $ 185.9$ & $ -1.9$ & 0.88 &  7 \\
KD2-07 & 20898                  & 10:08:44.30 & $+$12:19:56.3 & 19.89 & 1.78 & \phantom{$-$}0.03 & $285.3$ & $ 178.0$ & $ -4.1$ & 0.93 &  7 \\
KD2-08 & 20945                  & 10:08:43.74 & $+$12:19:57.9 & 19.84 & 1.70 & \phantom{$-$}0.03 & $296.9$ & $ 189.6$ & $ -3.8$ & 0.98 &  6 \\
KD2-09 & 23059                  & 10:08:27.55 & $+$12:20:20.9 & 19.81 & 1.65 &           $-0.02$ & $276.9$ & $ 169.6$ & $ -1.5$ & 0.80 &  7 \\
KD2-11 & 20198                  & 10:09:07.46 & $+$12:18:38.7 & 19.99 & 1.77 & \phantom{$-$}0.01 & $291.2$ & $ 183.9$ & $ -5.0$ & 0.88 &  6 \\
KD2-12 & 20231                  & 10:09:05.15 & $+$12:20:14.4 & 20.05 & 1.73 & \phantom{$-$}0.01 & $288.8$ & $ 181.6$ & $ -6.9$ & 0.94 &  6 \\
KD2-13 & 20345                  & 10:08:58.71 & $+$12:18:36.4 & 20.07 & 1.70 & \phantom{$-$}0.00 & $284.5$ & $ 177.2$ & $ -4.8$ & 0.88 &  7 \\
KD2-14 & 20982                  & 10:08:43.32 & $+$12:21:23.6 & 20.10 & 1.72 & \phantom{$-$}0.05 & $286.0$ & $ 178.8$ & $ -5.3$ & 0.80 &  6 \\
KD2-15 & 22396                  & 10:08:31.81 & $+$12:19:56.0 & 20.00 & 1.58 &           $-0.02$ & $287.5$ & $ 180.2$ & $ +0.6$ & 0.70 &  6 \\
KD2-16 & 20129                  & 10:09:14.89 & $+$12:18:35.0 & 20.10 & 1.65 & \phantom{$-$}0.01 & $275.5$ & $ 168.3$ & $ -8.3$ & 0.74 &  7 \\
KD2-17 & 20248                  & 10:09:04.17 & $+$12:20:46.0 & 20.13 & 1.68 & \phantom{$-$}0.03 & $292.0$ & $ 184.8$ & $-11.7$ & 0.93 &  6 \\
KD2-18 & 23131                  & 10:08:27.15 & $+$12:20:06.3 & 20.09 & 1.63 & \phantom{$-$}0.05 & $280.1$ & $ 172.8$ & $ -5.4$ & 0.86 &  6 \\
KD2-19 & 31135                  & 10:08:39.39 & $+$12:17:23.8 & 20.13 & 1.67 & \phantom{$-$}0.09 & $286.7$ & $ 179.3$ & $ +1.2$ & 0.81 &  6 \\
KD2-20 & 70295                  & 10:08:22.67 & $+$12:17:21.3 & 20.21 & 1.75 & \phantom{$-$}0.00 & $287.2$ & $ 179.7$ & $ -6.4$ & 0.42 &  5 \\
KD2-22 & 21617                  & 10:08:37.29 & $+$12:20:12.0 & 20.20 & 1.68 & \phantom{$-$}0.01 & $291.9$ & $ 184.6$ & $ -1.5$ & 0.90 &  6 \\
KD2-23 & 60019                  & 10:08:24.25 & $+$12:18:53.5 & 20.20 & 1.65 & \phantom{$-$}0.03 & $284.0$ & $ 176.6$ & $ +4.2$ & 0.84 &  6 \\
KD2-24 & 30123                  & 10:09:13.77 & $+$12:17:23.0 & 20.28 & 1.71 & \phantom{$-$}0.04 & $296.9$ & $ 189.6$ & $ -6.1$ & 0.75 &  6 \\
KD2-25 & 30522                  & 10:08:49.87 & $+$12:18:01.2 & 20.21 & 1.61 & \phantom{$-$}0.04 & $291.4$ & $ 184.0$ & $ +0.9$ & 0.79 &  6 \\
KD2-26 & 60310                  & 10:08:21.89 & $+$12:19:33.4 & 20.25 & 1.63 & \phantom{$-$}0.02 & $280.4$ & $ 173.0$ & $ +2.4$ & 0.79 &  6 \\
KD2-27 & 30102                  & 10:09:15.88 & $+$12:17:18.5 & 20.39 & 1.74 & \phantom{$-$}0.03 & $284.1$ & $ 176.8$ & $ -3.8$ & 0.80 &  6 \\
KD2-28 & 21961                  & 10:08:34.69 & $+$12:19:11.4 & 20.27 & 1.56 & \phantom{$-$}0.03 & $283.6$ & $ 176.3$ & $ -1.1$ & 0.82 &  6 \\
KD2-29 & 21005                  & 10:08:42.96 & $+$12:19:55.7 & 20.30 & 1.58 & \phantom{$-$}0.02 & $272.2$ & $ 164.9$ & $ -3.6$ & 0.79 &  6 \\
KD2-30 & 32615                  & 10:08:27.92 & $+$12:17:44.7 & 20.23 & 1.45 & \phantom{$-$}0.06 & $279.6$ & $ 172.1$ & $ +2.1$ & 0.82 &  6 \\
KD2-31 & 30710                  & 10:08:45.11 & $+$12:17:33.6 & 20.45 & 1.63 & \phantom{$-$}0.02 & $279.4$ & $ 172.0$ & $ -0.9$ & 0.68 &  6 \\
KD2-32 & 20514                  & 10:08:52.57 & $+$12:19:05.9 & 20.45 & 1.58 & \phantom{$-$}0.06 & $302.5$ & $ 195.2$ & $ -4.9$ & 0.78 &  6 \\
KD2-34 & 23305                  & 10:08:25.92 & $+$12:20:57.8 & 20.33 & 1.41 & \phantom{$-$}0.03 & $280.2$ & $ 172.9$ & $ -6.7$ & 0.81 &  6 \\
KD2-35 & 23364                  & 10:08:25.55 & $+$12:18:46.6 & 20.36 & 1.44 & \phantom{$-$}0.05 & $276.8$ & $ 169.4$ & $ -3.5$ & 0.72 &  6 \\
KD2-36 & 30941                  & 10:08:41.79 & $+$12:17:15.3 & 20.44 & 1.48 & \phantom{$-$}0.03 & $284.2$ & $ 176.8$ & $ -4.2$ & 0.69 &  7 \\
KD2-37 & 60143                  & 10:08:23.12 & $+$12:18:55.3 & 20.42 & 1.45 & \phantom{$-$}0.02 & $301.3$ & $ 193.9$ & $ +7.3$ & 0.65 &  6 \\
KD2-39 & 32487                  & 10:08:28.73 & $+$12:17:39.1 & 20.38 & 1.40 & \phantom{$-$}0.09 & $297.1$ & $ 189.6$ & $ -2.3$ & 0.73 &  6 \\
KD2-40 & 30156                  & 10:09:10.01 & $+$12:18:05.2 & 20.46 & 1.46 & \phantom{$-$}0.08 & $291.1$ & $ 183.8$ & $ -5.5$ & 0.61 &  6 \\
KD2-41 & 20491                  & 10:08:53.21 & $+$12:20:40.6 & 20.54 & 1.51 & \phantom{$-$}0.04 & $279.4$ & $ 172.2$ & $ -4.9$ & 0.81 &  7 \\
KD2-42 & 22143                  & 10:08:33.48 & $+$12:20:34.5 & 20.53 & 1.49 & \phantom{$-$}0.03 & $288.8$ & $ 181.5$ & $ -4.0$ & 0.82 &  6 \\
KD2-43 & 22640                  & 10:08:30.22 & $+$12:20:03.0 & 20.46 & 1.42 & \phantom{$-$}0.01 & $295.6$ & $ 188.3$ & $ -4.0$ & 0.77 &  6 \\
KD2-44 & 21074                  & 10:08:42.15 & $+$12:18:59.7 & 20.53 & 1.46 & \phantom{$-$}0.08 & $276.7$ & $ 169.4$ & $ -2.8$ & 0.60 &  6 \\
KD2-45 & 30471                  & 10:08:51.54 & $+$12:18:14.0 & 20.55 & 1.47 & \phantom{$-$}0.07 & $283.9$ & $ 176.6$ & $ +5.5$ & 0.71 &  6 \\
KD2-46 & 21281                  & 10:08:40.16 & $+$12:18:52.3 & 20.44 & 1.34 & \phantom{$-$}0.10 & $290.7$ & $ 183.4$ & $ -2.7$ & 0.71 &  6 \\
KD2-47 & 22567                  & 10:08:30.66 & $+$12:19:29.9 & 20.61 & 1.51 & \phantom{$-$}0.07 & $297.3$ & $ 190.0$ & $ -4.9$ & 0.59 &  7 \\
KD2-48 & 30891                  & 10:08:42.52 & $+$12:18:04.9 & 20.60 & 1.46 & \phantom{$-$}0.06 & $278.4$ & $ 171.0$ & $ +0.4$ & 0.68 &  7 \\
KD2-49 & 31576                  & 10:08:35.12 & $+$12:18:01.6 & 20.63 & 1.49 & \phantom{$-$}0.12 & $294.2$ & $ 186.8$ & $ -0.8$ & 0.75 &  6 \\
KD2-50 & 22692                  & 10:08:29.86 & $+$12:19:14.6 & 20.55 & 1.40 & \phantom{$-$}0.00 & $299.2$ & $ 191.8$ & $ +1.0$ & 0.79 &  6 \\
KD2-51 & 20691                  & 10:08:48.34 & $+$12:20:19.8 & 20.65 & 1.46 & \phantom{$-$}0.09 & $298.4$ & $ 191.2$ & $ -5.6$ & 0.45 &  6 \\
KD2-52 & 20592                  & 10:08:50.68 & $+$12:19:45.5 & 20.69 & 1.48 &           $-0.04$ & $286.4$ & $ 179.1$ & $ -4.0$ & 0.76 &  7 \\
KD2-53 & 21427                  & 10:08:38.99 & $+$12:18:56.7 & 20.54 & 1.33 & \phantom{$-$}0.01 & $294.4$ & $ 187.1$ & $ -3.4$ & 0.77 &  6 \\
KD2-54 & 20728                  & 10:08:47.38 & $+$12:19:05.7 & 20.58 & 1.37 & \phantom{$-$}0.03 & $283.0$ & $ 175.7$ & $ -5.6$ & 0.48 &  6 \\
KD2-55 & 30538                  & 10:08:49.45 & $+$12:17:14.2 & 20.68 & 1.46 & \phantom{$-$}0.07 & $292.6$ & $ 185.2$ & $ +3.3$ & 0.68 &  6 \\
KD2-56 & 30320                  & 10:08:57.42 & $+$12:17:20.3 & 20.72 & 1.49 & \phantom{$-$}0.06 & $304.7$ & $ 197.3$ & $ -9.3$ & 0.68 &  6 \\
KD2-57 & 22802                  & 10:08:29.10 & $+$12:19:23.8 & 20.76 & 1.43 & \phantom{$-$}0.02 & $290.3$ & $ 182.9$ & $ -4.5$ & 0.64 &  6 \\
KD2-58 & 22939                  & 10:08:28.28 & $+$12:20:17.2 & 20.80 & 1.46 & \phantom{$-$}0.05 & $288.8$ & $ 181.5$ & $ -5.0$ & 0.69 &  6 \\
KD2-59 & 70126                  & 10:08:23.68 & $+$12:17:20.3 & 20.82 & 1.45 &           $-0.02$ & $276.4$ & $ 168.9$ & $ -7.6$ & 0.30 &  7 \\
KD2-60 & 20354                  & 10:08:58.25 & $+$12:19:59.0 & 20.84 & 1.41 & \phantom{$-$}0.04 & $284.5$ & $ 177.3$ & $ -1.9$ & 0.56 &  6 \\
KD2-61 & 21243                  & 10:08:40.58 & $+$12:19:40.6 & 20.80 & 1.36 & \phantom{$-$}0.03 & $282.7$ & $ 175.4$ & $ -2.2$ & 0.57 &  6 \\
KD2-62 & 21147                  & 10:08:41.40 & $+$12:20:15.7 & 20.86 & 1.35 & \phantom{$-$}0.08 & $285.1$ & $ 177.8$ & $ +0.6$ & 0.50 &  5 \\
KD2-63 & 30033                  & 10:09:25.97 & $+$12:18:08.4 & 20.98 & 1.42 &           $-0.02$ & $287.1$ & $ 179.9$ & $ -3.2$ & 0.52 &  6 \\
KD2-64 & 20304                  & 10:09:00.64 & $+$12:19:03.5 & 21.04 & 1.47 & \phantom{$-$}0.04 & $298.4$ & $ 191.1$ & $ -3.5$ & 0.72 &  6 \\
KD2-65 & 21334                  & 10:08:39.76 & $+$12:19:17.1 & 20.91 & 1.34 & \phantom{$-$}0.02 & $286.8$ & $ 179.5$ & $ -1.0$ & 0.59 &  6 \\
KD2-66 & 22211                  & 10:08:33.04 & $+$12:20:41.7 & 20.98 & 1.40 & \phantom{$-$}0.08 & $283.5$ & $ 176.2$ & $ -6.1$ & 0.68 &  6 \\
KD2-67 & 30456                  & 10:08:52.02 & $+$12:17:57.0 & 20.98 & 1.37 & \phantom{$-$}0.01 & $288.2$ & $ 180.8$ & $ +1.4$ & 0.57 &  6 \\
KD2-68 & 20817                  & 10:08:45.52 & $+$12:20:01.9 & 20.98 & 1.34 & \phantom{$-$}0.06 & $290.1$ & $ 182.8$ & $ -4.7$ & 0.60 &  7 \\
KD2-69 & 30514                  & 10:08:50.25 & $+$12:17:23.7 & 20.96 & 1.28 & \phantom{$-$}0.04 & $280.0$ & $ 172.6$ & $ +0.6$ & 0.57 &  6 \\
KD2-70 & 20450                  & 10:08:54.48 & $+$12:20:13.6 & 21.06 & 1.37 & \phantom{$-$}0.04 & $276.3$ & $ 169.1$ & $ -3.0$ & 0.65 &  6 \\
KD2-71 & 20575                  & 10:08:51.05 & $+$12:18:52.6 & 21.03 & 1.34 & \phantom{$-$}0.02 & $290.4$ & $ 183.1$ & $ +0.8$ & 0.48 &  7 \\
KD2-72 & 22059                  & 10:08:33.97 & $+$12:19:23.0 & 21.04 & 1.33 & \phantom{$-$}0.02 & $304.8$ & $ 197.5$ & $ -8.3$ & 0.56 &  6 \\
KD2-73 & 22464                  & 10:08:31.42 & $+$12:21:53.0 & 21.02 & 1.30 & \phantom{$-$}0.04 & $308.5$ & $ 201.3$ & $ -2.1$ & 0.58 &  6 \\
KD2-74 & 31529                  & 10:08:35.57 & $+$12:17:47.9 & 21.09 & 1.36 & \phantom{$-$}0.11 & $273.6$ & $ 166.2$ & $ +2.1$ & 0.49 &  5 \\
KD2-75 & 22516                  & 10:08:31.05 & $+$12:19:46.4 & 21.14 & 1.35 & \phantom{$-$}0.04 & $298.1$ & $ 190.8$ & $ -5.7$ & 0.52 &  6 \\
KD2-76 & 30624                  & 10:08:46.92 & $+$12:17:44.6 & 21.14 & 1.31 & \phantom{$-$}0.04 & $283.3$ & $ 175.9$ & $ -0.9$ & 0.57 &  6 \\
KD2-77 & 20163                  & 10:09:10.54 & $+$12:18:44.2 & 21.25 & 1.34 & \phantom{$-$}0.04 & $284.3$ & $ 177.1$ & $ -6.0$ & 0.48 &  7 \\
KD2-78 & 21487                  & 10:08:38.49 & $+$12:18:56.8 & 21.26 & 1.34 & \phantom{$-$}0.05 & $313.8$ & $ 206.5$ & $ -4.0$ & 0.36 &  4 \\
KD2-79 & 22001                  & 10:08:34.33 & $+$12:19:15.1 & 21.18 & 1.25 & \phantom{$-$}0.05 & $322.9$ & $ 215.6$ & $ -7.6$ & 0.40 &  4 \\
KD2-81 & 20659                  & 10:08:48.95 & $+$12:19:22.0 & 21.34 & 1.35 & \phantom{$-$}0.04 & $316.3$ & $ 209.0$ & $ -5.3$ & 0.37 &  5 \\
KD2-82 & 20783                  & 10:08:46.14 & $+$12:20:34.3 & 21.34 & 1.34 & \phantom{$-$}0.06 & $295.8$ & $ 188.6$ & $ -6.3$ & 0.51 &  6 \\
KD2-83 & 32861                  & 10:08:26.35 & $+$12:17:59.8 & 21.23 & 1.21 & \phantom{$-$}0.13 & $276.2$ & $ 168.8$ & $ -4.4$ & 0.41 &  5 \\
KD2-84 & 20277                  & 10:09:02.40 & $+$12:19:23.9 & 21.35 & 1.31 & \phantom{$-$}0.04 & $283.5$ & $ 176.3$ & $ -4.7$ & 0.51 &  6 \\
KD2-85 & 30337                  & 10:08:56.69 & $+$12:17:32.3 & 21.38 & 1.33 & \phantom{$-$}0.09 & $294.0$ & $ 186.6$ & $ +1.7$ & 0.40 &  6 \\
KD2-86 & 31953                  & 10:08:32.18 & $+$12:18:19.0 & 21.47 & 1.37 & \phantom{$-$}0.11 & $272.7$ & $ 165.3$ & $ +1.5$ & 0.32 &  6 \\
KD2-87 & 21722                  & 10:08:36.57 & $+$12:19:02.4 & 21.44 & 1.33 & \phantom{$-$}0.04 & $288.9$ & $ 181.6$ & $ +1.0$ & 0.47 &  6 \\
KD2-89 & 30420                  & 10:08:53.58 & $+$12:18:05.3 & 21.48 & 1.34 & \phantom{$-$}0.08 & $295.7$ & $ 188.4$ & $ +1.6$ & 0.31 &  5 \\
KD2-90 & 21785                  & 10:08:36.04 & $+$12:19:28.3 & 21.45 & 1.28 & \phantom{$-$}0.06 & $294.4$ & $ 187.1$ & $ -1.3$ & 0.51 &  6 \\
KD2-91 & 30261                  & 10:09:01.03 & $+$12:17:26.0 & 21.63 & 1.38 & \phantom{$-$}0.10 & $290.9$ & $ 183.5$ & $ -0.5$ & 0.28 &  5
\enddata
\tablenotetext{a}{Helocentric radial velocities corrected for the Telurric offsets as described in text.}
\tablenotetext{b}{Radial velocities in Galactocentric standard of rest.}
\tablenotetext{c}{Telluric offsets -- see text.}
\tablenotetext{d}{Cross-correlation peaks.}
\tablenotetext{e}{Quality indicator -- see text.}
\tablenotetext{f}{Alignment stars observed with 4\arcsec wide slits.  All other stars were observed with 1\arcsec wide slits.}
\end{deluxetable}

\begin{deluxetable}{ccccccccc}
\tablewidth{0pt}
\tablecaption{ Parameters and results for adopted models.\label{t:modelparams}}
\tablehead{ \colhead{}      & \colhead{$m_{i}$\tablenotemark{a}}  & \colhead{$r_{0}$\tablenotemark{b}} & \colhead{$r_{\rm peri}$\tablenotemark{c}} & \colhead{$r_{\rm apo}$\tablenotemark{d}} & \colhead{$P$\tablenotemark{e}} & \colhead{$T_{\rm peri}$\tablenotemark{f}} & \colhead{$d{\rm f}/d{\rm t}$\tablenotemark{g}} & \colhead{$m_{f}$\tablenotemark{h}} \\
            \colhead{Model} & \colhead{($10^{7} {\rm M}_{\sun}$)} & \colhead{(kpc)}                    & \colhead{(kpc)}                           & \colhead{(kpc)}                          & \colhead{(Gyr)}                & \colhead{(Gyr)}                           & \colhead{($10^{-2}$ Gyr$^{-1}$)}               & \colhead{($10^{7} {\rm M}_{\sun}$)} 
          }
\startdata
111 & 3.0 & 0.3 & 15 & 450 & 6.33 & 0.96 & 1.3 & 2.3 \\
117 & 4.0 & 0.3 & 10 & 450 & 6.05 & 0.95 & 2.2 & 3.4
\enddata
\tablenotetext{a}{Initial mass of model satellite.}
\tablenotetext{b}{Scale length parameter in equation (5).}
\tablenotetext{c}{Perigalactic distance.}
\tablenotetext{d}{Apogalactic distance.}
\tablenotetext{e}{Radial period.}
\tablenotetext{f}{Time since last closest approach.}
\tablenotetext{g}{Fractional mass-loss rate.}
\tablenotetext{h}{Final mass of model satellite.}
\end{deluxetable}

\clearpage

\clearpage

\clearpage

\clearpage

\clearpage

\clearpage

\clearpage

\clearpage

\clearpage

\clearpage

\clearpage

\clearpage

\clearpage

\clearpage

\clearpage

\clearpage

\clearpage

\clearpage
      
\clearpage

\clearpage

\clearpage

\clearpage

\clearpage

\clearpage

\clearpage

\clearpage

\end{document}